\documentclass[onecolumn]{mn2e}  
\usepackage{mnras_cite}  
 \usepackage[dvips]{graphicx}   
\usepackage{amsmath}
\newcommand{\nc}{\newcommand}    

\nc{\de}{\delta} 
\nc{\tISW}{\triangle_T^{ISW}}
\nc{\hn}{\hat{n}}
\nc{\bH}{\bar{H}} 
\nc{\Ol}{\Om_{\Lambda}} 
\nc{\ul}{\underline} \nc{\al}{\alpha} \nc{\g}{\gamma}
\nc{\Del}{\Delta} \nc{\e}{\textrm{e}} \nc{\eps}{\epsilon}
\nc{\lam}{\lambda} \nc{\Om}{\Omega} \nc{\Omm}{\Omega_m}
\nc{\Oml}{\Omega_\Lambda} \nc{\LCDM}{$\Lambda$CDM~} %\nc{\CM}{CCM}
\nc{\ve}{\varepsilon} \nc{\mn}{{\mu\nu}} \nc{\vp}{\varphi}

\def\gsim{\; \raise0.3ex\hbox{$>$\kern-0.75em
\raise-1.1ex\hbox{$\sim$}}\; }

\nc{\Section}[2]{\section{#2}\label{#1}}    
\nc{\Bibitem}[1]{\bibitem{#1}}    
\nc{\Label}[1]{\label{#1}}    
 
\nc{\beq}[1]{\begin{equation}\label{#1}}      
\nc{\eeq}{\end{equation}}
\nc{\hq}{\hat{q}}
\nc{\hw}{\widehat{w}}

\def\ben{\begin{enumerate}}
\def\een{\end{enumerate}}
\def\bi{\begin{itemize}}
\def\ei{\end{itemize}}
\def\ee{\end{equation}}
\def\bea{\begin{eqnarray}}
\def\eea{\end{eqnarray}}

\nc{\Mpc}{Mpc/h}    
\nc{\vev}[1]{\langle #1 \rangle}    
    
\def\ltsima{$\; \buildrel < \over \sim \;$}    
\def\gtsima{$\; \buildrel > \over \sim \;$}    
\def\simlt{\lower.5ex\hbox{\ltsima}}    
\def\simgt{\lower.5ex\hbox{\gtsima}}    
\nc{\w}{$w_2(\theta)$\ }    
\nc{\ie}{i.e.}     
\nc{\eg}{e.g.}

\input epsf

\usepackage{amsmath}
\usepackage{amssymb}
\def\xisp{\xi(\sigma, \pi)}
\def\xis{\xi(s)}
\def\xir{\xi(r)}
\def\xips{\xi(\pi,\sigma)}

\nc{\be}[1]{\begin{equation}\mbox{$\label{#1}$}
             
            \end{equation}
}

\begin{document}

\title{Clustering of luminous red galaxies I:\\
large scale redshift space distortions}

\author[Cabr\'e \& Gazta\~{n}aga]{Anna Cabr\'e, Enrique Gazta\~{n}aga\\ 
Institut de Ci\`encies de l'Espai, CSIC/IEEC, Campus UAB, F. de Ci\`encies, Torre C5 par-2, Barcelona 08193, Spain}

% e mail??

\twocolumn

\maketitle

\begin{abstract}
This is the first paper of a series where we study 
the clustering of LRG galaxies in the latest spectroscopic SDSS 
data release, DR6, which has 75000 LRG galaxies covering over 1 $Gpc^3/h^3$
at $0.15<z<0.47$.
Here we focus on modeling redshift space distortions in $\xips$,
the 2-point correlation in separate line-of-sight and 
perpendicular directions, on large scales.
% and  away from the line-of-sight. 
We use large mock simulations to study the validity of models and 
errors. We show that errors in the data are dominated by a shot-noise
term that is $40\%$ larger than the Poisson error commonly used.
We first use the normalized
quadrupole for the whole sample
(mean z=0.34) to estimate $\beta=f(\Omega_m)/b=0.34 \pm 0.03$, where
$f(\Omega_m)$ is the linear velocity growth factor
and $b$ is the linear bias parameter that relates galaxy to 
matter fluctuations on large scales. 
We next use the full $\xips$ plane to find $\Omega_{0m}= 0.245 \pm 0.020$ (h=0.72)
and the biased amplitude $b \sigma_8 = 1.56 \pm 0.09$. For standard gravity,
we can combine these measurements to break degeneracies 
and find $\sigma_8=0.85 \pm 0.06$, $b=1.85 \pm 0.25$
and $f(\Omega_m)=0.64 \pm 0.09$.
We present constraints for modified theories of gravity and find
that standard gravity is consistent with data as long as $0.80<\sigma_8<0.92$.
We also calculate the cross-correlation with WMAP5 and show how both
methods to measure the growth history are complementary to constrain
non-standard models of gravity. Finally, we show results for different 
redshift slices, including a prominent BAO peak in the monopole
at different redshifts.  The $\xips$ data on large scales
is shown to be in remarkable agreement with predictions 
and shows a characteristic large region of negative correlation
in the line of sight, a BAO ring and a prominent radial BAO peak. 
The significance of this is presented in paper IV of this series.
We include a study of possible systematic effects in our analysis to find
that these results are quite robust.

\end{abstract} %\pacs{PACS numbers: 98.70.Vc}

%\keywords{large-scale structure of Universe, cosmological parameters, cosmic microwave background, cosmology: observations}

\maketitle

%%%%%%%%%%%%%%%%%%%%%%%%%%%%%%%%%%%%%%%%%%%%%%%%%%%%%%%%%%%%%%%%%%%%%%%%

\section{Introduction} \label{sec:intro}

Galaxy clustering allows the study of different physical
phenomena at each scale. On large scales density fluctuations are
 small and can be modeled by linear theory to
constrain cosmological parameters. 
We need large surveys with galaxy positions
to do this.

The measured redshift distance $s$ of a galaxy differs from the
true radial distance $r$ by its peculiar velocity along the 
line-of-sight $v_r$. These
displacements lead to redshift distortions, with two important contributions.
The first, on large scale fluctuations, is caused by coherent bulk motion. We see
walls denser and voids bigger and emptier, with a squashing effect in the
2-point correlation function along the line-of-sight, known as 
the Kaiser effect \cite{kaiser}. At small scales, random velocities 
inside clusters and groups  of galaxies produce a radial
stretching pointed at the observer, known as fingers of God (FOG) \cite{jackson}.

Although such distortions complicate the interpretation of redshift maps as
positional maps, they have the advantage of bearing unique information about the
dynamics of galaxies. In particular, the amplitude of the quadrupole redshift
distortion on large scales yields a measure of the linear redshift distortion parameter $\beta$,
which is related to the matter density, and gives direct information on
the growth of the newtonian gravitational potential $f(\Omega)$ and
bias $b$, which describes how the number density of galaxies traces matter density (see Eq.(\ref{fb})).
We will also show how one can use the anisotropic redshift distortions, 
broken into line of sight and perpendicular directions, to find  separate
 measurements for $f(\Omega)$ and $b$.

In this work we use the most recent luminous red galaxies (LRGs) from the
spectroscopic SDSS public data release DR6 \cite{dr6}, and perform studies of
linear bias on large scales to obtain cosmological parameters. As we will see,
we can break the degeneracy between $bias$ and $\sigma_8$ present in the
correlation function thanks to redshift distortions anisotropies, and look at
the growth history and possible modifications of the gravity 
%(see \cite{linder1}, \cite{guzzo} and the recent \cite{huetsi2008}). 
(see Linder 2005, Guzzo et al 2008 and the recent Yamamoto et al 2008).

A dark energy dominated universe causes gravitational potentials to evolve, 
changing the energy of CMB photons passing through them. This is the so called  Integrated Sachs-Wolf (ISW) effect.
Here, we have also cross-correlated LRGs with WMAP \cite{wmappaper} in order to investigate this ISW effect,
reproducing a high signal as seen in recent studies  (see Gaztanaga etal 2006, Giannantonio et al 2008), 
as compared to current $\Lambda CDM$ model.  In this way,
we can also break the
$bias-\sigma_8$ degeneracy  and study the growth history using the
cross-correlation between temperature of CMB and galaxies $w_{TG}$, which give independent constraints on bias and $\sigma_8$ compared with 
those obtained from redshift space distortions.

%both from
%another point of view than the used in redshift distortions.

The same LRGs (but with reduced area) have been studied from different points of
view. 
%\cite{cosmoconstraints}
Tegmark et al (2006) have done an analysis of the power spectrum at
large scales to obtain cosmological parameters. 
%\cite{intermediate} 
Zehavi et al (2005) study LRGs
at intermediate scales (0.3 to 40Mpc/h), where they calculate the projected
correlation function, the monopole and real-space correlation function to study
mainly the linear high bias, the non-linear bias and the differences between
luminosities, remarking that there are differences from a power law for scales
smaller than 1Mpc/h.

While we were working on our results, a very interesting paper by
%\cite{okumura} 
Okumura et al (2008) showed the first application of the anisotropy in the 2-point
correlation function including the baryonic features, to obtain constraints on
cosmological parameters. They use the DR3 spectroscopic sample of LRGs to
calculate the 2-point correlation function with a different definition of what parallel and perpendicular directions mean (the line of sight is defined in their paper as the direction toward the galaxy 1, Matsubara 2004). They fit $\xisp$ for large scales using the
linear Kaiser model, from 40Mpc/h to 200Mpc/h, excluding the FOG zone. 
They
point out that a direct measurement of the growth function can be obtained from the distortion to the clustering 
signal caused by bulk flows (Kaiser anisotropy).
%, while $D_a(z)$ and $H(z)$ can be found from the anisotropic baryon acoustic peak, with improved LRG data. 
Although constraints are weak by now, the
fitting of all the anisotropic 2-point correlation function, including the
baryonic feature, would enable us to divide the effect of redshift distortions into
dynamical and geometrical components. The anisotropy due to geometric distortion
contributes to a better estimation of the equation of state of the dark energy.
Here we try to obtain new constraints, encouraged by the fact that DR6 has doubled the volume of DR3.

There has already  been significant progress in this direction using LRG.
%\cite{detection} 
Eisenstein et al (2005) detected the baryon acoustic peak in the 2-point correlation
function using LRGs. 
%\cite{hutsia,hutsib} 
H{\"u}tsi (2006 a,b) use LRGs to constrain cosmological
parameters in the power spectrum, including the baryonic peak. 
%\cite{percival}
Percival et al (2007)
have analyzed also the LRGs using both 2dF and SDSS. 
%\cite{padmanabhan2007}
Padmanabhan et al (2007)
used the photometric catalog to work with a bigger set of LRGs with photometric
redshifts, obtaining also cosmological constraints, the same as 
%\cite{blake},
Blake et al (2007),
which work with the MegaZ-LRG, a photometric-redshift catalog of luminous red
galaxies based on the imaging data of the SDSS DR4.

This is the first and main paper of a series on clustering of LRG. Here we will 
focus on redshift space distortions in $\xips$ on larges scales. 
In Paper II \cite{paper2} we will look
at small scales in $\xips$ and address the issue of the FOG. 
In Paper III \cite{paper3} we look at the 3-point function and 
in Paper IV \cite{paper4} we will focus on the significance of the
BAO detections.  In this first paper we present the basis 
for the error analysis
and the study of systematics effects that will be used 
in the rest of the series.

The paper is organized as follows.
Section II gives a summary of the theory involved in redshift space distortions, the model used for the 2-point correlation function in separate line of sight and perpendicular directions, and the different multipoles. Section III presents an accurate description of the data selected to work with, LRG galaxies. Section IV is the main part of this paper, where we explain the analysis done and our main results. We use the distortions in redshift space to obtain the distortion parameter $\beta$, $\Omega_m$, $\sigma_8$ and modifications to gravity. We also look at the ISW effects for this LRG data, and explore the differences between redshift slices.
As usual, we end with discussion and future work in section V. The Appendix A is devoted to present the mock simulations that we use to validate the errors used, Monte Carlo, jackknife and a new theoretical approach. In this section, we also validate the methods used to obtain constraints (section IV) in real data with our mocks. In the Appendix B we present a study of possible systematic effects.

\section{Theory}

\label{sec:model2point}

%\cite{Kaiser} 
Kaiser (1987) pointed out that, in the large-scale linear regime, and in the
plane-parallel approximation (where galaxies are taken to be sufficiently far
away from the observer that the displacements induced by peculiar velocities are
effectively parallel), the distortion caused by coherent infall velocities takes
a particularly simple form in Fourier space:
\begin{equation}\label{eq:kaiserpoint} P_s(k) = (1 + \beta\mu_k^2)^2 P(k).
\end{equation} where $P(k)$ is the power spectrum of density fluctuations
$\delta$, $\mu$ is the cosine of the angle between $k$ and the line-of-sight,
the subscript $s$ indicates redshift space, and $\beta$ is proportional to
the velocity growth rate in linear theory.

If the galaxy overdensity $\delta$ is linearly biased by a factor $b$ relative
to the underlying matter density $\delta_{m}$ of the Universe, \begin{equation}
\label{eq:bias} \delta = b \delta_m, 
\end{equation} 
then the observed value of
$\beta$ is 
\begin{equation} 
\label{fb} \beta = {f(\Omega_m) \over b} \equiv {1 \over b} 
\frac{d\;ln\;D}{d\;ln\;a} 
\end{equation} 
where $D$ is the linear density growth
factor and $f$ the velocity growth rate, which we can write as:
 \begin{equation}
\label{eq:fgamma} f(\Omega_m)=\Omega_m(a)^\gamma 
\end{equation} where $\gamma$ is the
gravitational growth index \cite{linder1}, $\Omega_m(a)$ is the matter density
at a redshift z where $a=1/(1+z)$,

\begin{equation}\label{eq:omegaa}
\Omega_m(a)=\frac{H_0^2\Omega_{0m}a^{-3}}{H^2(a)} \end{equation}

where $\Omega_{0m}$ is the matter density at z=0 (which is often just called $\Omega_m$) and

\begin{equation} H(a)=H_0\sqrt{\Omega_{0m}a^{-3}+(1-\Omega_{0m})a^{-3(1+w)}}
\end{equation}

where $w$ is dark energy equation of state parameter.

By construction \cite{linder1}, the growth index formalism separates out two
physical effects on the growth of structure: $\Omega(a)$ involves the expansion
history and $\gamma$ focuses on the gravity theory. The value $\gamma=0.55$
corresponds to standard gravity, while $\gamma$ is different for modified
gravity, for example $\gamma=0.68$ in the braneworld cosmology \cite{linder1}. The linear
density growth factor can be found using Eq.(\ref{fb}): 
\begin{equation} D(a)=e^{\int_0^a
d\ln a\,[\Omega(a)^\gamma-1]} \label{eq:dgamma} 
\end{equation}

\subsection{Modelling $\xisp$}

%\cite{hamilton1992} 
Hamilton (1992) translated Kaiser's results into real space,

\begin{equation} \xi'(\sigma, \pi) = \xi_0(s)P_0(\mu) + \xi_2(s)P_2(\mu) +
\xi_4(s)P_4(\mu), \end{equation} where $\pi$ is the separation along the
line-of-sight (LOS) and $\sigma$ is the separation in the plane of the sky
,the absolute distance of
separation is $s=\sqrt{\sigma^2+\pi^2}$, $\mu$ is the cosine of the angle
between $s$ and the line-of-sight and $P_\ell$ are Legendre polynomials.

In general, we can define the multipoles of $\xips$ (see also Eq.(\ref{eq:moment})), \begin{equation} \xi_0(s) = \left(1
+ \frac{2\beta}{3} + \frac{\beta^2}{5}\right)\xi(r), \end{equation}
\begin{equation} \xi_2(s) = \left(\frac{4\beta}{3} +
\frac{4\beta^2}{7}\right)[\xir-\overline{\xi}(r)], \end{equation}
\begin{equation} \xi_4(s) = \frac{8\beta^2}{35}\left[\xir +
\frac{5}{2}\overline{\xi}(r) -\frac{7}{2}\overline{\overline{\xi}}(r)\right],
\end{equation} and \begin{equation} \overline{\xi}(r) =
\frac{3}{r^3}\int^r_0\xi(r')r'{^2}dr', \end{equation} \begin{equation}
\overline{\overline{\xi}}(r) = \frac{5}{r^5}\int^r_0\xi(r')r'{^4}dr'.
\end{equation}

We use these relations to create a model $\xi'(\sigma, \pi)$. 
In this paper, which analyzes large scales, we will use a linear real space correlation function $\xi(r)$ as input \cite{eisensteinhu}.\footnote{We have also tested the inclusion of the effect of non-linear growth
(with the halofit model) and non-linear bias. We find no changes on our fit to large scales, as expected.} 
Then we convolve
it with the distribution function of random pairwise velocities, $f(v)$, to give
the final model $\xisp$ \cite{peebles1980}:

\begin{equation} \label{eq:hamiltonmethod} \xisp =
\int^{\infty}_{-\infty}\xi'(\sigma, \pi - v/H(z)/a(z))f(v)dv \end{equation}

where we divide peculiar velocities by $a(z)$ to translate to comoving
distances, since velocities are defined in physical coordinates.

We represent the random motions by an exponential form, \begin{equation} f(v) =
\frac{1}{\sigma_{v}\sqrt{2}}\exp\left(-\frac{\sqrt{2}|v|}{\sigma_{v}}\right)
\label{e:fv} \end{equation}

where $\sigma_{v}$ is the pairwise peculiar velocity dispersion. An exponential
form for the random motions has been found to fit the observed data better than
other functional forms \cite{ratcliffe1998,landy2002}. We have also checked
this relation with our simulations, which will be described later on.

%\cite{matsubara2004} and \cite{scoccimarro} 
Matsubara (2004) and Scoccimarro (2004) have presented different models
for the 2-point correlation function in redshift space. 
%\cite{tinker1} and \cite{tinker2} 
Tinker et al (2006) and Tinker (2007)
model redshift space distortions in the context of
halo occupation distribution model (HOD). These models are complementary to the
one studied here. In most situations the differences are small and we will show
that our modeling gives good agreement with simulations and real data.

\subsection{Multipoles of $\xisp$}

We can define the multipoles of $\xi(\pi,\sigma)$ as
\begin{equation}\label{eq:moment} \xi_{\ell}(s) =
\frac{2\ell+1}{2}\int^{+1}_{-1}\xi(\pi,\sigma) P_{\ell}(\mu)d\mu. \end{equation}

where $\mu$ is cosine of the angle to the line-of-sight $\pi$.

The normalized quadrupole is defined as \cite{hamilton1992}

\begin{equation}\label{eq:quadru}
Q(s)=\dfrac{\xi_2(s)}{\xi_0(s)-(3/s^2)\int_0^s{\xi_0(s')s'^{2} ds'}}
\end{equation}

Linear bias in Eq.(\ref{eq:bias}) cancels in the  quadrupole, and $Q(s)$ is only
very slightly dependent on the shape of the correlation function, such as 
changes on large scales due to varying $\Omega_m$ 
 or scale dependent non-linear bias for
small scales.

In the Kaiser approximation, ie at large scales, the quadrupole is directly
related to $\beta$:
 \begin{equation}\label{eq:quadrub} Q(s) =
\frac{\frac{4}{3}\beta + \frac{4}{7}\beta^2}{1 + \frac{2}{3}\beta +
\frac{1}{5}\beta^2} \end{equation}
 The validity of this approximation is studied using simulations in section \ref{sec:errors}.
At small scales, the quadrupole depends strongly on the random pairwise
velocities, represented by $\sigma_{v}$, but it does not depend much on
non-linear bias, as we will show.

\subsection{The real-space correlation function $\xi(r)$}\label{sec:realspace}

We can estimate the real-space correlation function by calculating the projected
correlation function, $\Xi(\sigma)$, integrating the redshift distorted $\xisp$
along the line-of-sight $\pi$.

\begin{equation} 
\Xi(\sigma) = 2\int^{\pi_{max}=\infty}_{\pi=0}\xisp~d\pi
\label{e:Xi} 
\end{equation} 
We would like $\pi_{max}=\infty$, however, with real
data, we can not integrate until infinity. Here we will use  $\pi_{\rm{max}} =
80Mpc/h$. The result does not change when we change the upper limit of the
integral for $\pi_{\rm{max}}>60Mpc/h$ and large $\sigma$ in the data.

%\cite{davpeeb1983} 
Davis and Peebles (1983) show that $\Xi(\sigma)$ is directly related to the
real-space correlation function.

\begin{equation} 
\Xi(\sigma)= 2\int^{\infty}_{\sigma}\frac{r \xir dr}{(r^2 -
\sigma^2)^{\frac{1}{2}}}. \label{e:proj} 
\end{equation}

It is possible to estimate $\xir$ by directly inverting $\Xi(\sigma)$
\cite{saunders1992} 
\begin{equation} \xir = -\frac{1}{\pi}\int^{\infty}_r
\frac{(d\Xi(\sigma)/d\sigma)}{(\sigma^2 - r^2)^{\frac{1}{2}}}d\sigma.
\label{eq:xirr} 
\end{equation} 
Assuming a step function for $\Xi(\sigma)=\Xi_i$
in bins centered on $\sigma_i$, and interpolating between values,
\begin{equation} \label{eq:xirr2} \xi(r) = -\frac{1}{\pi}\sum_{j\geq
i}\frac{\Xi_{j+1}-\Xi_j}{\sigma_{j+1}-\sigma_j}\ln\left(\frac{\sigma_{j+1}+\sqrt
{\sigma^2_{j+1} - \sigma^2_i}}{\sigma_j + \sqrt{\sigma^2_j - \sigma^2_i}}\right)
\end{equation} for $r = \sigma_i$. \\

\begin{figure}
\centering{ \epsfysize=5cm\epsfbox{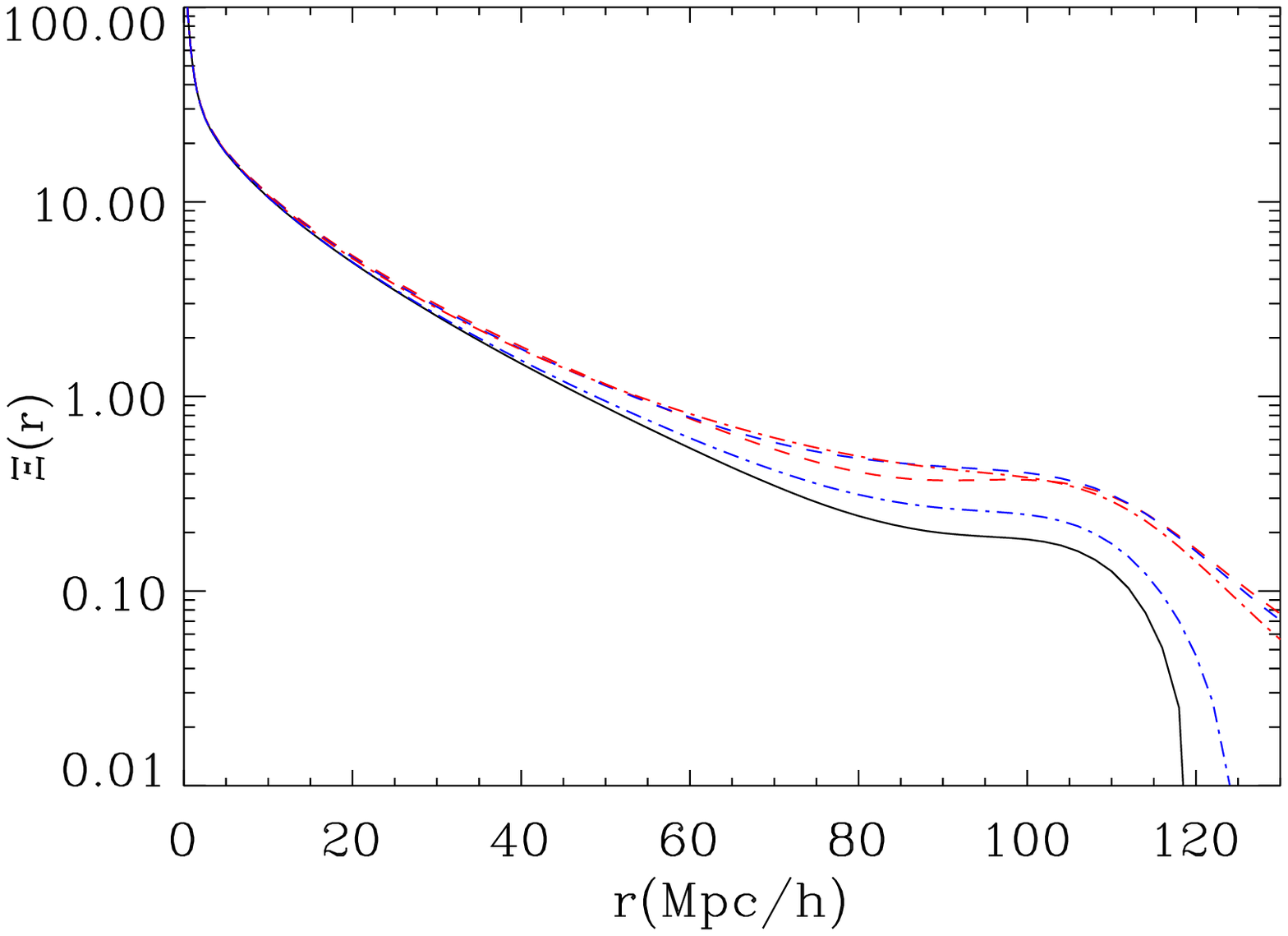}}
\centering{\epsfysize=5cm\epsfbox{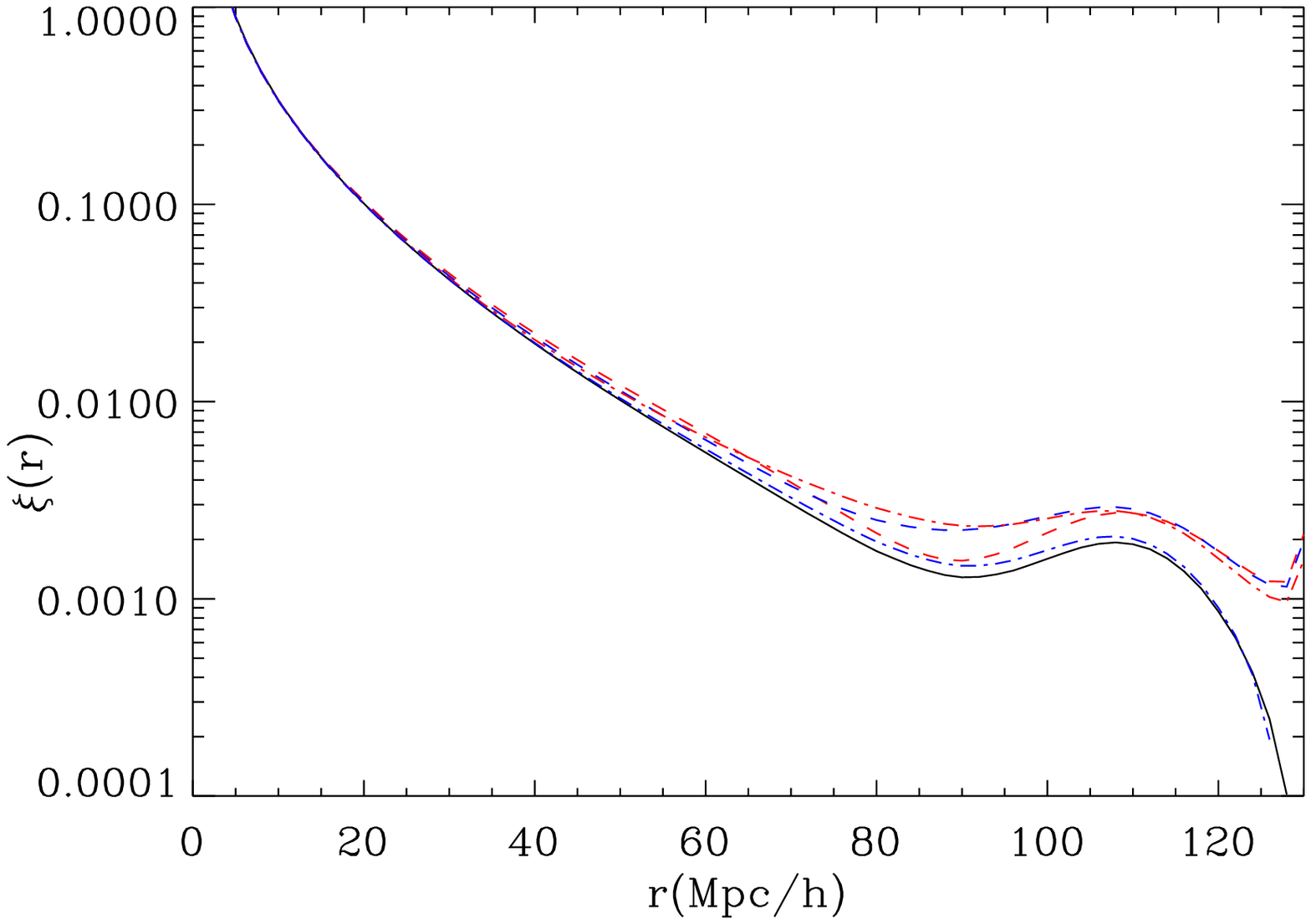}}
\caption{{\it Top:} Solid line shows the projected perpendicular correlation function
$\Xi(\sigma)$
calculated theoretically from $\xi(r)$ in Eq.(\ref{e:proj}), for a model with
$\beta=0.35$ and $\sigma_v=400km/s$. This is compared to integral of $\xips$ 
in Eq.(\ref{e:Xi}) with
$\pi_{max}$=60.(red dashed), 80.(blue dashed), 100.(red dash-dot), 200.(blue dash-dot). Bottom: Estimation of
the real space correlation function by deprojecting the color lines of
$\Xi(r)$ in the top panel using Eq.(\ref{eq:xirr}). We can see good agreement on
scales smaller than 30Mpc/h.
\label{fig:perpint} }
\end{figure}

We take the redshift space anisotropic model $\xisp$ in
Eq.(\ref{eq:hamiltonmethod}), with a fixed $\beta=0.35$ and
$\sigma_{v}=400km/s$, and we use Eq.(\ref{e:Xi}) to obtain the projected
correlation function $\Xi(\sigma)$ as if it was data. We use different values to
fix the upper limit in the integral: $\pi_{max}=$ 60Mpc/h, 80Mpc/h, 100Mpc/h and
200Mpc/h (color lines in top panel of Fig.\ref{fig:perpint}), and compare the
"true" result obtained from Eq.(\ref{e:proj}) (solid line in top panel of
Fig.\ref{fig:perpint}). As we increase $\pi_{max}$ we approach the true result,
but we can not integrate until 200Mpc/h in real data because, as will be shown
in the following sections, data is quite noisy for $\pi>60Mpc/h$. The
implication of these results is that we will not be able to recover  the true
values of $\Xi(\sigma)$ from $\xips$. But we can use this method at smaller
scales (see Paper II).

In the bottom panel of Fig.\ref{fig:perpint} we see the real-space correlation
function recovered from the previously calculated projected correlation function
with different $\pi_{max}$, Eq.(\ref{eq:xirr}). We obtain a good estimation of
$\xir$ below 30Mpc/h, where we will study the non-linear bias
(see Paper II).\\

This analysis has been done for different values of $\beta$ and $\sigma_{v}$ and
 we find very similar conclusions. For illustration, we have shown in our plots
the values that are more in concordance with real LRG SDSS data.\\

Once we recover the real-space correlation function, we can also estimate the
ratio of the redshift-space correlation function, $\xis$, to the real-space
correlation function, $\xir$, which gives an estimate of the redshift distortion
parameter, $\beta$, on large scales:

\begin{equation} \label{eq:xisr} \frac{\xis}{\xir} = 1 + \frac{2\beta}{3} +
\frac{\beta^2}{5}. \end{equation}

As we have shown in Fig.(\ref{fig:perpint}), $\xir$ will be in general slightly
overestimated at large scales when we estimate it from the projected correlation
function $\Xi(\sigma)$, so the expression $\frac{\xis}{\xir}$ will in general be
slightly lower than expected on large scales.

%%%%%%%%%%%%%%%%%%%%%%%%%%%%%%%%%%%%%%%%%%%%%%%%%%%%%%%%%%%%%%%%%%%%%%%%
\section{The Data} \label{sec:data}

\begin{figure} 
\centering{ \epsfysize=6cm\epsfbox{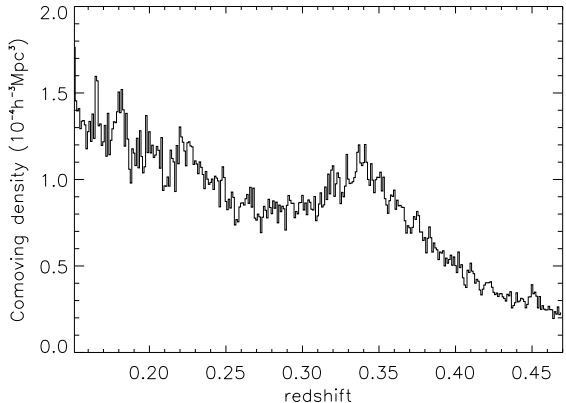}}
\caption{Comoving density vs redshift for our selected LRGs\label{fig:comoving}}
\end{figure}

%\begin{figure} %	\centering{ \epsfysize=7cm\epsfbox{figures/plot81.3.ps}}
%	\caption{Density distribution of galaxies in the main catalog, z=0.15-0.47
%\label{fig:plot81}} %\end{figure}

\begin{figure*} \centering{ \epsfysize=10cm\epsfbox{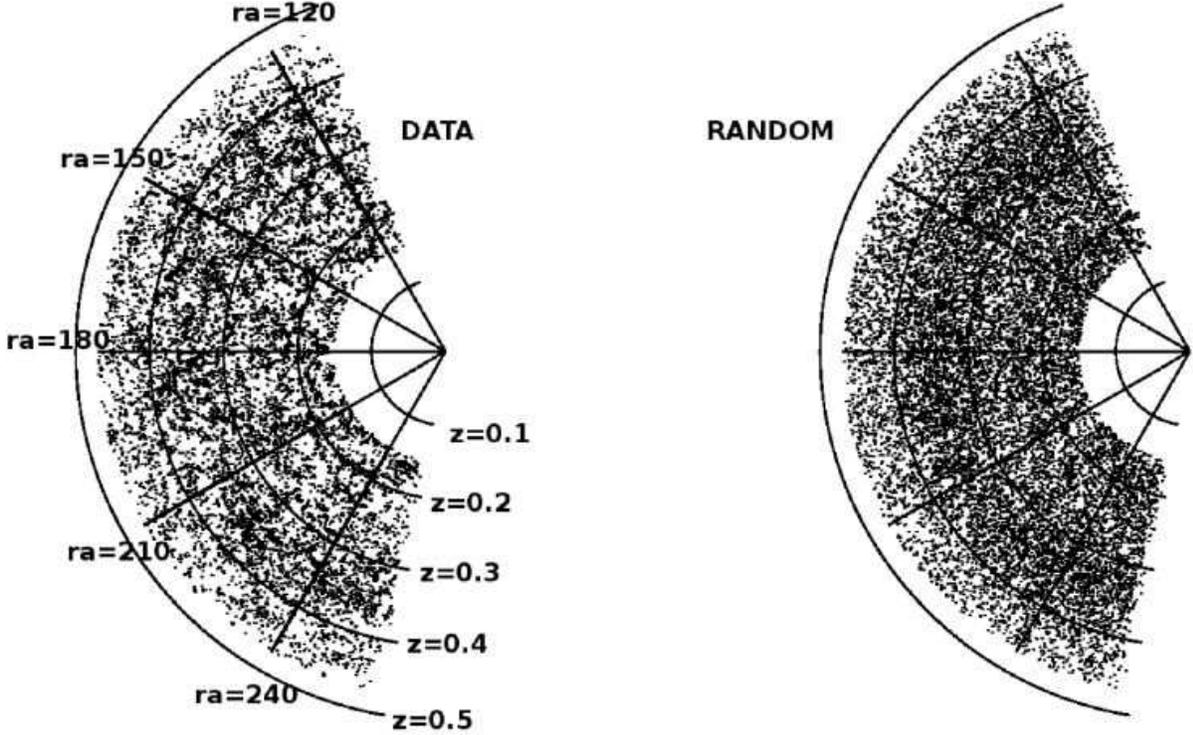}}
\caption{Slice in dec = 32-40 deg showing ra vs redshift, in order to see the
structure of the data (left panel). We also plot the random distribution for
this slice (right panel)\label{fig:plot82}} \end{figure*}

The luminous red galaxies (LRGs) are selected by color and magnitude to obtain
intrinsically red galaxies in Sloan Digital Sky Survey (SDSS). See
%\cite{eisenstein2001}
Eisenstein et al (2001) or http://www.sdss.org for a complete description of the
color cuts. In this paper, we work with the last public catalog, DR6, which covers a solid angle of $4\pi/7$ in the sky. These galaxies trace a big volume, around $1Gpc^3h^{-3}$, which make
them perfect to study large scale clustering.  LRGs are supposed to be red old
elliptical galaxies, which are usually passive galaxies, with relatively low
star formation rate. 
%They have steeper slopes in the correlation function than
%the rest of galaxies, since they reside in the centers of big halos, inducing
%non-linear bias dependent on scale, for small scales. They are well known
%galaxies, so they represent a good chance to use it as dark matter clustering
%tracers.

They have steeper slopes in the correlation function than the rest of galaxies,since in some cases, 
there is more than one LRG per halo. 
They trace a big volume, so they represent  good candidates to use as dark matter clustering tracers.

LRG's are targeted in the photometric catalog, via cuts in the (g-r, r-i, r) 
color-color-magnitude cube. Note that all colors are measured using model 
magnitudes, and all quantities are corrected for Galactic extinction 
following Schlegel et al (1998). %\cite{ext1998}.
The galaxy model colors are rotated first to a basis that is aligned with 
the galaxy locus in the (g-r, r-i) plane according to:

$c_{\perp}$= (r-i) - (g-r)/4 - 0.18

$c_{||} $= 0.7(g-r) + 1.2[(r-i) - 0.18]\\

Because the 4000 Angstrom break moves from the g band to the r band 
at a redshift z $\simeq$ 0.4, two separate sets of selection 
criteria are needed to target LRGs below and above that redshift:\\

Cut I for z $<$ 0.4

    rPetro $<$ 13.1 + $c_{||}$ / 0.3

    rPetro $<$ 19.2

    $|c_{\perp}|$ $<$ 0.2

    $mu_{50}$ $<$ 24.2 mag $arcsec^{-2}$

    $r_{PSF}$ - $r_{model}$ $>$ 0.3 \\

Cut II for z $>$ 0.4

    $r_{Petro}$ $<$ 19.5

     $|c_{\perp}|$ $>$ 0.45 - (g-r)/6

    g-r $>$ 1.30 + 0.25(r-i)

    $mu_{50}$ $<$ 24.2 mag $arcsec^{-2}$

    $r_{PSF}$ - $r_{model}$ $>$ 0.5 \\

where $mu_{50}$ is the mean surface brightness within the radii containing 50\% of Petrosian flux.

Cut I selection results in an approximately volume-limited LRG sample 
to z=0.38, with additional galaxies to z $\simeq$ 0.45. Cut II selection adds 
yet more luminous red galaxies to z $\simeq$ 0.55. The two cuts together 
result in about 12 LRG targets per $deg^2$ that are not already 
in the main galaxy sample (about 10 in Cut I, 2 in Cut II).
The radial distribution and magnitude-redshift diagrams are plotted
in Fig.\ref{fig:dndzrandoms} and Fig.\ref{fig:slicesplot1}.

We k-correct the r magnitude using the Blanton program 'kcorrect'
\footnote{http://cosmo.nyu.edu/blanton/kcorrect/kcorrect\_help.html}. We need to
k-correct the magnitudes in order to obtain the absolute magnitudes and
eliminate the brightest and dimmest galaxies. We have seen that the previous
cuts limit the intrinsic luminosity to a range $-23.2<M_r<-21.2$, and we only
eliminate from the catalog some few galaxies that lay out of the limits. Once we
have eliminated these extreme galaxies, we still do not have a volume limited
sample at high redshift  but we will account for this using a random catalog
with identical selection function.

We show the comoving density in Fig.\ref{fig:comoving} once we have removed the
brightest and dimmest galaxies, and in Fig.\ref{fig:plot82} we show the redshift
space distribution in a slice of dec = 32-40 deg. We compare it with the same
slice of random points. We see clear evidence of clustering in the data.

We have masked the catalog using at the first step the photometric DR6 mask,
based on the number of galaxies per pixel. In previous works \cite{Cabre1} we 
showed that the
mask that we obtain statistically by dropping out the pixels with small number
of galaxies (see \S B3 for details) gives identical correlation function that the one 
obtained when
extracting the polygons masked by the SDSS team. After that, we compare our
masked catalog to the LRG spectroscopic catalog, and we extract the galaxies
that lay outside from the ``good'' plates. Fig.\ref{fig:platesplot} illustrates
the result.

The mask could imprint spurious effects at very small scales. But we are
not focused in such small scales, because fiber collisions in the redshift catalog
are limiting our analysis. This is for distances less than 55arc sec, ie less than 0.3Mpc/h
at the mean redshift of LRG data, z=0.34 (see Fig.17 below). We obtain 75,000 galaxies for the
final catalog, from z=0.15 to z=0.47, ie over one cubic Gigaparsec 
($Gpc^3/h^3$).

\section{Results}\label{sec:results}

\begin{figure} 
\centering{\epsfysize=7cm\epsfbox{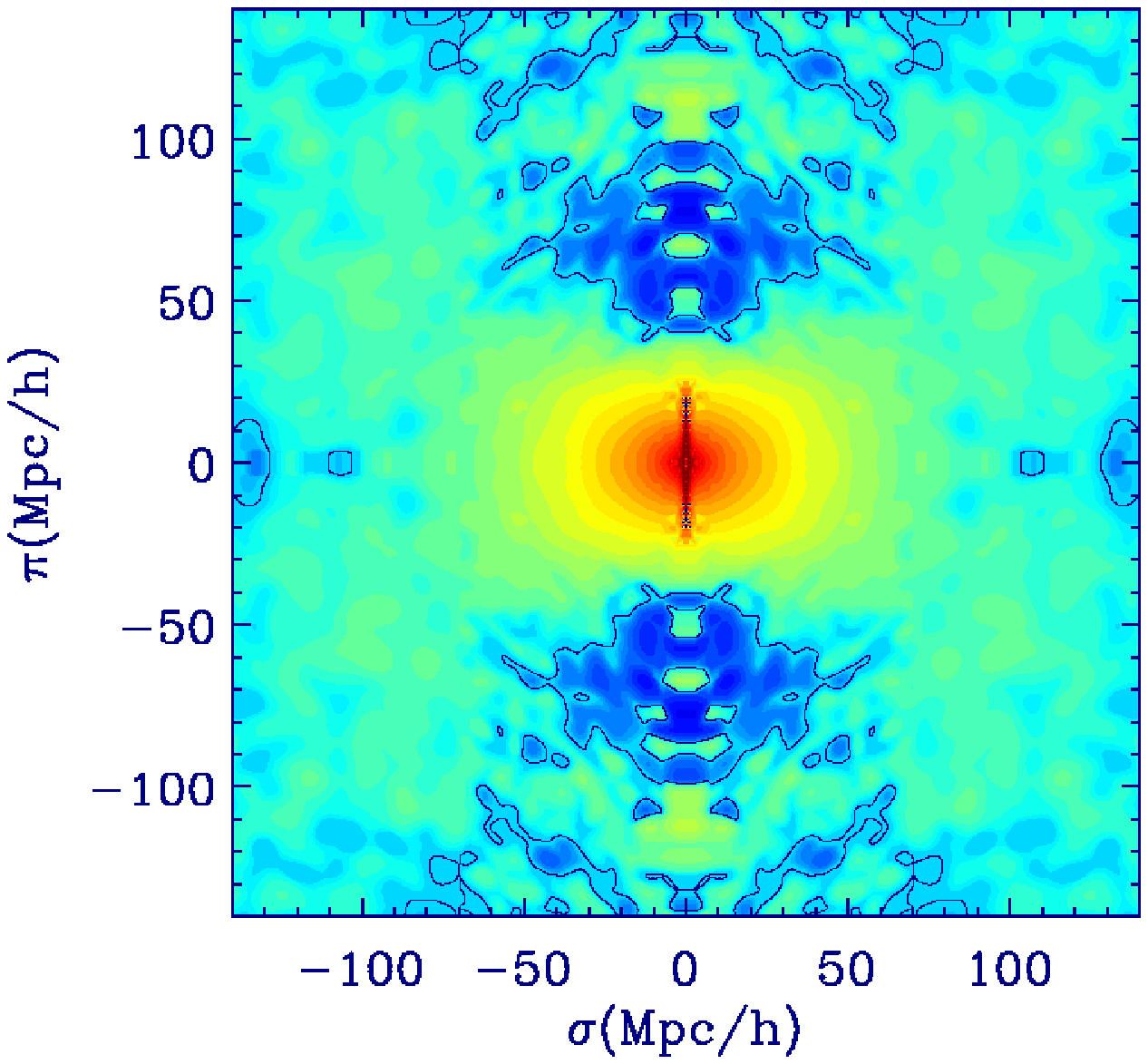}}
\centering{ \epsfysize=7cm\epsfbox{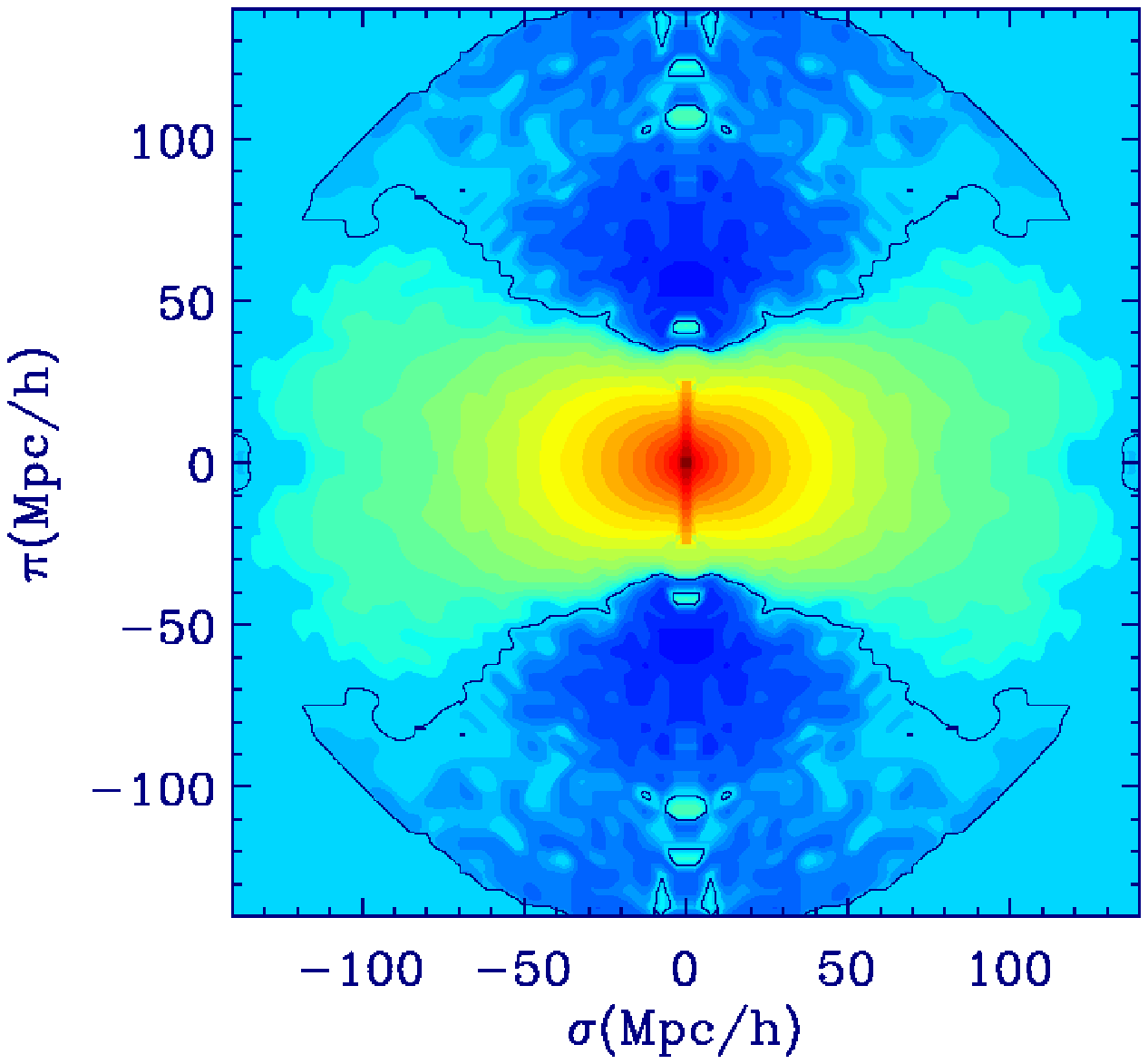}} \caption{2-point
anisotropic redshift correlation function $\xisp$ for LRG galaxies in DR6
catalog (top panel) and mock MICE simulations (bottom panel, with a linear bias b=2,
in order to be similar in amplitude to real LRG). Contours are -0.5 to -0.004 with logarithmic bin
of 0.4, 0. (over-plotted as a line), 0.003 to 40. with log bin=0.4. We see how the real
LRG have a less flattened shape around the center, since $\beta$ is smaller because of higher
bias \label{fig:pisigmaDR6} } 
\end{figure}

In this section we analyze the LRGs from SDSS DR6. First, we have used credible simulations to study the errors carefully as well as the methods that we use here (see Appendix A for an extensive description).  

Here we define the parameters that we assume during all this work, which are
motivated by recent results of WMAP, SNIa and previous LSS analysis \footnote{see http://lambda.gsfc.nasa.gov/product/map/dr3/parameters.cfm}: $n_s=0.98$,
$\Omega_b=0.045$, $h=0.72$. We will use the power spectrum analytical form for
dark matter by 
%\cite{eisensteinhu}, 
Eisenstein and Hu (1998), and the non-linear fit to halo theory by
%\cite{smith2003}. 
Smith et al (2003). For part of our analysis we have followed the method explained in
%\cite{hawkins}, 
Hawkins et al (2003), an extensive analysis of redshift distortions in the 2dF
catalog.

To estimate the correlation $\xisp$, 
we use the $\xi$ estimator of 
%\cite{landyszalay},
Landy and Szalay (1993) as in Eq.(\ref{eq:dddrrr}),
with a random catalog $N_R=20$ times denser than the SDSS catalog. 
The random catalog has the same redshift (radial) distribution as the data, 
but smoothed with a bin $dz=0.01$ to avoid the elimination of
intrinsic correlations in the data. Randoms also have the same mask. We count the pairs
in bins of separation along the line-of-sight (LOS), $\pi$,  and across the sky,
$\sigma$. The LOS distance $\pi$ is just the difference between the radial comoving distances
in the pair. The perpendicular distance between the two particles corresponds
approximately to the mean redshift. It is exactly defined here as $\sigma=\sqrt{s^2-\pi^2}$,
where $s$ is the distance between the particles. We use the wide-angle
approximation, as if we had the catalog at an infinite distance, which is accurate
until the angle that separates the galaxy pair in the sky 
is larger than 15 degrees  for the quadrupole and about 10 degress for the
$\xisp$ \cite{szapudiwide,matsuwide}. This later restriction
corresponds to scales  $\sigma >80Mpc/h$ for our 
mean catalog.

In Fig.\ref{fig:pisigmaDR6} we can see our $\xisp$ estimation for LRG DR6 catalog
and for MICE simulations (with a linear bias b=2 to see similarities visually). Here
we can clearly see the FOG at small $\sigma$ as we have increased the pixel resolution when
we approach the central part of the image (pixel size varies from 0.2 Mpc/h in the
center to 10 Mpc/h at large scales). As
expected, simulations have less noise because we have done the mean over 216
mocks. Moreover, the distortion parameter $\beta$ is higher in the simulations
than in data because of the high bias of LRG, as can be seen in the shape of
the 2-point correlation function, which gets more flattened as $\beta$ increases. We see
clearly the baryonic peak, as a ring in the correlation function $\xisp$. 
This feature is studied in detail in Paper IV of this series. 
In next sections we will
analyze all the information hidden in this figure, using the analytical form of
the error (see \S\ref{sec:errorsxips}).

\subsection{Quadrupole and $\beta$ estimation}\label{sec:secquadru}

\begin{figure}
\centering{ \epsfysize=6cm\epsfbox{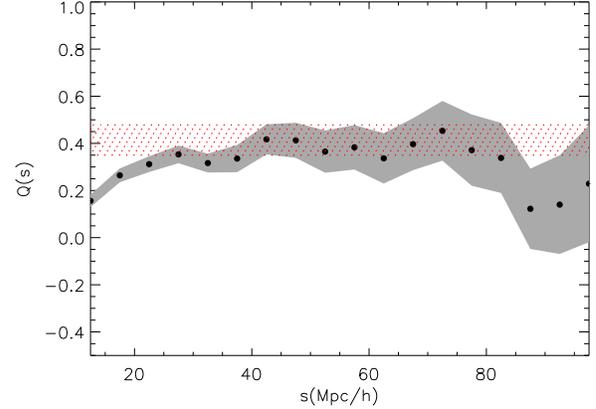}} \caption{Q(s)
 (points with errors) and best fit asymptotic value of $\beta$ at large scales
(Eq.(\ref{eq:quadrub})) translated to the quadrupole (red dotted)
\label{fig:qukaiser}} \end{figure}

\begin{figure}
\centering{ \epsfysize=6cm\epsfbox{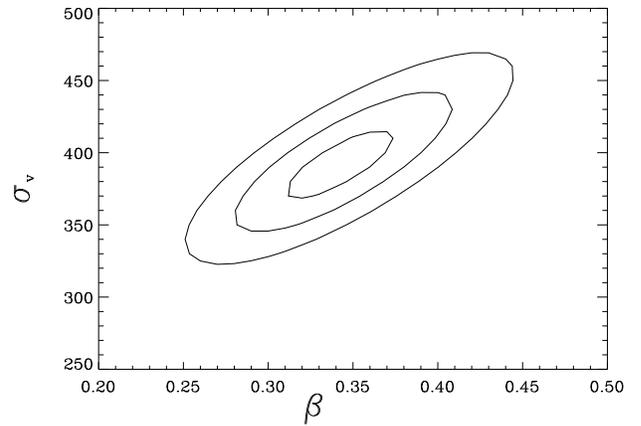}}
\centering{ \epsfysize=6cm\epsfbox{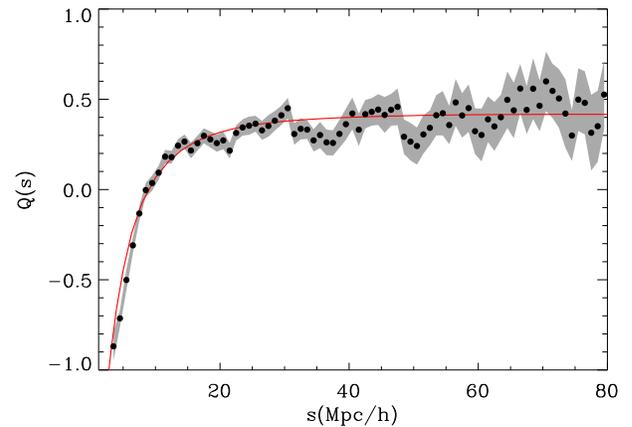}}
\caption{Top panel: Contour for $\sigma_v$ and $\beta$ obtained from the
quadrupole Q(s) for the slice z=0.15-0.47 at distances between 5-60Mpc/h. Bottom
panel: Measured Q(s) (dots with error in gray) and overplotted the best fit
(solid red) \label{fig:quadrupole1Mpc} } 
\end{figure}

We use the normalized 
quadrupole Q(s) to obtain
$\beta-\sigma_v$ (extensive explanation can be found in \S\ref{sec:validity}). But first, we can measure $\beta$ using the asymptotic
large scale value in
the quadrupole where there is no dependence on $\sigma_{v}$ (see Eq.(\ref{eq:quadrub})). We find 
$\beta=0.34 \pm 0.06$ in the range $40-80
Mpc/h$; $\beta=0.32 \pm 0.08$ for $50-80 Mpc/h$ and 
$\beta=0.34 \pm 0.05$ for $40-100 Mpc/h$. In Fig. \ref{fig:qukaiser} we see the
quadrupole with jackknife errors at large scales and the error obtained in
$\beta$, translated to the quadrupole as a band of red dots.

The quadrupole only depends strongly on $\beta$ and
$\sigma_v$. It does not depend on linear bias because it cancels out in
the ratio, and it
does not depend much on the shape of the 2-point correlation function
($\Omega_m$ and other parameters) or in the non-linear bias for small scales.  We
have tried to fit $\beta$ and $\sigma_v$ using all the scales in the quadrupole,
fixing $\Omega_m$ to 0.25 (which in our model means fixing the shape of the
real-space correlation function) and we use a power law form for the non-linear
bias. When we change the shape of the $\xir$ in the model (that is $\Omega_m$
for large scales and non-linear bias for small scales) , we obtain the same
contours for $\beta-\sigma_v$, so we arrive to the conclusion that the
normalized quadrupole is a good estimator to find $\beta$ separately from the
other parameters, which are degenerate with them in the $\xi(\pi,\sigma)$
plane.

We can obtain better errors in $\beta$ by doing a joint fit
with $\sigma_v$ and using smaller scales, as we do with simulations in Appendix A.
We will fit the quadrupole above 5Mpc, because this seems 
a reliable scale according to simulations and results do not
change much when using a cut on larger scales. Below this minimum scale,
$\sigma_v$ starts increasing, and even when this has only a small effect 
in the quadrupole at larger scales, it can bias the value of $\beta$
when doing a fit to all scales.

In Fig.\ref{fig:quadrupole1Mpc} we show the fit as $\Delta\chi^2$
contours in the $\beta-\sigma_v$ plane
(top panel) and the best fit to Q(s) for the mean slice (bottom panel). We have fitted
$\beta-\sigma_v$ for different slices in redshift. First, we divide the catalog
in 3 redshift slices: z=0.15-0.3, z=0.3-0.4, z=0.4-0.47. And then, we divide it
in 2 redshift slices: z=0.15-0.34, z=0.34-0.47. The fitted values $\sigma_v$ and
$\beta$ are similar in all the redshift slices (see table
\ref{tab:tabbetasigma}). This is a bit surprising because 
we expect $f=\Omega_m(z)^{0.55}$ 
to increase slightly with $z$, for this redshift range.
The similarity in the value of the distortion
parameter $\beta$ means that the bias must also increase
as a function of redshift, roughly in
the same way as $f=\Omega_m(z)^{0.55}$ so that the
effect cancels out (see Eq.(\ref{fb})). This is not totally surprising,
because we expect bias $b$ to scale with redshift as $1/D(z)$,
the inverse of the growth factor, $D(z)$, which is proportional to $f$, in order to
 have stable clustering, ie $b(z)D(z)$ constant along redshift (Peebles 1980). 
%But note that there is an anomaly in sample $z=0.30-0.40$ that we will
%adress below.

\begin{table} \begin{center} \begin{tabular}{|c|c|c|} \hline \textbf{Sample}&
\boldmath{$\beta$}& \boldmath{$\sigma_v$} \textbf{(km/s)} \\ 
\hline z=0.15-0.47& 0.310-0.375  & 365-415  \\ 
\hline 
z=0.15-0.34 & 0.280-0.365 &  320-410 \\
z=0.34-0.47 & 0.305-0.405 &  345-420\\ 
\hline 
z=0.15-0.30 & 0.280-0.395 &  305-435 \\
z=0.30-0.40 & 0.285-0.365 &  335-390  \\
z=0.40-0.47 & 0.270-0.415 &  305-420\\
\hline
\end{tabular} \end{center} \caption{Marginalized values for $\beta$ and
$\sigma_v$ to 1-$\sigma$ errors for each sample in redshift obtained from the
study of the quadrupole Q(s). We can see that $\beta$ is nearly constant along
the different redshift slices \label{tab:tabbetasigma}}

\end{table}

\subsection{Fitting $\xir$}\label{sec:large}
\begin{figure} 
\centering{ \epsfysize=6cm\epsfbox{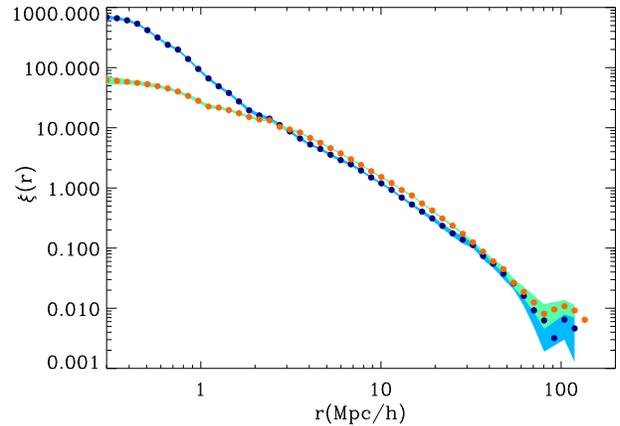}}
\caption{Real-space correlation function $\xi(r)$ (blue dots) and monopole in
redshift space $\xi(s)$ for the main slice (orange dots). We see clearly how the redshift-space
correlation function is the real-space correlation function biased by a constant
factor that represents gravitational infall (dependent on $\beta$ in Kaiser
approximation) at scales above 4Mpc/h.
However, for small scales, the redshift-space $\xis$ is strongly suppressed
compared to $\xir$ due to random peculiar velocities. \label{fig:corrplot}}
\end{figure} 

We have calculated the real-space correlation function integrating through the
LOS direction (Eq.(\ref{eq:xirr})). This method is complementary to 
the estimation from angular correlation \cite{baugh1996}.
In Fig.\ref{fig:corrplot} we compare $\xir$
(in blue) with the monopole $\xis$ (orange) for the main slice in redshift. The difference at large 
scales (r $>$ 5Mpc/h)  is a constant value, at least until $r\le30Mpc/h$ where we can trust the recovered $\xir$. 
Note that, although we expect the recovered $\xir$ to be systematically
 biased at large scales (eg see Fig\ref{fig:perpint}), we can still
see the baryonic peak at $\xir$, in real space! So this is a strong
feature in the LRG data.

The large
differences at small scales are due to random velocities, which we will study in
more detail in Paper II of this series. 

\begin{figure} \centering{
\epsfysize=6cm\epsfbox{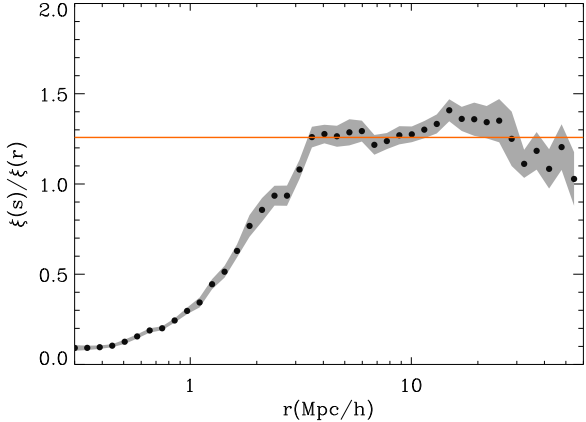}} \caption{We plot here the ratio
$\xi(s)/\xi(r)$, and its prediction for large scales
(Eq.(\ref{eq:xisr}), solid red line) using $\beta=0.34$  obtained from the quadrupole Q(s)
 \label{fig:corrscorrplot}} \end{figure}

In Fig.\ref{fig:corrscorrplot} we show the ratio
$\xis/\xir$ and  we have overplotted
the constant value in the Kaiser approximation (Eq.(\ref{eq:xisr})), using
the value of $\beta$ obtained with Q(s), ie $\beta=0.34$. This gives
a good fit but note how it would be quite difficult to estimate $\beta$
from this ratio given how noisy the data becomes when we reach an
asymptotic constant value in $\xis/\xir$.

\subsection{Fitting large scales in $\xips$}

We obtain $\Omega_m-Amp$ from the
anisotropic redshift-space correlation function $\xisp$ using the
full covariance matrix from the MICE group mocks (more details in \S\ref{sec:validity}). Here
$Amp$ refers to the factor $b(z)\sigma_8$. The growth factor $D(z)$,
which is a function of $\Omega_m$, has been included in the model of $\xips$.
We do not find any difference when fitting $\xisp$ with models
 with non-linear bias (as used in Paper-II) or linear bias.
This is because we restrict our analysis to
scales from 20Mpc/h to 60Mpc/h and for angles above 30-40 deg away from the
line-of-sight in the $\sigma-\pi$ plane.

\begin{figure}
%\centering{ \epsfysize=6.5cm\epsfbox{figures/contour.z0.34.20.40.40.3.ps}}
\centering{
\epsfysize=6.5cm\epsfbox{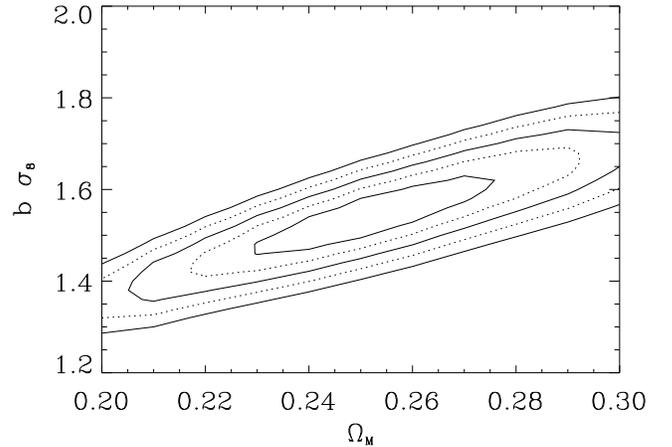}}
%omampfitLRGb1z0.34.20-40.3.40.1.ps}}
\caption{Contours for $\Omega_m$ and $Amp=b\sigma_8$, 
 from a fit to $\xips$
on large scales, for the slice z=0.15-0.34. We have marginalized for
$\beta-\sigma_v$, using priors from the quadrupole Q(s).
Solid lines are $1-\sigma$, $2-\sigma$ and $3-\sigma$ (1 dof), and
dotted lines $1-\sigma$ and $2-\sigma$ (2 dof) \label{fig:largescale34}}
\end{figure}

In Fig.\ref{fig:largescale34}  we show constraints
on $\Omega_m-Amp$ for the slice z=0.15-0.34. We have used the
$\beta-\sigma_v$ obtained with the quadrupole Q(s) fit to marginalize the space of
parameters $\Omega_m-Amp-\beta-\sigma_v$ in the $\xips$ fit, although
this only makes a slight difference. The values of
$\beta$ and $\sigma_v$ are strongly degenerate in the range where we fit large
scales, but $\sigma_v$ is not strongly degenerate with the other parameters,
since we are located away from the fingers of God, where we have the strongest
effect of $\sigma_v$.  When we marginalize, we assume that the likelihood 
scales as $exp(-\chi^2/2)$.

We have done the same analysis for different redshift slices. We have summarized
the results in table \ref{tab:largescales} where we have annotated the
marginalized 1-$\sigma$ errors for the amplitude Amp=$b \sigma_8 $ 
and for $\Omega_m$. Note that we use a fixed value of $h=0.72$ to model
the shape the linear power spectrum.

\begin{table} 
\begin{center} 
\begin{tabular}{|c|c|c|c|c|} \hline 
\textbf{Sample}& \textbf{$Amp=b \sigma_8$} & \boldmath{$\Omega_m$}  & \boldmath{$b$}  & \boldmath{$f(\Omega_m)$} \\ 
\hline 
z=0.15-0.47 & 1.47-1.65  & 0.225-0.265 & 1.73-1.94 & 0.54-0.73\\  \hline
z=0.15-0.34 & 1.45-1.62 &  0.230-0.275 & 1.71-1.91 & 0.48-0.70\\
z=0.34-0.47 & 1.55-1.82 &  0.215-0.285 & 1.82-2.14 & 0.56-0.87\\ \hline
z=0.15-0.30  & 1.45-1.80   & 0.240-0.320 & 1.71-2.11 & 0.48-0.83\\ 
z=0.30-0.40  & 1.42-1.60   & 0.210-0.260 & 1.67-1.88 & 0.48-0.69\\ 
z=0.40-0.47  & 1.60-2.00   & 0.195-0.305 & 1.88-2.35 & 0.51-0.98\\ 

\hline
\end{tabular} 
\end{center} 
\caption{Marginalized 1-sigma intervals for 
$Amp=b \sigma_8 $, $\Omega_m$, $b$ and $f(\Omega_m)$
for each redshift sample. Here for $b$ we used the best fit values
of $\sigma_8=0.85$ for the whole sample. 
\label{tab:largescales}}

\end{table}

If we try to fit to even larger scales, we always obtain a slightly 
biased low $\Omega_m$
compared to the one we obtain in the scales before the acoustic peak, probably
because wide angle effects that we are not taking into account in the modelization (see \S\ref{sec:wideangleeffect})
but possibly also because non-linear
effects on the BAO peak and large sampling errors. We have also
seen this effect in the simulations, where we have checked that the best 
region to obtain parameters is at
intermediate to large scales, where we do our fit.

Note how the best fit values of $Amp=b\sigma_8$ seem to change from sample to sample. This
could be due to bias, which is both a function of $z$ and luminosity. 
The values of $\Omega_m$ agree within $1-\sigma$ for 2
degrees of freedom (dotted lines).

\begin{figure} \centering{ 
\epsfysize=8cm\epsfbox{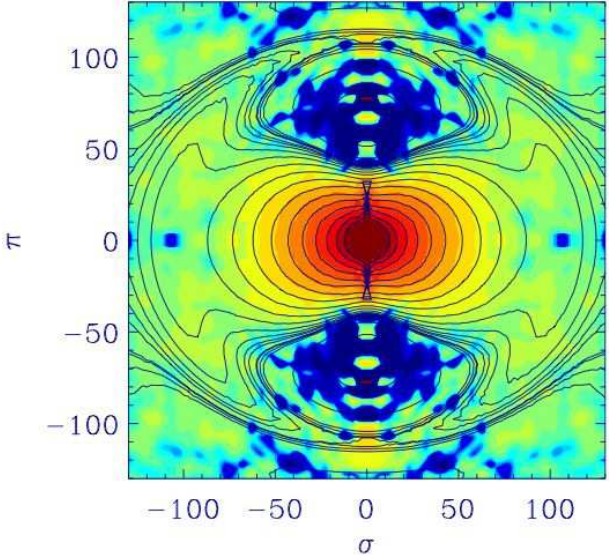}}
\caption{Best fit (solid lines) to the 2-point anisotropic
redshift correlation function $\xisp$ for LRG galaxies (colors). 
Colors are as in  Fig.\ref{fig:corrmice}).
 The agreement with the model is excellent
at all levels.
 \label{fig:pisigmaDR62} } 
 \end{figure}

In Fig.\ref{fig:pisigmaDR62} we have plotted the $\xisp$ with the best model
overplotted in solid lines. We can see how this simple Kaiser model can explain
most features in the observations. There is a very good agreement on the 
region of negative correlation (in blue) which is a very good tracer
of $\Omega_m$ and $\Omega_b$ (see Paper IV). All these features are very
significant given the errors and their coherence over large regions.
We can clearly see the FOG at small $\sigma$ as we have increased the pixel resolution when
we approach the central part of the image (pixel size varies from 0.2 Mpc/h in the
center to 5 Mpc/h at large scales). The baryonic peak can also be
 seen in both the data and in the models.
The significance of this is studied in detail in Paper IV, while the smaller
scales and the prominent fingers of God are presented in Paper II \cite{paper2}.

%\begin{figure}

%\centering{ \epsfysize=5cm\epsfbox{figures/contour.z0.47.20.40.40.3.ps}} %
%\centering{ \epsfysize=5cm\epsfbox{figures/omampfitLRGb1z0.47.20-40.3.40.1ps}}
%	\caption{z=0.4-0.47. Best fit and contour $\Omega_m-Amp$ once marginalized
%for $\beta-\sigma_v$ from the quadrupole Q(s) \label{fig:largescale47} }
%\end{figure}

%\clearpage

\subsection{The value of $\sigma_8$}

We first try to obtain a fit to the parameter $\sigma_8$, which we can separate
from the bias $b(z)$ thanks to redshift distortions, following Eq.(\ref{fb}) and
Eq.(\ref{eq:fgamma}). We use our previous estimation of $\Omega_m-Amp$ from large scales 
(ie Table \ref{tab:largescales}) and
the value of $\beta$ from Q(s) (ie in Table \ref{tab:tabbetasigma}), 
using $\gamma=0.55$, for standard gravity. 
As $Amp=b(z) \sigma_8 $, we obtain $\sigma_8$ from:

\begin{equation} \sigma_8=\frac{\beta Amp}{\Omega_m(z)^{0.55}} \end{equation}

where $\Omega_m(z)$ is given in Eq.(\ref{eq:omegaa}).
We also assume a flat universe (WMAP results motivated) with a constant dark
energy equation of state characterized by $w=-1$ (ie as in
the cosmological constant model).\\

\begin{figure}
\centering{ \epsfysize=6cm\epsfbox{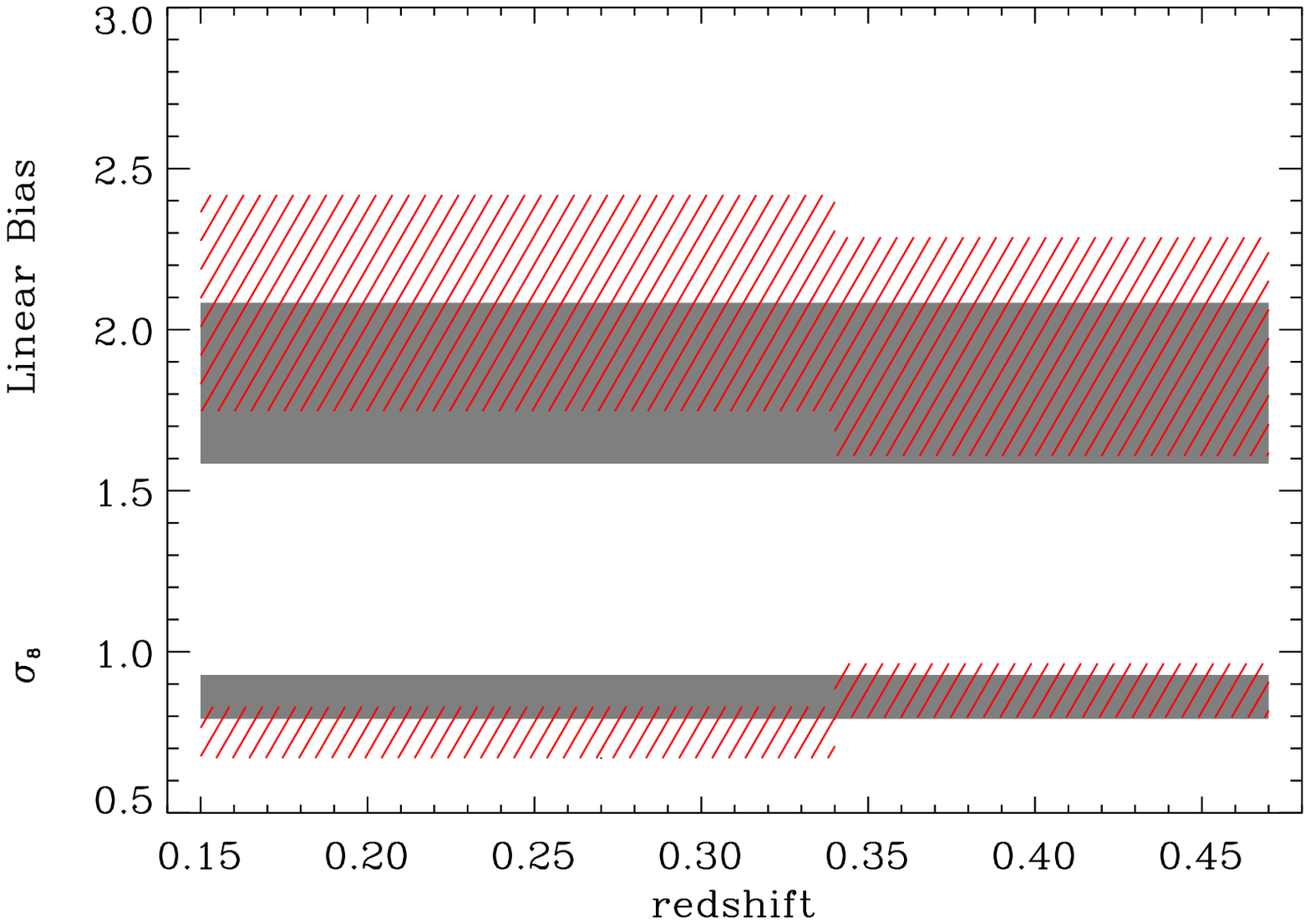}} 
\centering{ \epsfysize=6cm\epsfbox{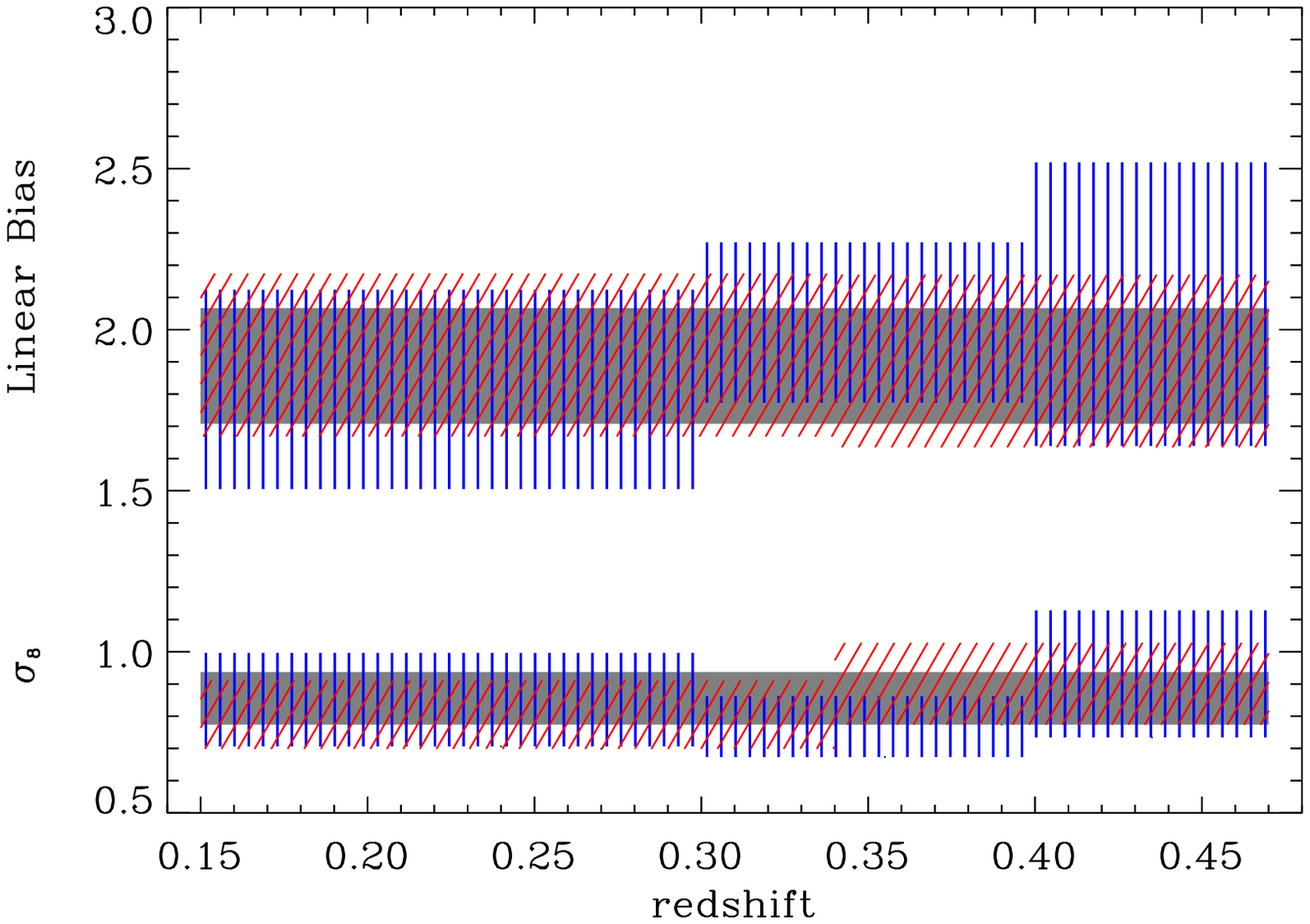}} 
\caption{Estimation
of linear bias, b(z), and
 $\sigma_8$ for each slice in redshift.
Gray: All the catalog, Red: z=0.15-0.34, z=0.34-0.47, Blue: z=0.15-0.3,
z=0.3-0.4, z=0.4-0.47. In the top panel we have marginalized the values over
the values of $\beta$, $\Omega_m$ and $Amp=b(z)\sigma_8$ in each of the
3 samples shown.
In the bottom panel we used a fixed best fit value of $\Omega_m=0.25$
and show results for all subsamples.
  \label{fig:s8found}} 
\end{figure}

In the top panel of Fig.\ref{fig:s8found} we show $\sigma_8$,
marginalized over $\beta$, $Amp$ and $\Omega_m$ in each sample.
We also show results using a fixed value of
$\Omega_m=0.25$ for all samples (bottom panel).
There seems to be a lower $\sigma_8$ detection for the middle slice of
z=0.3-0.4, but it is consistent with the others at $2-\sigma$ level. 
In the same figures we also show
the estimated bias $b(z)$ from the amplitude $Amp$ and the values
$\sigma_8$. For a fixed $\Omega_m$,
the linear bias $b(z)$ seems higher as we move to higher redshifts, and it is
consistent with previous results found also with LRG \cite{intermediate,cosmoconstraints,okumura}. Note that the
luminosity of the galaxies also increases with redshift and this is probably
the main reason for the increase in $b$ (more luminous galaxies trace
higher density peaks and therefore are more biased). There is one exception
to this tendency in slice $z=0.30-0.40$. The values of $b$ for the best fit 
$\sigma_8 = 0.85$ are shown in Table \ref{tab:largescales}, were we can see
how $b$ is lower for  $z=0.30-0.40$. This is due to the LRG selection
explained in \S\ref{sec:data}, which includes a population
of less luminous galaxies $M_r>-21.5$ at $z\simeq 0.35$ (see Fig.\ref{fig:slicesplot1})
resulting in a lower bias. 

Once we know the bias we can use the values of $\beta$ to estimate
the linear velocity growth function $f(\Omega_m)$ for each slice, which are also shown
in Table \ref{tab:largescales}.

\subsection{Modified gravity}

We can also use redshift distortions to constrain modified gravity.
For standard gravity, the linear theory growth factor, $D(a)=\delta/\delta(0)$, 
depends purely on the expansion history $H(a)$, $w(a)$, $\Omega_m(a)$. 
Any discrepancy found between the observed growth factor and predictions 
based on the expansion history can be used to test the gravity.

The idea of using the growth of structure to test gravity has a long
history and dates back to Brans-Dicke (BD) model (see 
%\cite{brans}
Brans 2005 for a
historical review). More recently, after cosmic acceleration, 
the BD and variations have been revisited to explore structure
formation outside general relativity \cite{lobo,lue1,lue2,acquaviva}.
%\cite{zhang} 
Zhang et al (2007) propose a test to discriminate between 
different models for gravity on cosmological scales
based on lensing. 
%\cite{nesseris} 
Nesseris and Perivolaropoulos (2008) 
%have compiled a data set of various data
%points at a redshift range that can be used to constrain the linear perturbation
%growth rate $f$ through redshift distortions or indirectly through the rms mass
%fluctuation $\sigma_8(z)$ inferred from $Ly-\alpha$. 
have compiled a data set of various observations of the growth rate $f$ at different redshifts, obtained through redshift distortions or indirectly through the rms mass
fluctuation $\sigma_8(z)$ inferred from $Ly-\alpha$. These data points are used to constrain and parametrize the linear perturbation
growth rate $f$.  %\cite{wang}
Wang (2007) does a
prediction of the characteristics that a survey must accomplish to be able to
rule out the DGP gravity model (an extra-dimensional modification of gravity),
where the idea is to calculate H(z) from the baryon acoustic peak and f(z) from
redshift distortions.  
% \cite{guzzo} 
Guzzo et al (2008) test the nature of cosmic acceleration
using galaxy redshift distortions at z=0.8, obtaining $f$, but errors are still
too high to distinguish between different theories. 
%\cite{acquaviva} 
Acquaviva et al (2008) have
recently done a new compilation of results. See 
%\cite{zukin}
Bertschinger and Zukin (2008) for a theoretical
approach to modified gravity.

\begin{figure} \centering{ \epsfysize=6cm\epsfbox{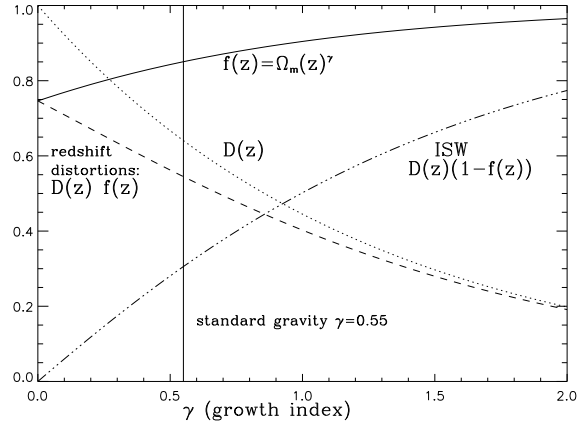}}
\caption{We see here the change of $f$ and $D$ with $\gamma$ and the factors
involved in redshift distortions and ISW effect. We have fixed $\Omega_m=0.25$
and $z=0.34$, as in the main LRG catalog  \label{fig:gammachange}} \end{figure}

If we fix a value for $\sigma_8$, which can also be known from other
observations, we can assume that the changes in the $\xips$
amplitude with redshift could  be  explained
by changes in the growth factor $D(z)$ due to a different
law of gravity at cosmological scales \cite{linder1}. This
can be represented by the growth index $\gamma$. Both $f(z)$ and the growth
factor $D(z)$ changes with $\gamma$. We have plotted in Fig.\ref{fig:gammachange}
how $f$ and $D$ change with $\gamma$  for z=0.34 (mean redshift in LRG) and
$\Omega_m=0.25$.  The measured $\xips$ amplitude on large
scales depends linearly on the product $\sigma_8  b(z)  D(z)$. If
we now fix $\sigma_8$ we can have an estimation of $b(z)D(z)$ as well as our
separate estimation of $\beta$, from the quadrupole. 
We therefore can use this combined data to produce an
estimate of $f(z)D(z)$.

\begin{figure} \centering{ \epsfysize=5cm\epsfbox{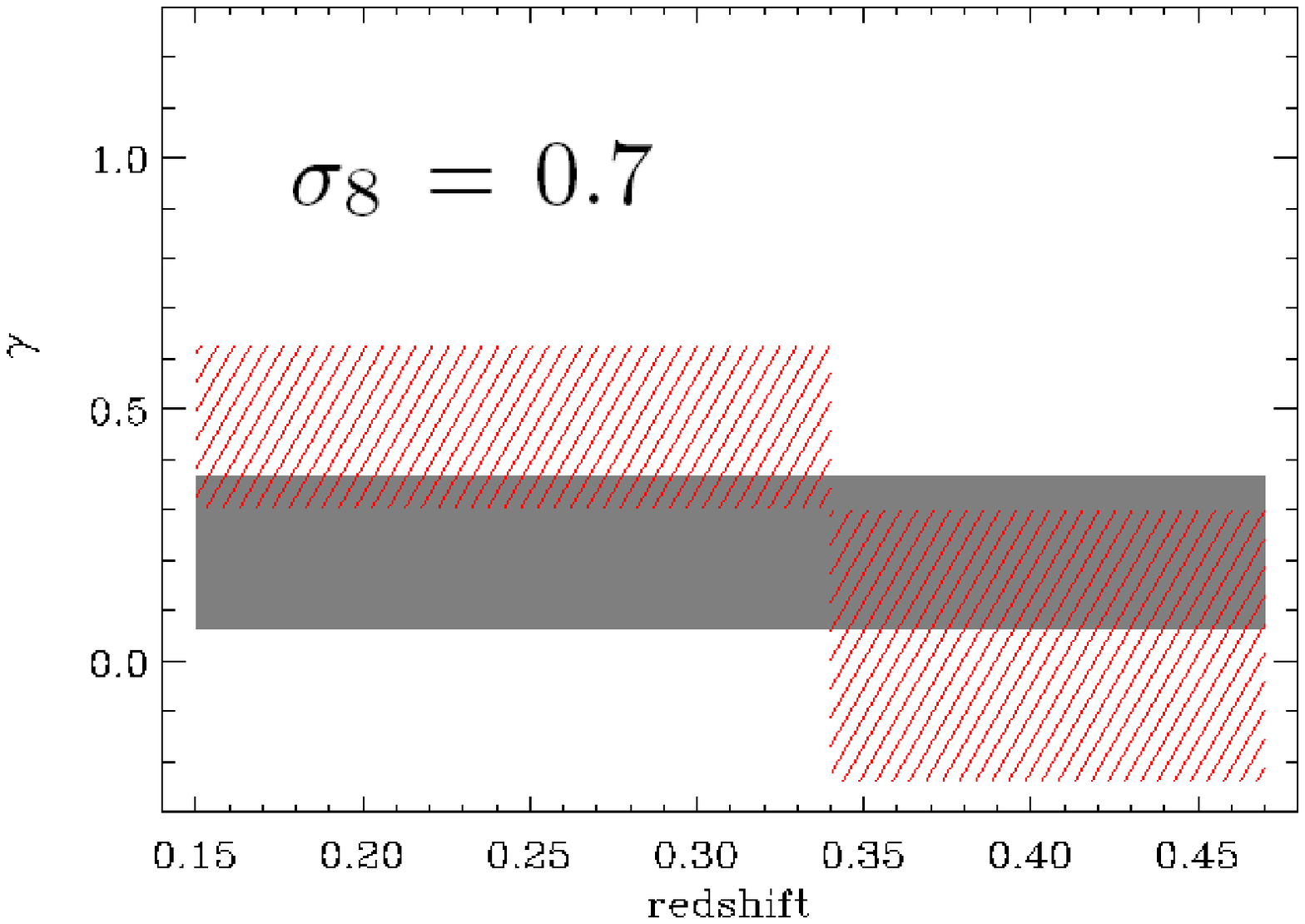}}
\centering{ \epsfysize=5cm\epsfbox{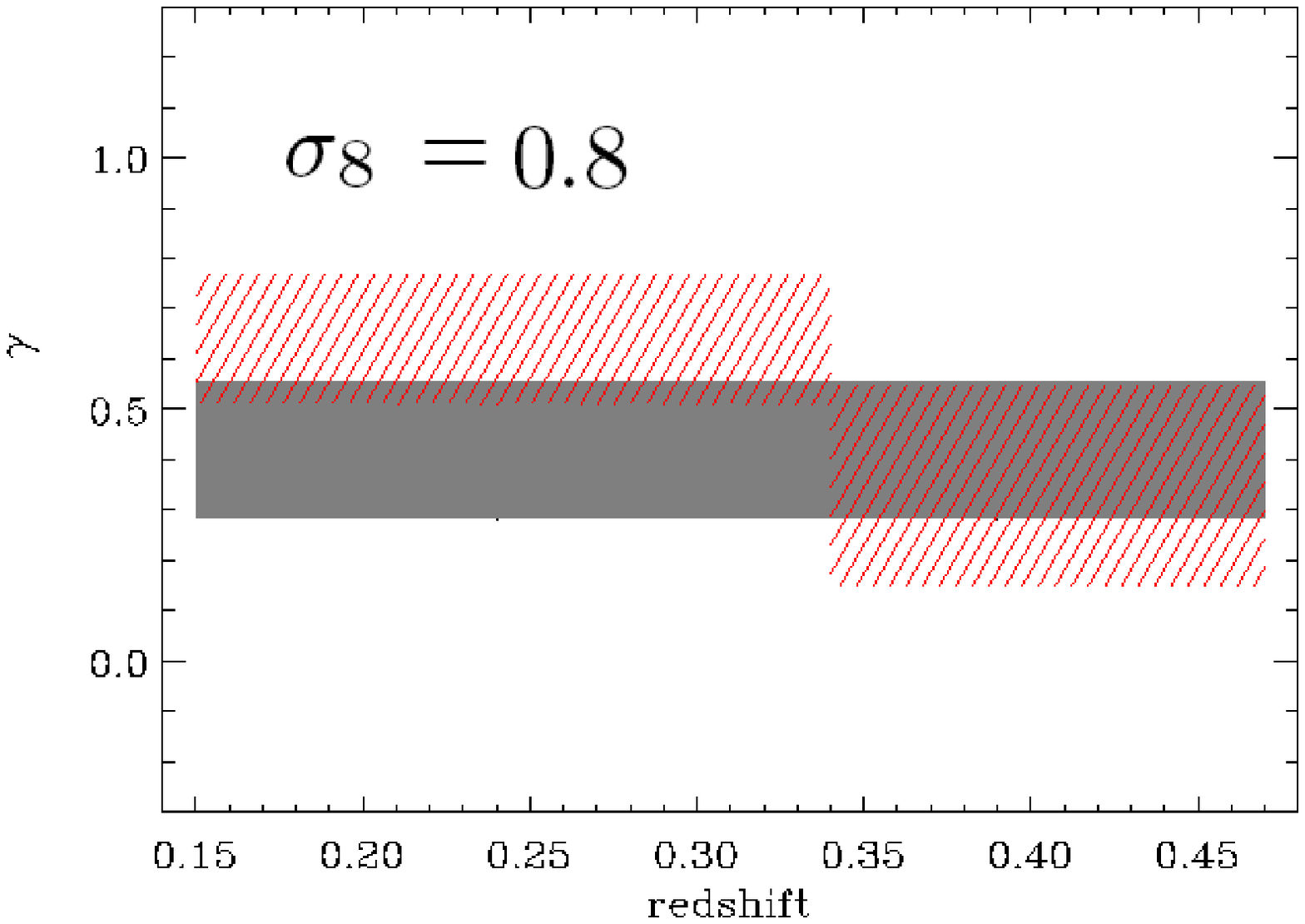}} 
\centering{\epsfysize=5cm\epsfbox{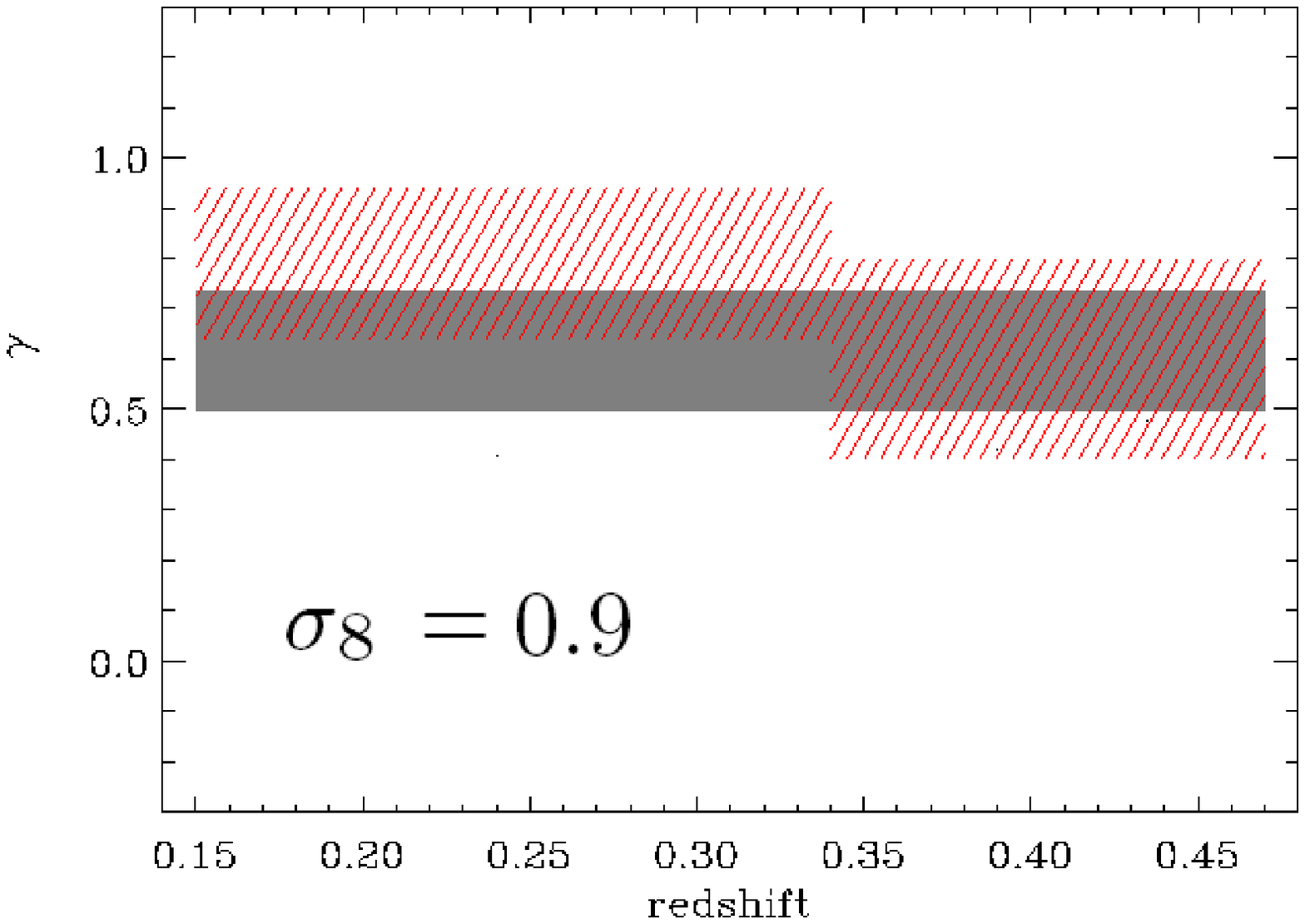}} 
\caption{Growth index
$\gamma$ for different redshift slices as in top panel of Fig.\ref{fig:s8found} when we fix
$\sigma_8=0.7,0.8,0.9$ (from top to bottom). As we increase $\sigma_8$ we
find that $\gamma$ also wants to be higher, in favor of modified gravity,
such as DGP, where $\gamma=0.68$ (in standard
gravity $\gamma=0.55$) 
\label{fig:gammaplot2}} \end{figure}

We have assumed that the observations can be explained by varying 
$\gamma$, for a fixed  dark energy equation of state $w$. This is a good approximation
because $w$ only depends slightly on $\gamma$ \cite{linder3}. We show in
Fig.\ref{fig:gammaplot2} our estimation of $1-\sigma$ errors for the growth
index $\gamma$ once we fix $\sigma_8$=0.7, 0.8, 0.9 
and marginalized over $\beta$, $Amp$ and $\Omega_m$.
As shown in
Fig.\ref{fig:gammachange}, the product $f(\gamma)D(\gamma)$, at a given
redshift, decreases with $\gamma$. If we change the value $\sigma_8$ to higher
values, the factor $\beta \;b(z_{slice})\;D(z_{slice})$ obtained from observations will
be lower, and this needs to be compensated by reducing $f \; D$, thus
increasing $\gamma$. In next section (\S\ref{sec:angular}) we will show
how the argument goes in the
opposite way when we work with ISW effect. At $2-\sigma$ and for all the
redshift slices, $\gamma$ is consistent with a standard gravity, except for
$\sigma_8=0.7$ where we need $3-\sigma$ for the last slice, favoring a
$\sigma_8$ clearly higher than 0.7, which is in agreement with recent observations.

Following this argument, we can calculate which is the range of $\sigma_8$ that
is consistent with standard gravity, $\gamma=0.55$. We find $0.80 \le\sigma_8\le0.92$ at
1-$\sigma$. In the modified gravity DGP model, $\sigma_8$ needs to be higher
since $\gamma$ is also higher ($\gamma=0.68$). We could say that DGP model is
inconsistent with data if $\sigma_8 < 0.84$. These results are in agreement
 with the recent study of %\cite{huetsi2008}
Yamamoto et al (2008), which appeared as we were writing
 this paper.

\subsection{Angular correlation and cross-correlation with WMAP}
\label{sec:angular}

\begin{figure}
\centering{ \epsfysize=5cm\epsfbox{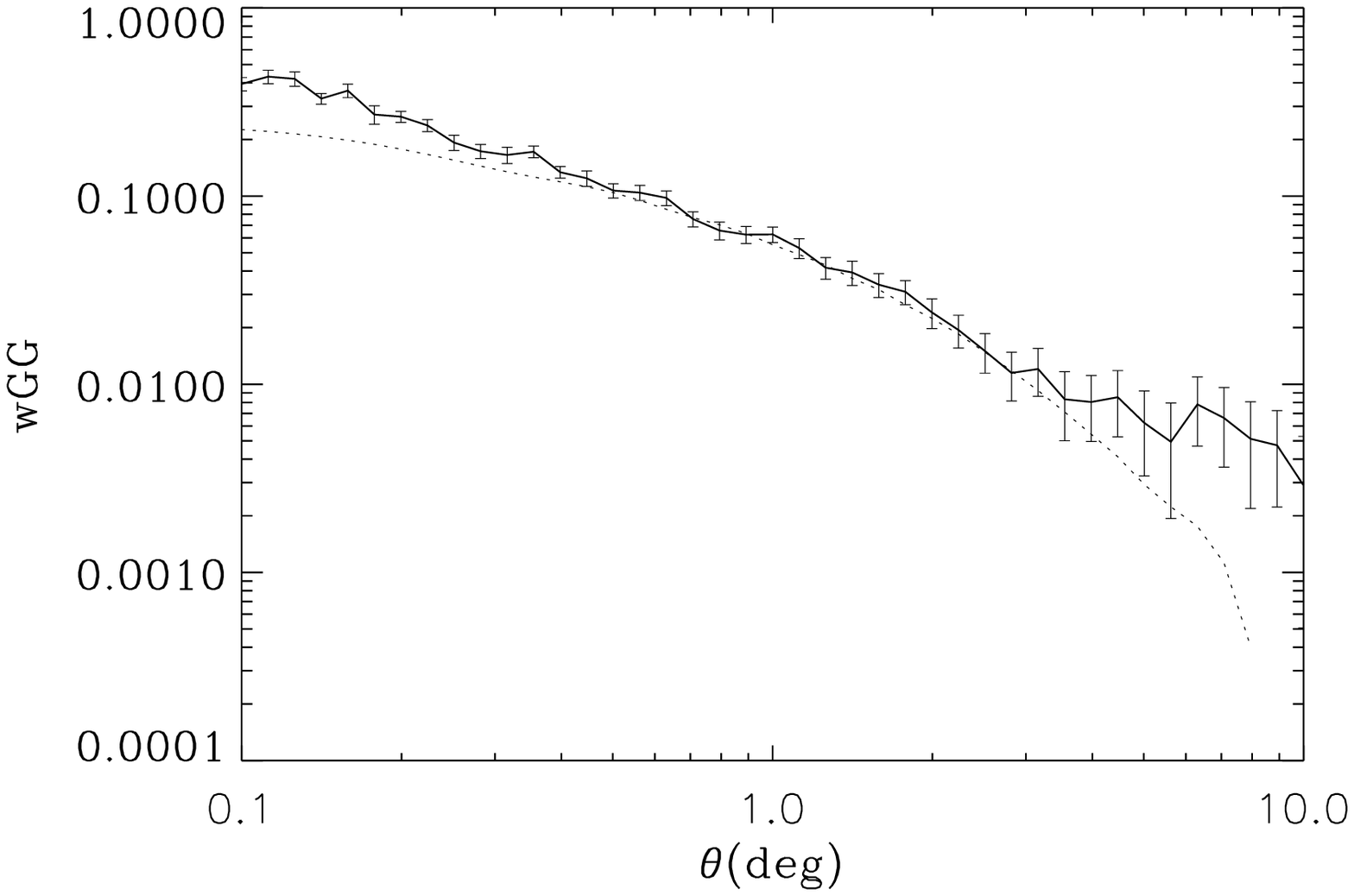}} 
\centering{\epsfysize=5cm\epsfbox{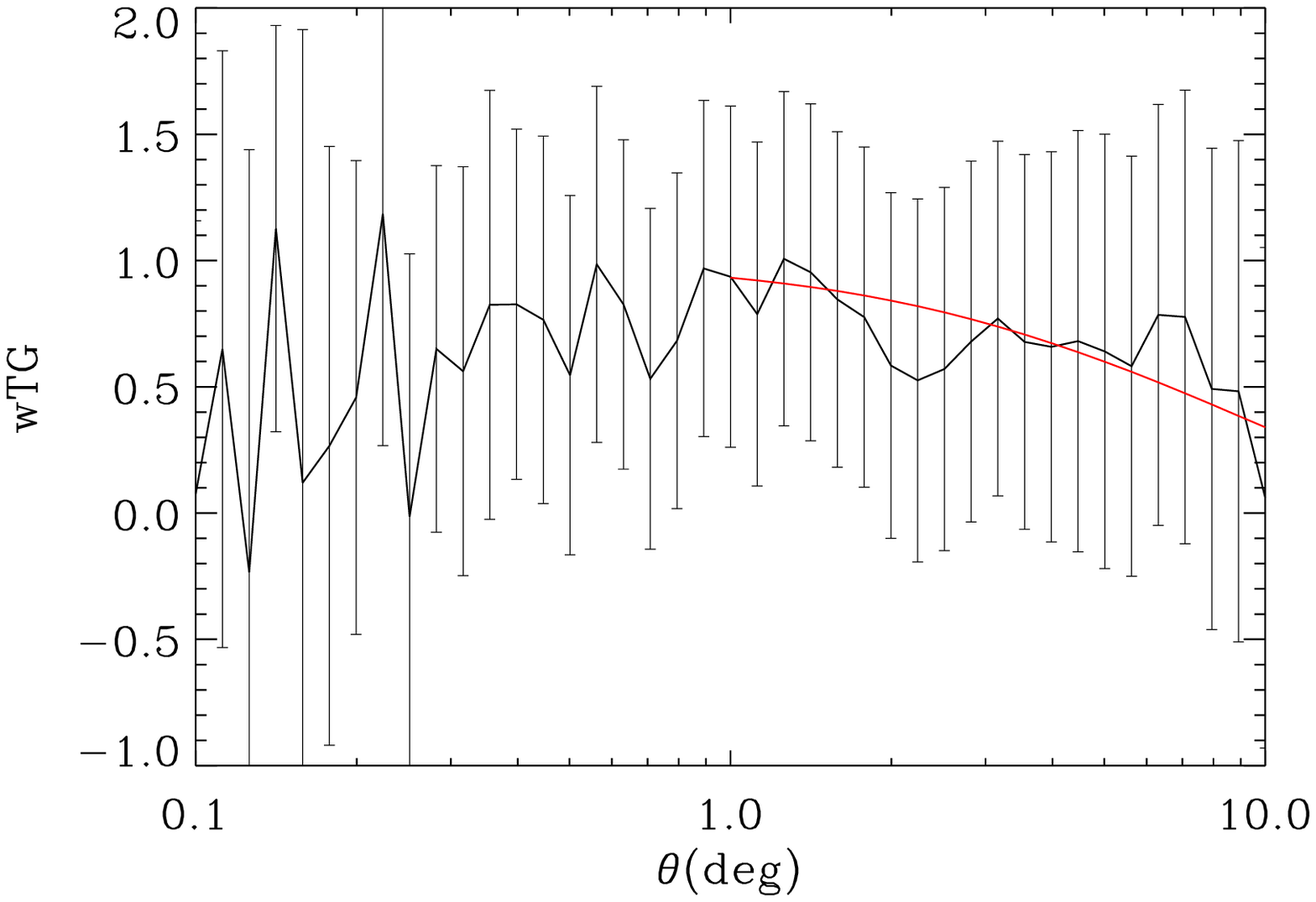}} 
%\centering{\epsfysize=5cm\epsfbox{figures/Chi2sigma8.z0.34.ps}} 
%\centering{\epsfysize=5cm\epsfbox{figures/Chi2gamma.z0.34.ps}} 
\caption{{\it Top
panel}: $w_{GG}$ of LRGs (solid with errors) and best linear model (dotted line) for the slice in redshift z=0.15-0.34.
{\it Bottom panel}: $w_{TG}$ (solid black with errors) and best model when fitting
$\sigma_8$, for a  $\sigma_8=1.2$ and standard gravity, $\gamma=0.55$ (solid red
line). \label{fig:GGTG34}} 
\end{figure}

We can also learn about the growth of fluctuations  using the ISW effect. This method
is independent to the one explained before with redshift distortions (see
%\cite{Cabre2}
Cabr\'e et al 2007 and references therein for an explanation of 
the ISW effect in the context of SDSS galaxies).
In this section we explore the angular correlation function and also the ISW
effect through the cross-correlation between galaxies and fluctuations of
temperature in WMAP. Fig.\ref{fig:GGTG34} shows the results for
$w_{GG}$ and $w_{TG}$ at the redshift slice z=0.15-0.34. This is the same data that
we have used before, but we now look at angular clustering.
The angular auto-correlation function GG scales as:

\begin{equation} w_{GG} ~\propto~ \sigma_8^2 \phi_G(z)^2b(z)^2D(z)^2, 
\end{equation}

while the cross-correlation function between galaxies and CMB temperature
fluctuations $w_{TG}$ is proportional to

\begin{equation} w_{TG} ~\propto~ \sigma_8^2 \phi_G(z) b(z)D(z)\frac{d[D(z)/a]}{dz},
\end{equation}
where

\begin{equation} \frac{d[D(z)/a]}{dz}=D(z)(1-f) \end{equation}
and $\phi_G(z)$ is the galaxy selection function.
Both $w_{GG}$ and $w_{TG}$ are proportional to $\sigma_8^2$ because
this factor comes from the normalization of the power spectrum, but $w_{GG}$ is
proportional to $(\phi_G(z)b(z)D(z))^2$ while $w_{TG}$ is proportional to
$(\phi_G(z)b(z)D(z))$ (from the clustering of galaxies) and
$\frac{d[D(z)/a]}{dz}$ (from the evolution of gravitational potentials).
Thus, this allows for a separate estimation of these quantities \cite{F03}.

We find that the measured signal of $w_{TG}$ in Fig.\ref{fig:GGTG34}
is higher than expected,
 a clear tendency that has been seen before (see
 Gaztanaga etal 2006,
%\cite{giannancompilation} 
Giannantonio et al 2008 for a compilation of ISW observations). The high
signal $w_{TG}$ could be due to: sampling variance,
 higher $\sigma_8$, lower $\Omega_m$, non-linear
effects, bias between matter and galaxies different from the one obtained from
galaxies-galaxies, non-linear bias, different form of dark energy as $w>-1$ (see
%\cite{Cabre1}
Cabr\'e et al 2006 for some hints in this direction), modified gravity at
cosmological scales, or non-linear magnification (linear magnification is not
expected to affect ISW at low redshifts, 
%\cite{Lo06}).
Loverde et al 2007).

The signal to noise is not very high so we can not obtain tight constraints, but
we can explore what could be creating this high signal. We  study two reasons
here: a change in $\sigma_8$ or a change in the growth index $\gamma$. We have also
studied these two parameters in the section above of redshift distortions and we want
to see if results are compatible.

We can break the degeneracy between $b(z)$ and $\sigma_8$ in the
auto-correlation function $w_{GG}$, which is proportional to $b^2\sigma_8^2$, by
combining the result with $w_{TG}$, which goes  as $b\sigma_8^2$. We 
will assume that 
$b(z)D(z)=b(\bar{z}_{slice})D(\bar{z}_{slice})$ is constant through all the redshift
slice, to be consistent with the previous section of redshift distortions.

We fix the shape ($\Omega_m=0.25$ and flat universe) of $w_{GG}$ and use the amplitude
to find the factor $b(\bar{z}_{slice})D(\bar{z}_{slice})\sigma_8$. This should be equal
to the amplitude that we found in
previous sections, when analyzing 3D redshift distortions, since we 
are working with the same LRG galaxy samples. We find that this is the
case: both measurements of the amplitude are consistent within the
errors. There is an extra power at large scales ($>4$ degrees)
as in the 3D case that can be due to sampling variance.
 This is explained with more detail in  \S\ref{sec:systematicr}. 
We next fix the dark energy equation of state
parameter to $w=-1$ and $\gamma=0.55$ to standard gravity. We now try
to explain the observed amplitude of $w_{TG}$ by just changing $\sigma_8$.
We therefore break the degeneracy of $\sigma_8$ with bias $b(z)$ in a way 
that is completely independent from the one used in the previous section, based on
3D redshift distortions. In bottom panel of Fig.\ref{fig:GGTG34}, 
we have compared $w_{TG}$ to the best model (in red) which
corresponds to $\sigma_8=1.2$. This value is quite high
but the errors in $w_{TG}$ are also large, so that
 at 1-$\sigma$ (or $2-\sigma$ for some of the redshift
slices), the values of $\sigma_8$ are consistent
with our previous estimation.

We  next see how we could explain the observed high ISW signal if it is due to a 
modification of gravity. We fix $\sigma_8=0.8$ and vary
$\gamma$. We assume that $w(z)$ changes only slightly with $\gamma$ and that
almost all the variation with $\gamma$ in ISW comes from the factor $D(1-f)$.
Both $D$ and $(1-f)$ grow with $\gamma$, so  that $D(1-f)$ also grows with $\gamma$
(see Fig.\ref{fig:gammachange}). Thus if we fix $\sigma_8$ to a low
value we need larger values of $\gamma$ from the ISW effect, but
lower values of $\gamma$ from redshift space distortions.
Thus, this seems to be a promising test
because both ways to obtain the growth histories seem to be constrained by
data in opposite directions (ie see Fig.\ref{fig:gammachange}),
and we therefore can use these test to break degeneracies in future surveys with better ISW
signal-to-noise. For the slice z=0.15-0.34 and for a $\sigma_8=0.8$, 
we find that $\gamma$ needs to be $\gamma=0.8$, but errors are in this case really big, 
fully consistent with standard gravity.

From these observations, we conclude a preference for either a higher $\sigma_8$ or
equivalently, a higher $\gamma$ than in standard gravity. But this is only a 
2-sigma effect. High power could also be due to cosmic variance. 
Fundamentally,
we want to remark that ISW can provide independent and complementary information
to the growth parameter $\gamma$. This issue will be resolved with future surveys.

%\begin{figure}
%\centering{ \epsfysize=5.cm\epsfbox{figures/GG.ps}} %\centering{
%\epsfysize=5.cm\epsfbox{figures/TG.ps}} %\centering{
%\epsfysize=5.cm\epsfbox{figures/Chi2sigma8.ps}} %\centering{
%\epsfysize=5.cm\epsfbox{figures/Chi2gamma.ps}} %  \caption{z=0.15-0.47.
%$w_{GG}$, $w_{TG}$, $\chi^2$ for $\sigma_8$ and $\chi^2$ for $\gamma$ once we
%assume $\sigma_8=0.8$  \label{fig:GGTG}} %\end{figure}

%\begin{figure}

%\centering{ \epsfysize=5cm\epsfbox{figures/GG.z0.47.ps}} %\centering{
%\epsfysize=5cm\epsfbox{figures/TG.z0.47.ps}} %\centering{
%\epsfysize=5cm\epsfbox{figures/Chi2sigma8.z0.47.ps}} %\centering{
%\epsfysize=5cm\epsfbox{figures/Chi2gamma.z0.47.ps}} %  \caption{z=0.34-0.47.
%$w_{GG}$,$w_{TG}$, $\chi^2$ for $\sigma_8$ and $\chi^2$ for $\gamma$ once we
%assume $\sigma_8=0.8$ \label{fig:GGTG47}} %\end{figure}

\subsection{Different redshift slices}

\begin{figure} \centering{ \epsfysize=5.5cm\epsfbox{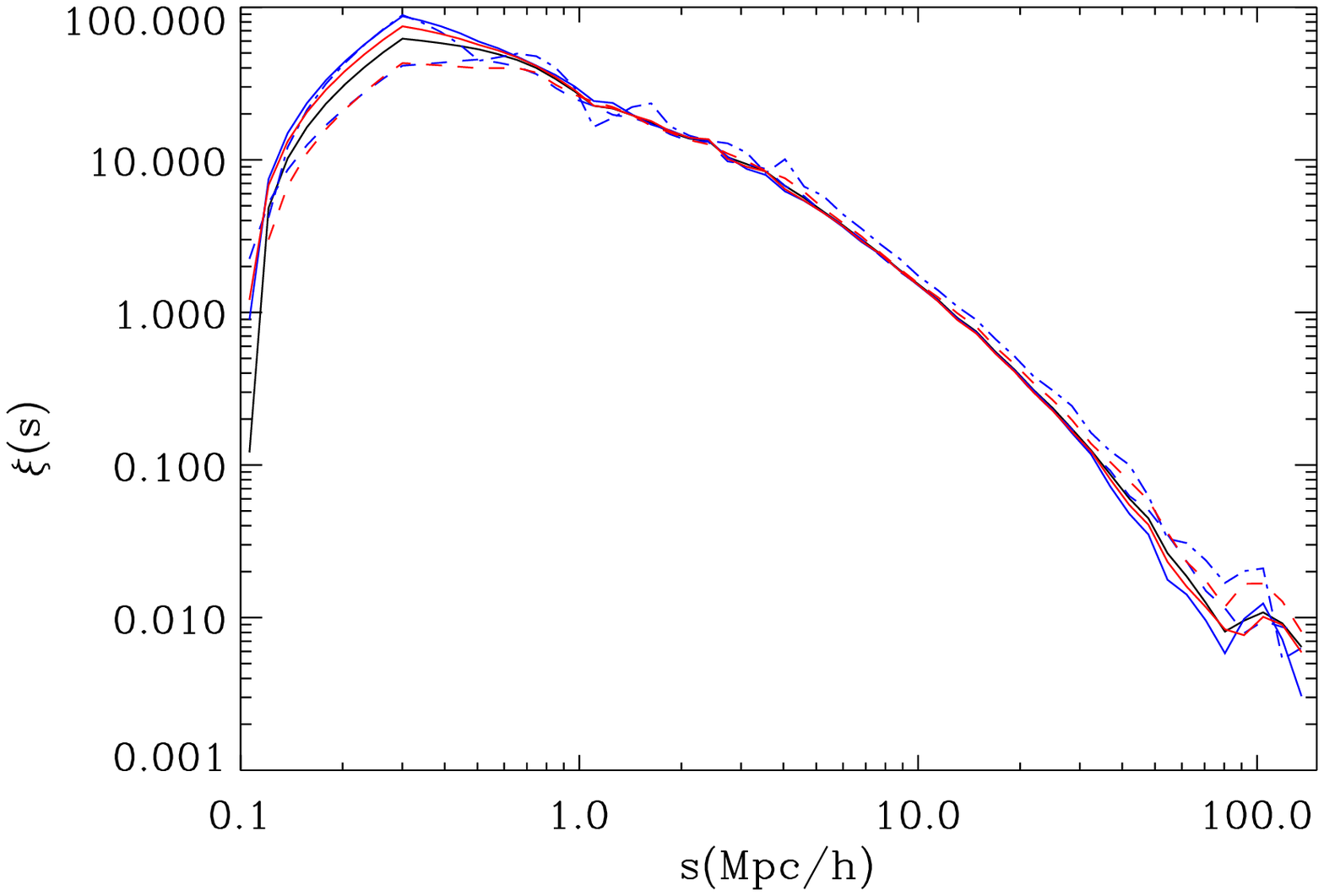}}
\caption{Comparison between $\xi(s)$ in redshift space for different slices in
redshift: All (black), z=0.15-0.3 (solid blue), z=0.3-0.4 (dashed blue),
z=0.4-0.47 (dashed-dotted blue); z=0.15-0.34 (solid red), z=0.34-0.47 (dashed
red)  \label{fig:monoslices}} 
\end{figure}

\begin{figure}
\centering{ \epsfysize=5.5cm\epsfbox{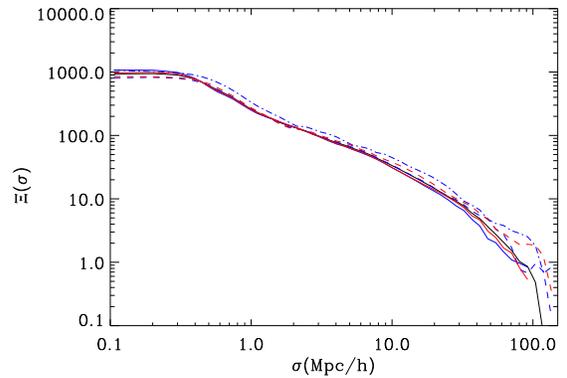}} \caption{Comparison
between the projected correlation function $\Xi(\sigma)$ for different slices in
redshift. All: black, z=0.15-0.3 (solid blue), z=0.3-0.4 (dashed blue),
z=0.4-0.47 (dashed-dotted blue); z=0.15-0.34 (solid red), z=0.34-0.47 (dashed
red)  \label{fig:perpslices}} 
\end{figure}

\begin{figure} \centering{ \epsfysize=5.5cm\epsfbox{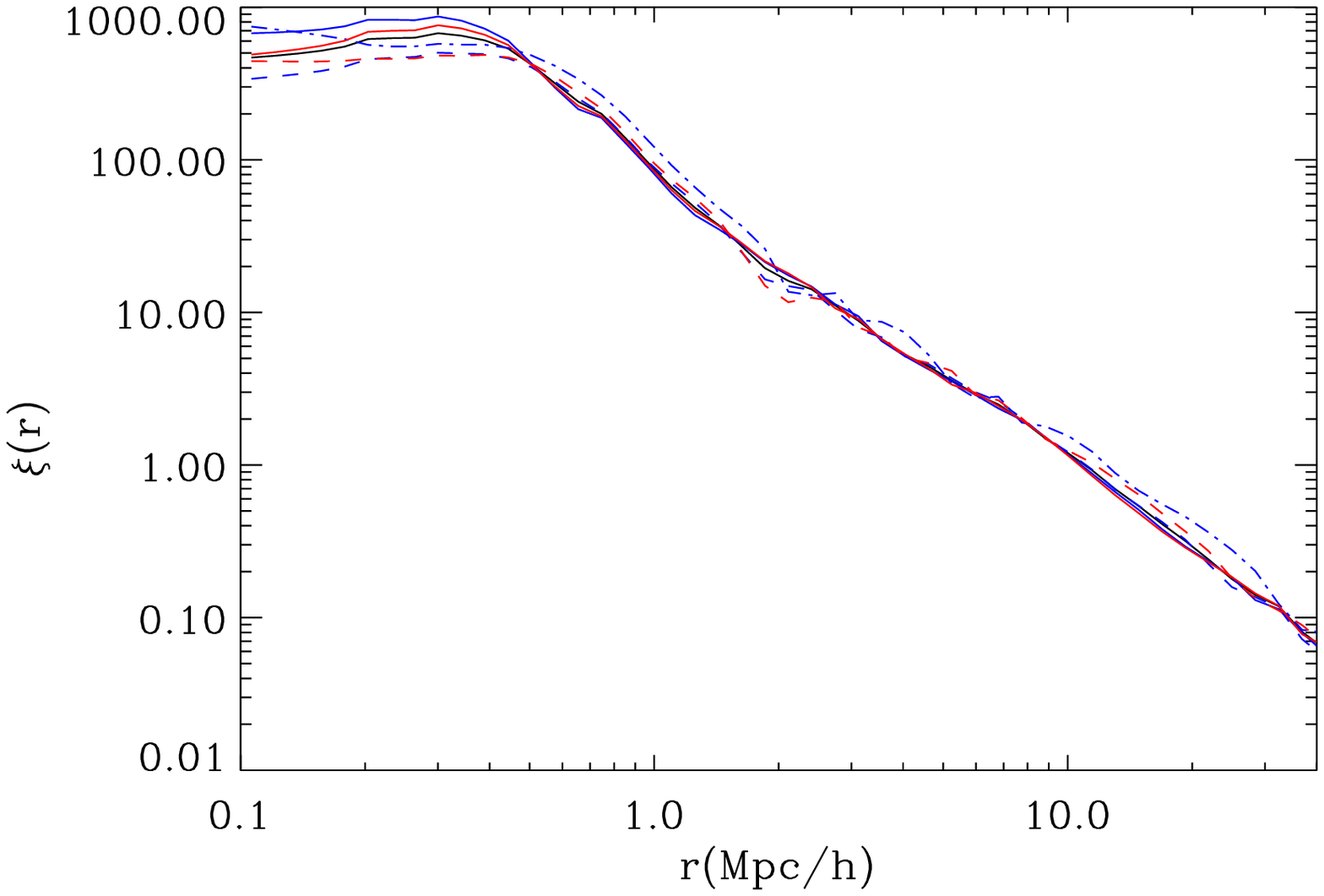}}
\caption{Comparison between $\xi(r)$ in real space for different slices in
redshift. All: black, z=0.15-0.3 (solid blue), z=0.3-0.4 (dashed blue),
z=0.4-0.47 (dashed-dotted blue); z=0.15-0.34 (solid red), z=0.34-0.47 (dashed
red)  \label{fig:corrslices}} 
\end{figure}

\begin{figure}
\centering{ \epsfysize=5.5cm\epsfbox{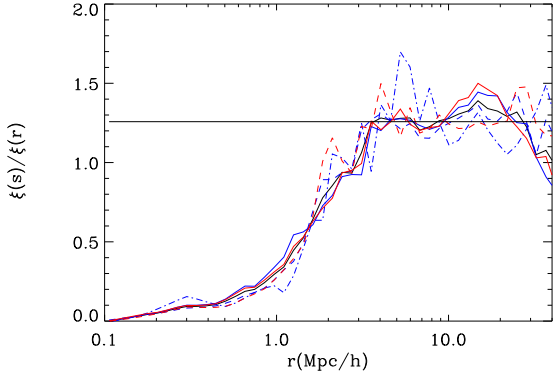}}
\caption{Comparison between $\xi(s)/\xi(r)$ for different slices in redshift.
All: black, z=0.15-0.3 (solid blue), z=0.3-0.4 (dashed blue), z=0.4-0.47
(dashed-dotted blue); z=0.15-0.34 (solid red), z=0.34-0.47 (dashed red). The
constant solid black line shows the Kaiser expression for $\xir/\xis$ at large
scales (as a function of $\beta$, see Eq.(\ref{eq:xisr})) for $\beta=0.34$
\label{fig:corrscorrslices} } 
\end{figure}

We next look at the differences between the redshift slices in the monopole
(Fig.\ref{fig:monoslices}), in the projected correlation function $\Xi(\sigma)$
(Fig.\ref{fig:perpslices}), in the real-space correlation function $\xir$
(Fig.\ref{fig:corrslices}) and in the ratio $\xis/\xir$
(Fig.\ref{fig:corrscorrslices}).

As we have seen in section \S\ref{sec:secquadru}, $\beta$ is similar for all the
redshift slices. We can also see this in Fig.\ref{fig:corrscorrslices},
which shows that the ratio $\xis/\xir$ is quite similar at large scales,
indicating similar values of $\beta$ (Eq.(\ref{eq:xisr})) in all samples. 
In some of the
cases, there seems to be a turning down of the $\xis/\xir$ ratio at 
scales larger than 30 Mpc/h. This is due to the bias in the recovered
$\xir$  which does not work well for scales larger than 30Mpc/h
(ie see Fig.\ref{fig:perpint}).

The monopole is approximately a measure of the shape of real-space correlation function for
large scales, but scaled up by a function of $\beta$ similar for all the slices.
Looking at the monopole (Fig.\ref{fig:monoslices}) and also at the projected
correlation function (Fig.\ref{fig:perpslices}) and at the real-space
correlation function (Fig.\ref{fig:corrslices}) we see that all the slices
except from the further one (blue dash-dot) lay in the same line, meaning that
$D(z)b(z)$ is almost constant with redshift, what is sometimes called stable clustering.

In Fig.\ref{fig:corrslices} we also see that the change of the slope at small
scales moves to larger scales as we more to more distant slices. We think that this
effect is probably not physical but due to geometry since it corresponds 
to a fixed angular scale, ie the mask or the fiber collision problem.

\begin{figure}
\centering{ \epsfysize=5.5cm\epsfbox{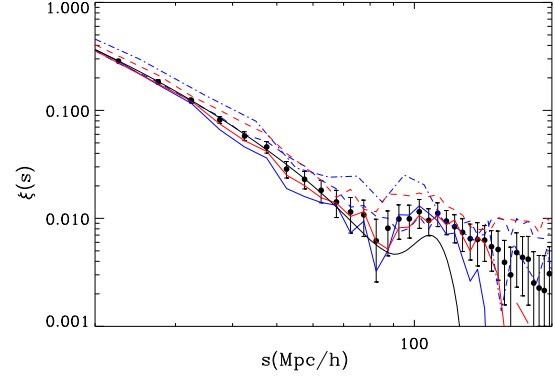}}
  \caption{Comparison between the BAO peak in $\xi(s)$ for different
 slices in redshift. All: black dots with errors, z=0.15-0.3 (solid blue), 
z=0.3-0.4 (dashed blue), z=0.4-0.47 (dashed-dotted blue);
 z=0.15-0.34 (solid red), z=0.34-0.47 (dashed red). 
 These measurements are compared to 
the best fit linear model for the principal redshift slice 
(solid line)  \label{fig:baoslices}}
\end{figure}

Fig.\ref{fig:baoslices} shows a zoom over
the BAO peak region, as measured in the monopole.
This peak is roughly in agreement with
predictions and is dominated by modes in the
perpendicular direction. The BAO peak is detected
in all slices and our best fit model (shown as
a continuous line in the Figure) seems a bit lower
than the data. This can be fixed by exploring 
the full parameter space of cosmological models
and will be done in Paper IV of this series.
This is an important test
because it reproduces previous results and indicates
that systematic effects can not be too important in our
analysis on scales of 100 Mpc/h or smaller. On larger
scales, some of the slices show more power than
predicted by models. This is compatible with the errors,
but we wonder if there could be some hidden systematics
here. We will explore this further in the  Appendix B
which is dedicated to the study of systematic effects.

%\clearpage

\section{Discussion and Conclusion} 
\label{sec:discuss}

In this paper we work with luminous red galaxies (LRGs) from the catalog SDSS
DR6 to obtain cosmological parameters from large scales. These are good galaxies
to study this range of scales since they trace a big volume of the universe. We
use a simple approximation to model the redshift space distortions (see
\S\ref{sec:model2point}), which work incredibly well with LRGs on large
scales, as explained in this paper, but also on small scales, shown
in Paper II \cite{paper2}  of this series.

We have used one of the largest N-body simulation run to date (with box
size of L=7680Mpc/h with $2048^3$ particles) to create realistic
mock LRG galaxy catalogs to analyze the data (Appendix A). We have checked our
results to some extent with smaller simulations. This is a crucial step given the
complications in the data and the accuracy needed for its interpretation.
We have validated all the methods that we use during the paper with credible 
mock  simulations. We also test the validity of the models and the errors: 
Monte Carlo, jackknife and a theoretical approach for $\xisp$. We find that JK 
errors does not work properly for large perpendicular distance $\sigma$.

There are usually two important sources of statistical errors in clustering
analysis: Shot-noise error, due to the finite number of galaxy pairs, and
sampling errors, due to the finite amount of volume in the survey. We
have shown with our mock simulations that the LRG samples in SDSS DR6 
are dominated by the shot-noise term in the 2-point $\xisp$. Usually, this term is assumed to follow
the Poisson distribution, with an error given by the inverse of the square
root of the number of measurements (ie see Eq.(\ref{eq:Poisson})). This is in fact
the case to a good accuracy in our dark matter mocks. But this is not the
case for groups of dark matter particles (halos) which are selected in a way that,
by construction, the detection of a group creates a large excluding region
around the group position.  This produces what we have called super-Poisson errors,
which are about $40\%$ larger than Poisson errors. Jack-knife errors are shown 
to be good at detecting this additional shot-noise component and we find
that this effect is also present in the real LRG data. This is interpreted here as 
direct evidence that the dominant fraction of the LRG galaxies populates 
separate dark matter halos. But this does not exclude a minor fraction of them
%to populate more than one halo. 
to be found with other LRGs within the same halo.
In fact, in Paper III, \cite{paper3}, of this
series we find evidence for a large quadratic bias which indicates that this
is the case \cite{bhalo}.

We conclude our discussion on errors by emphasizing the importance of using
realistic mocks to assess our results. We have demonstrated here that dark matter
mocks are not good enough. Diagonal, sampling errors can in principle be scaled from
dark matter to real data because the relative error only depends in the
sampling volume. But this is not the case for the shot-noise or for
the covariance. The difference is important. Errors and covariances
are underestimated with dark matter mocks and results will be wrongly
interpreted.

We have calculated the distortion parameter $\beta=[0.310-0.375]$ using the
normalized quadrupole Q(s), which also provides us of an effective value 
for the dispersion of random peculiar velocities $\sigma_v=[365-415]$ km/s (see Table
\ref{tab:tabbetasigma} for more slices in redshift). The great advantage of the
normalized quadrupole introduced by 
%\cite{hamilton1992}
Hamilton (1992) is that it can measure
the squashing produced by bulk galaxy motions with independence of the overall
amplitude or shape of the galaxy 2-point correlations. We have also checked
if this value for $\beta$ gives the correct ratio $\xis/\xir$ on large scales.
We have shown that with actual
data, we can only recover the real-space correlation function $\xir$ for scales
smaller than 30Mpc/h, and we see that the ratio between the redshift-space
monopole $\xis$ and real-space correlation function $\xir$, $\xis/\xir$, can be
described by the approximated Kaiser value at large scales, with the same
value of $\beta$ found in the quadrupole.

We obtain a value for $\Omega_m$ from the shape  of the
redshift-space correlation function in the anisotropic plane $\pi-\sigma$
at large scales. The value is $\Omega_m=[0.225-0.265]$ for the mean slice 
(see Table \ref{tab:largescales} for other redshift slices). We also obtain
 a value for the amplitude $b\sigma_8$ in the linear regime. We can therefore
break the
degeneracy between $\sigma_8$ and $b$ thanks to redshift distortions through the
parameter $\beta$, which results in $\sigma_8=[0.79-0.91]$ and $b$ around 2.

We next look at the growth history and possible modifications of gravity
through the parameter $\gamma$ which is 0.55 for standard gravity and 0.68 for
the modified gravity DGP model. We find consistence with standard gravity for
$0.8\le\sigma_8\le0.92$ at 1-$\sigma$. DGP model is inconsistent with our
results if $\sigma_8 < 0.84$.  

We have also cross-correlated LRGs with WMAP in order
to investigate the ISW effect, obtaining a high signal, as in other studies,
compared to current $\Lambda CDM$ model. The degeneracy  $b-\sigma_8$ and
the growth history can also be studied using the ISW effect which is 
independent from the study of redshift distortions. The cross-correlation
indicates a preference for higher $\sigma_8$.

In the Appendix B we look for systematic errors, and arrive to the conclusion 
that our results are robust in the scales we are doing the analysis.

\section*{Acknowledgments}

We would like to thank Pablo Fosalba, Francisco Castander, Marc Manera
and Martin Crocce for their help and support at different stages of
this project.
We acknowledge the use of simulations from the MICE consortium 
(www.ice.cat/mice) developed at the MareNostrum supercomputer
(www.bsc.es) and with support form PIC (www.pic.es),
 the Spanish Ministerio de Ciencia
y Tecnologia (MEC), project AYA2006-06341 with
EC-FEDER funding, Consolider-Ingenio CSD2007-00060
and research project 2005SGR00728
from Generalitat de Catalunya. AC acknowledge support
from the DURSI department of the Generalitat de
Catalunya and the European Social Fund.

%\bibliographystyle{mn2e}
%\bibliography{tesi2}

\appendix

\section{Simulations and errors}\label{sec:errors}

We first present the simulations used in this
paper. Next we show how simulations have been used to provide an
error estimate for the different statistics used in this paper.
Finally we use the simulations to validate the methods that
we will  apply to real data. We will conclude that the methods 
and errors that we use to obtain parameters from LRG data are all 
well validated by simulations.
We follow similar methodology and notation as in 
%\cite{Cabre1}.
Cabr\'e et al (2006).

\subsection{Simulations}

We have used different comoving outputs
(between z=0.0 and z=0.5) of a MICE simulation, run in the
supercomputer Mare Nostrum in Barcelona by the MICE consortium (www.ice.cat/mice) in
order to study the errors and validate the models used \cite{onion}. The  simulation contains
$2048^3$ dark matter particles, in a cube of side 7680Mpc/h, $\Omega_M=0.25$,
$\Omega_b =0.044$, $\sigma_8=0.8$, $n_s=0.95$ and $h=0.7$, with a flat cosmology. We have divided this
big cube in $3^3$ cubes of side 2 x 1275Mpc/h, and taking the center of these
secondary cubes as the observation point (as if we were at z=0), we apply the
selection function of LRG, which extends to z=0.47 (r=1275Mpc/h). We can
obtain 8 octants from the secondary sphere included in the cube, so at the end
we have 8 mock LRG catalogs from each secondary cube, which have the same
density per pixel as LRG in order to have the same level of shot noise, and the
area is slightly smaller (LRG occupies 1/7 of the sky with a different shape).
The final number of $M$ independent mock catalogs is 216 (27x8).

We use both dark matter particles and groups in this simulation.
By definition, there is no
bias in the dark matter particles, so
$\beta=\frac{\Omega_m(z)^{0.55}}{b}=0.62$, where $\Omega_m(z)=\Omega_m
(1+z)^3/(\Omega_m (1+z)^3+1-\Omega_m)$ and $b=1$.
To simulate biasing we 
select groups of particles using friend-of-friends with linking scale of $0.20$
(we have also used $0.16$ with consistent results).
At $z=0$ we find a total of 107 million groups with more than 5 particles 
($M>1.87 \times 10^{13}$). When the number of particles in the group is
larger than about 40 or so, they usually correspond to DM halos
 and have similar mass function \cite{gus08} and bias. We find that 
 $50-80\%$  of groups with 5-40 particles (respectively) correspond to
 DM halos. The rest of the groups are just high density regions, which
 in any case are also candidate sites for bias galaxy formation. We
 therefore use groups of different masses as biased tracers and use the mass threshold to
choose the group subsample that better matches the number density and clustering
amplitude in the real LRG data.
We have also used a MICE simulations with $2048^3$ dark matter particles, 
in a cube of side 3072Mpc/h (which we call MICE3072, same parameters as MICE7680)
which has 15 times better mass resolution to check for mass resolution effects.
We find very similar results in both cases to the extent that we can compare
the smaller number statistics.
When we  select groups with $M > 2-4 \times 10^{13}$,
both the clustering amplitude ($b_1 \simeq 1.9-2.2$ for $\sigma_8=0.8$) 
and the number density ($\bar{n} \simeq 4-6 \times 10^{-5}$) 
are similar to the real LRG galaxies in our SDSS sample (the range reflects
the fact that the actual number depends on LRG sample used, ie redshift
and selection). We have explored a number of cases to make sure we understand
how errors depend on number density and bias. We will focus in presenting
results for 3 different
cases: a) dark matter particles at $z=0.3$ ($b=1$, $\beta=0.62$), diluted
to fit LRG number density b) groups with $M>2.2 \times 10^{13}$
($b=1.9$, $\beta=0.25$) at $z=0$, c)  groups with $M>3.74 \times 10^{13}$
($b=3$, $\beta=0.23$) at $z=0.5$, which cover
the range $b$, $\beta$, $z$ and densities in the real LRG in our samples.

We generate the mock catalogs applying redshift distortions in the line-of-sight
direction,

\begin{equation} s=r+v_{r}/H(z)/a(z) \end{equation}

We obtain the Monte Carlo (MC) error from the dispersion of $M$ independent
realizations of our universe. For M realizations, the MC covariance is

\begin{equation}\label{eq:covMC} 
C_{ij} =  \frac{1}{M}
\sum_{k=1}^{M}(\xi(i)^{k}-\widehat\xi(i))(\xi(j)^{k}-\widehat\xi(j))
\end{equation}

where $\xi(i)^{k}$ is the measure in the k-th simulation (k=1,...M) and
$\widehat\xi(i)$ is the mean over M realizations. The case i=j gives the
diagonal error (variance). Typically, we need $M=100$ independent simulations
for the diagonal error (in order to obtain 5\% accuracy), and more for the covariance matrix, depending on the
case. The MC error is an estimation of the true error,  but it takes lots of
computational time and it also requires simulations with the same
particularities of the data analyzed. When we refer to off-diagonal 
elements of  $C_{ij}$ we use the normalized covariance:

\begin{equation}\label{eq:ncovMC} 
\hat{C_{ij}} \equiv  {C_{ij}\over{\sqrt{C_{ii}C_{jj}}}}
\end{equation}
which ranges from -1 to +1. Values of $|\hat{C_{ij}}|<0.2$
usually have little impact on the error analysis.

\begin{figure} 
\centering{\epsfysize=7cm \epsfbox{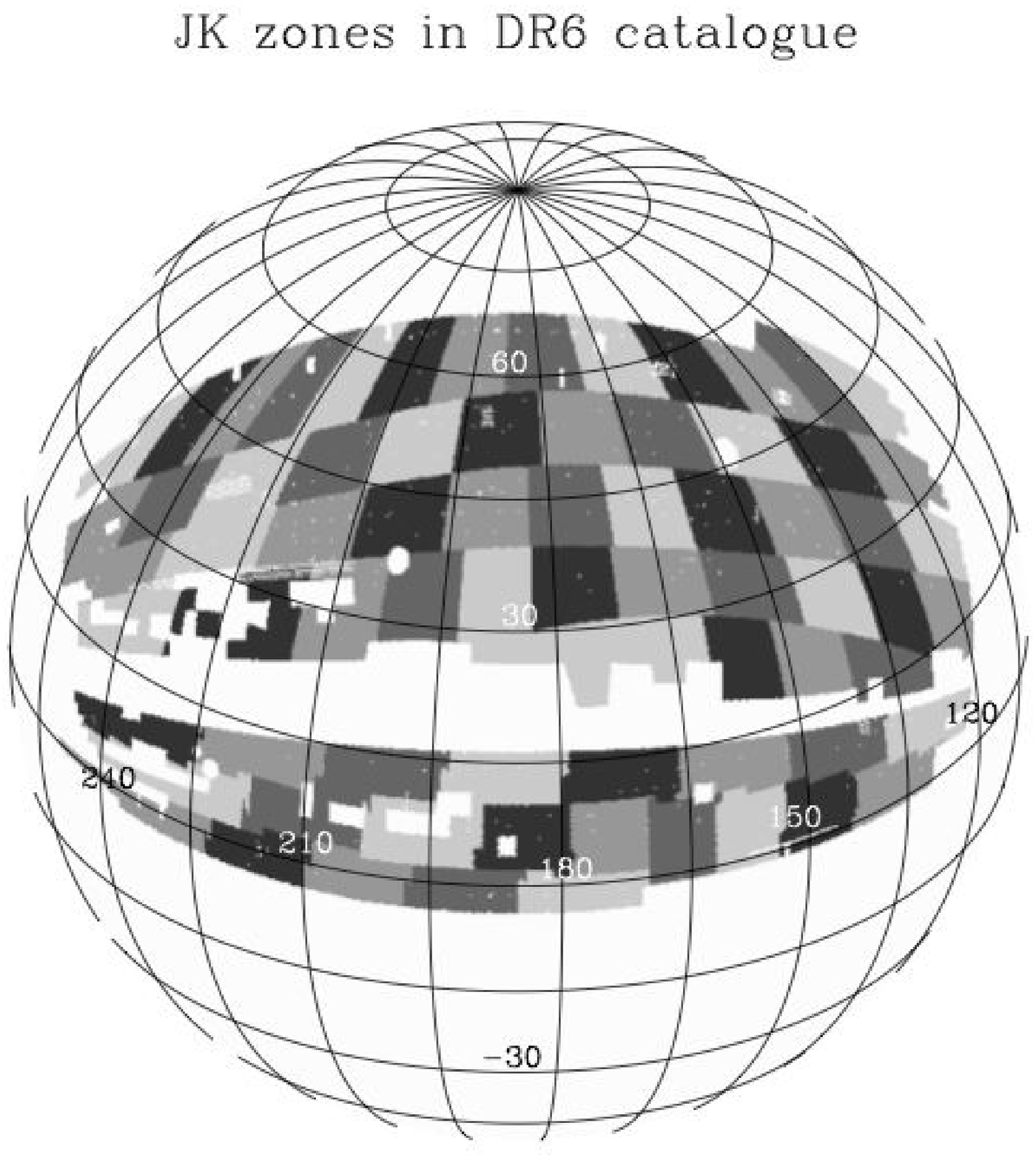}}
\centering{\epsfysize=7cm \epsfbox{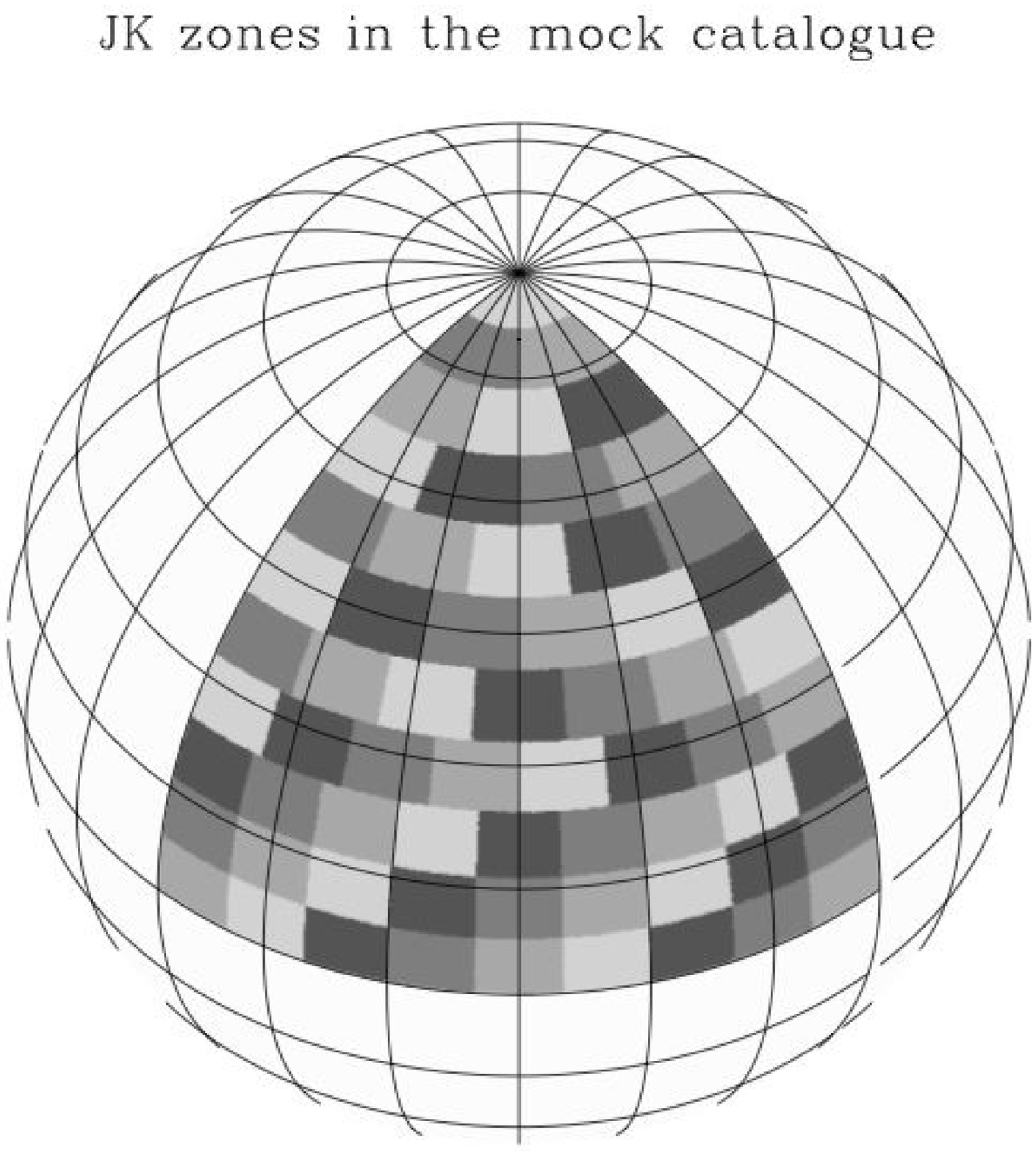}} \caption{Top panel: JK
zones for the LRG catalog (equatorial coordinates). Bottom panel: the mock
catalog with its JK zones\label{fig:jkzones}}
\end{figure}

In the case of the jackknife (JK) error, we obtain the different realizations
that we need to compute the error from the same data (or from a single 
realization in the case of simulations).
This is done by dividing the sample in M zones. Each JK subsample consists in
the whole catalog except from one of these M zones. In this case, as the
realizations are clearly not independent, we apply a multiplicative factor of
$(M-1)$ to the previous covariance to account for this effect.
Fig.\ref{fig:jkzones} shows the octant catalog used for the galaxy mocks divided
in 63 jackknifes zones, with similar area and shape; and the SDSS DR6 real
catalog divided in 73 jackknife zones. 
%%%%%%%%%%%%%%%%%%%%%%%%%%%%%%%%%%%%%%%%%%%%%%%%%%%%%%%%%%%%%%%%%%%%%%%%%%%%%%%%%%%%%%%%%%%%%%%%%%%%%%%%%%
Note how the real mask is different than the mask in the mocks, but note that
the JK regions are similar in shape.
We have tested that this difference in
mask does not make a large difference in the conclusions of our error analysis, so we find adequate to work with an octant which allows us to obtain a higher number of mocks, necessary for a good determination of the error. We have studied with a reduced set of mocks with the proper LRG mask that the general shape of the mask is not important for the large scale analysis as far as 
 it remains compact. Small holes in the mask do not affect the errors at large scales.
 In this way we can apply the conclusions obtained from our mocks to the real LRG data. 
We use the dark matter and group mock simulations to probe the limit in which 
the JK errors represent a good approximation.  We  then use the JK error from the real LRG data when adequate.
%%%%%%%%%%%%%%%%%%%%%%%%%%%%%%%%%%%%%%%%%%%%%%%%%%%%%%%%%%%%%%%%%%%%%%%%%%%%%%%%%%%%%%%%%%%%%%%%%%%%%
 We study the errors 
in the redshift space correlation function $\xi(\pi,\sigma)$, the monopole
 $\xi(s)$ and the quadrupole $Q(s)$. Finally, we use the simulations 
to validate the models that we will use with real data.\

\begin{figure*}
\centering{
\epsfysize=6cm\epsfbox{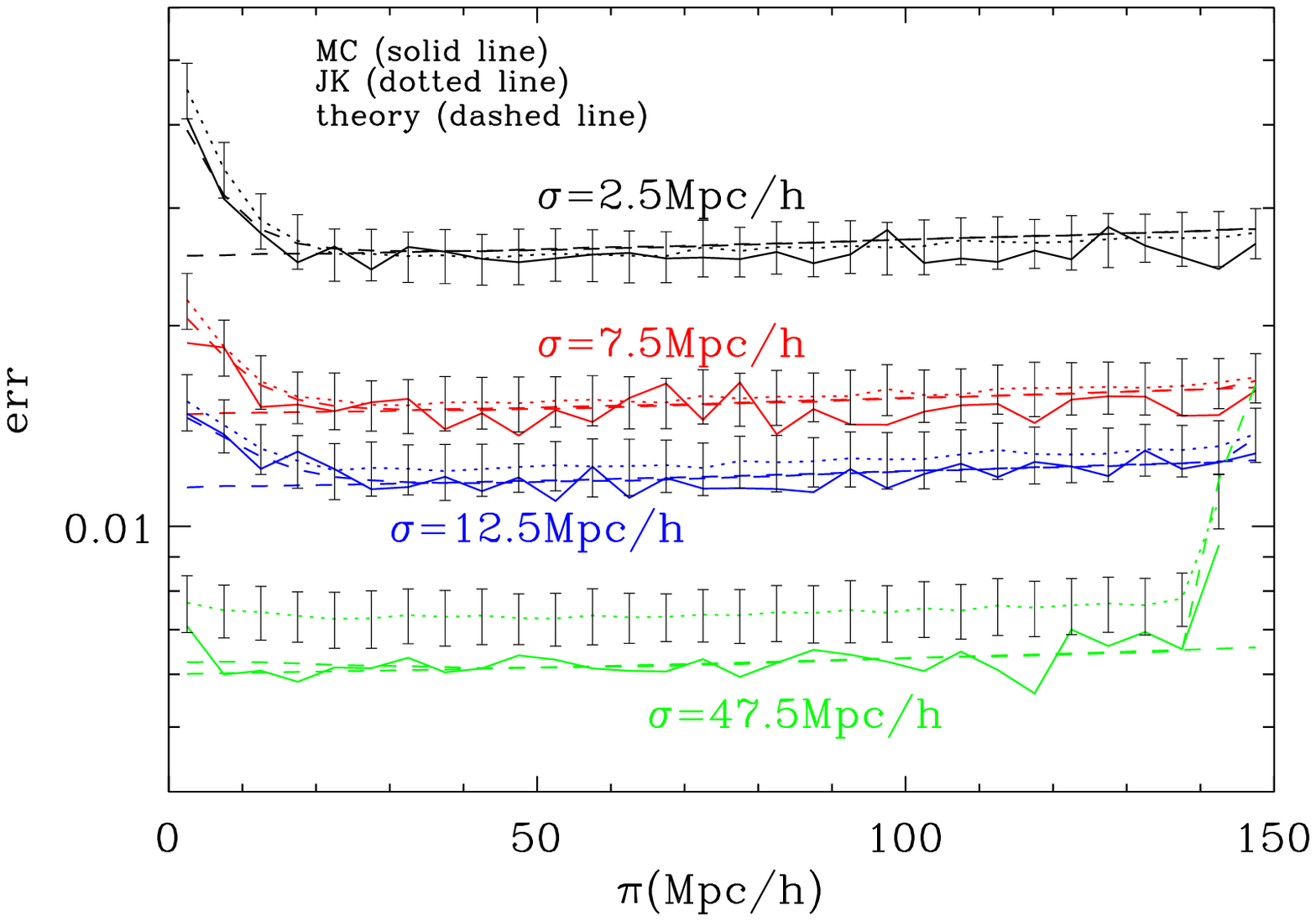}
\epsfysize=6cm\epsfbox{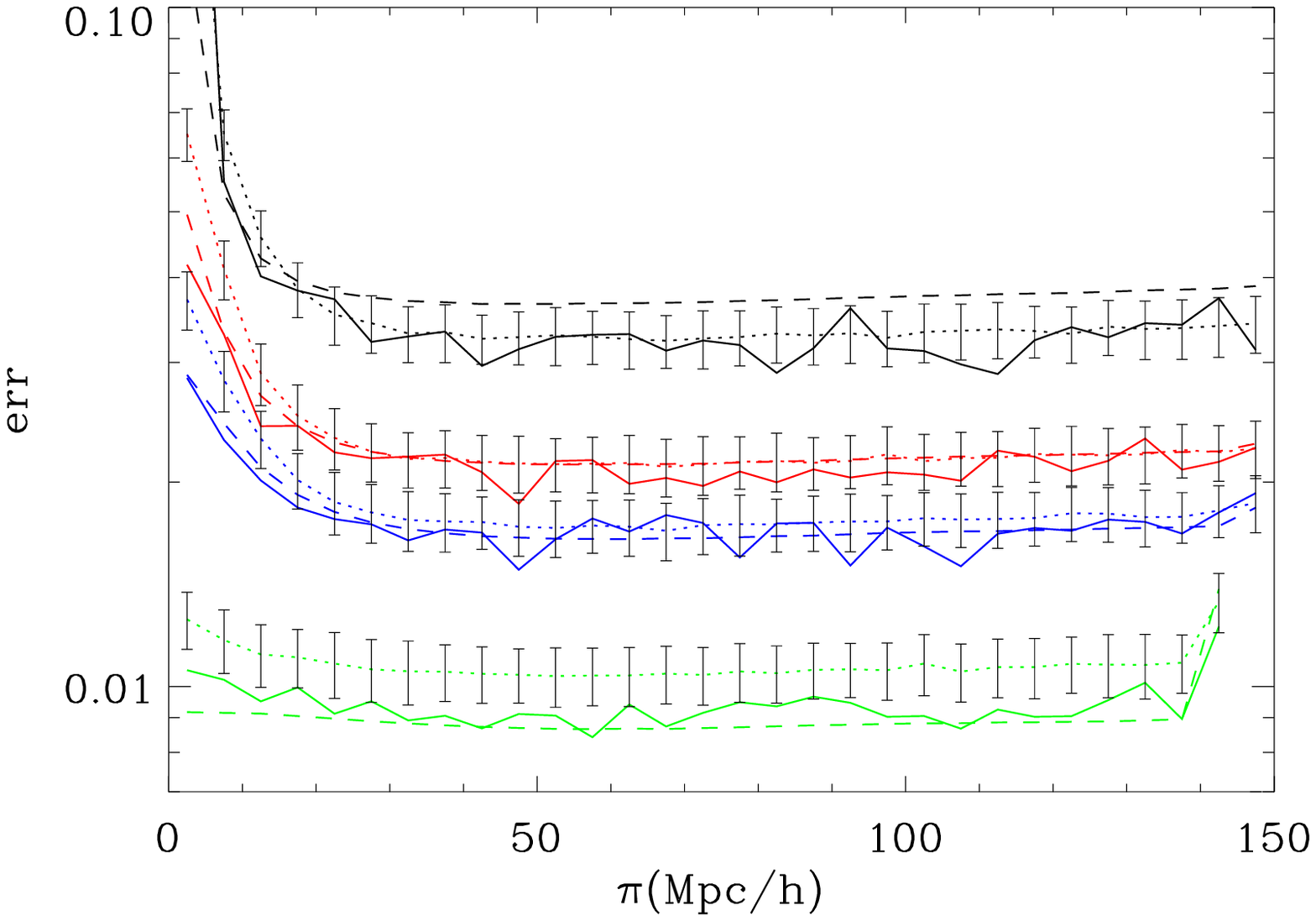}}
\caption{
We compare different error estimators for the 2-point correlation function 
in redshift space $\xisp$ for MICE dark matter mocks (left panel)
and group mocks with $b=1.9$ (right panel). We fix the 
perpendicular distance $\sigma$ and move along the line-of-sight $\pi$ 
as indicated in the figure. The errors are: MC (solid line), theory (dashed line) 
with and without the  signal part of the error, and JK (dotted line with errorbars). 
As we move to higher $\sigma$ (lower amplitudes), JK starts to fail, but the 
analytical and MC errors always agree well. On the radial direction, top lines
for $\sigma=2.5$Mpc/h, the JK and MC agree well, but theory model only works
in the dark matter case.
\label{fig:errorvalid2}} 
\end{figure*}

\subsection{Errors in $\xisp$}\label{sec:errorsxips}

We study here the errors in the anisotropic redshift-space correlation function
$\xisp$. The JK and MC errors only seem to agree at small perpendicular scales 
$\sigma< 20Mpc/h$, which means that we can not use the JK error
for large $\sigma$ values. This is illustrated in Fig.\ref{fig:errorvalid2}.
The reason for this behaviour could partially be related to the fact
that we have defined the JK
zones in angular space, ie in the $\sigma$ direction. As we increase $\sigma$  the
different JK zones are less independent. This could explain why JK errors do not work
so well. In the radial direction, the different JK zones are more independent
from each other and the JK errors give a better agreement to MC errors.

\subsubsection{Error model}

We have found a phenomenological model (we sometimes call it "theory" or
"analytical" error)
for the errors that match well the MC errors at large $\sigma$, even at large scales. 
The advantage of using a model is that is smoother than MC errors and can also
be used to calibrate JK errors. We will 
first test this model in different situations.

Our "theory error" has a dominant shot-noise contribution, which scales as 
$1/\sqrt{number\:of\: pairs}$,
and a part proportional to the signal. This works well in this case because
our LRG sample is shot-noise dominated, specially at large scales, and 
we are able to separate both 
contributions, as if we were in a diagonal space.  In general, the analytical 
derivation of the error is more complicated that this simple modeling.
We propose the error to have the following form
$\Delta\xi= \Delta\xi_{shot-noise} + \Delta\xi_{signal} $, with
two arbitrary coefficients $\alpha_{noise}$ and $\alpha_{signal}$ , so that:

\begin{equation}
\Delta\xi= 
 \alpha_{noise} ~\Delta\xi_{Poisson} + \alpha_{signal} ~\xi
\label{eq:errth}
\end{equation}
adding these two terms linearly fits the simulations better than in quadrature.
For Poisson shot-noise we have by definition that $\alpha_{noise}=1$. 
This works very well for
dark matter mocks, but we will show that groups do not follow the Poisson distribution
and $\alpha_{noise}=1.4$.
We can associate $\alpha_{signal}$ to the number of independent nodes in the catalog
and we therefore expect $\alpha_{sigmal}$ to scale with the square root of the
volume of the sample used. We also expect  $\alpha_{signal}$ to depend on the binning used.
We use the $\xi$ estimator of 
%\cite{landyszalay} 
Landy and Szalay (1993) to calculate the correlation
function,

\begin{equation} \label{eq:dddrrr}
\xisp  = \frac{DD - 2DR + RR}{RR} 
\end{equation}

In the case of dark matter mocks,
the error that comes from having a limited number of 
data-data (DD) pairs (Poisson shot-noise) is:
$err(DD)=1/\sqrt{RR/N_R^2}$, and for DR pairs: $err(DR)=2/\sqrt{RR/N_R}$, where the 
random catalog is $N_R$ times denser than the data catalog.
The error in random-random is insignificant, because we are using a denser
random catalog.  We add the different errors in quadrature to find:

\begin{eqnarray} \Delta\xi_{Poisson}=\sqrt{\frac{N_R^2+4N_R}{RR}} 
\label{eq:Poisson}
\end{eqnarray}

\begin{figure*} 
\centering{
\epsfysize=5.5cm\epsfbox{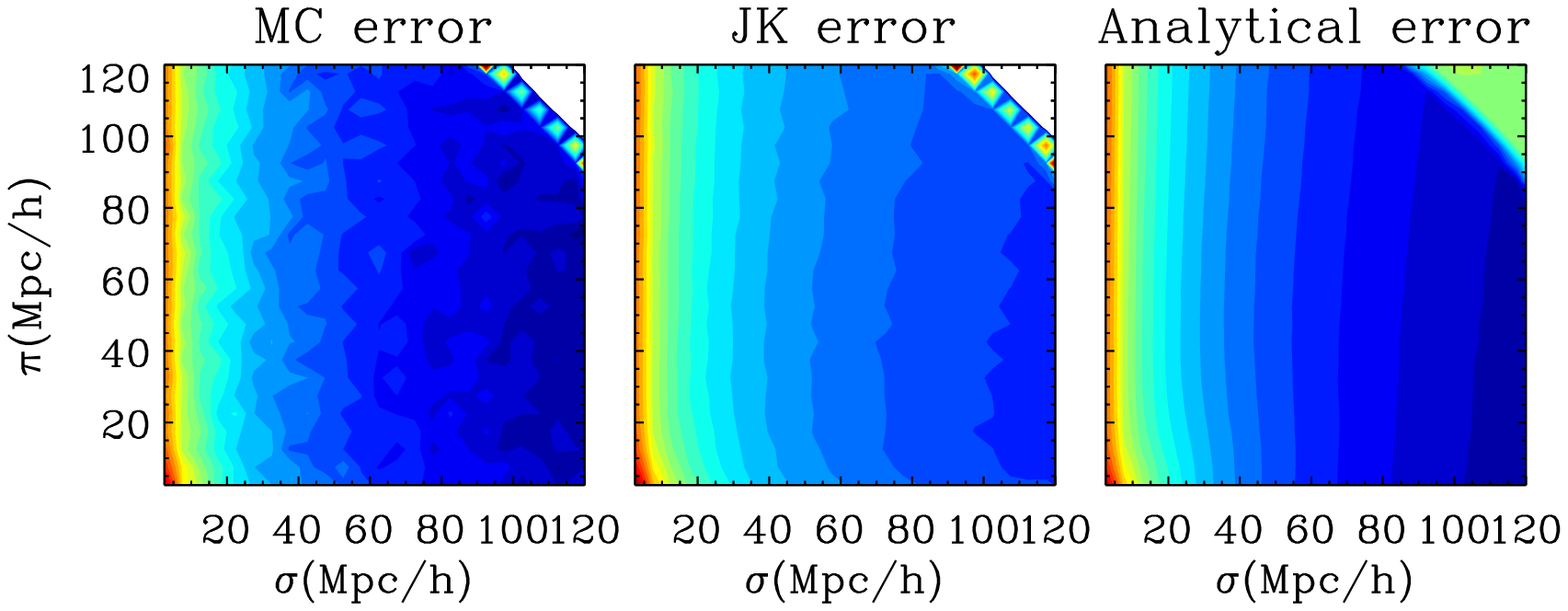}
\epsfysize=5.5cm\epsfbox{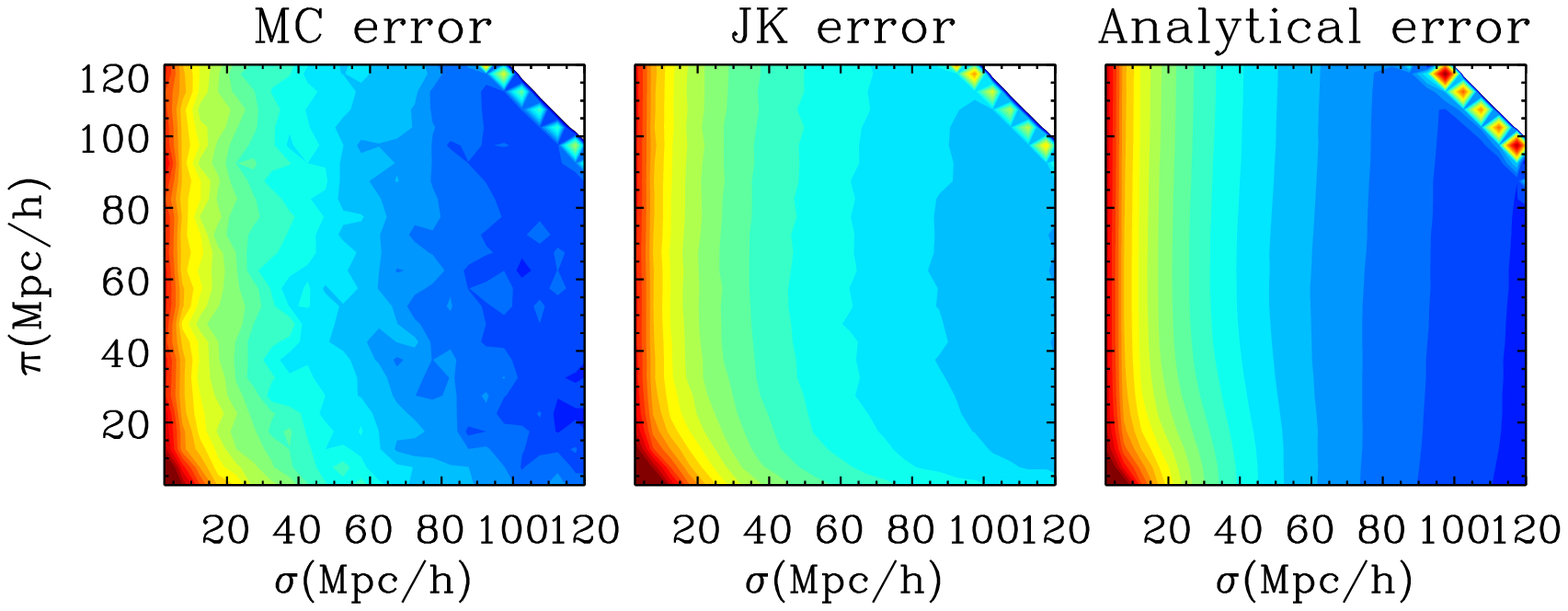}}
\caption{Diagonal error in $\xisp$ in redshift space for MICE dark matter simulations
(top panel)  and groups with $b=1.9$ (bottom panel)
with contours $\xi=0.0035-0.05$ (log increment=0.1)
\label{fig:errps5mice}} \end{figure*}

We can reduce the error in $\xi(\pi,\sigma)$ by increasing the number of particles in the
random catalog, modifying the error coming from DR, which is inversely
proportional to the square root of the pairs data-random. The first part of the
error is always the same, because it depends only on the data.
For $N_R=10$, as in the simulations, $\Delta\xi_{Poisson}=\sqrt{140./RR}$, so the DR
part is 40\% of the DD part. For $N_R=20$, as in the analysis of LRGs,
$\Delta\xi_{Poisson}=\sqrt{480./RR}$, so the DR part is now 20\% of the DD part.
For a binning of 5Mpc/h, we find $\alpha_{signal}=1/95$ and for binning of 1Mpc/h, 
$\alpha_{signal}=1/25$.
For $\sigma>20$ Mpc/h, the signal part in $\xips$
is not very significant. This can be seen in Fig.\ref{fig:errorvalid2}, where
the linear dashed line correspond to the shot-noise term and the curved dashed
line (only important on scales $\pi<20$Mpc/h) also includes
the signal part in Eq.(\ref{eq:errth}).  For small $\sigma$ scales, JK and MC coincide,
so we could just use JK error from the actual LRG data. In practice we
prefer to use our model error to the JK error because this avoids introducing
noise in the error analysis, but we have checked that none of our results depends
on this choice.

\subsubsection{Super-Poisson errors}

In the case of groups, the Poisson model for the shot-noise does not work
well. The reason for this is that groups are selected using 
friend-of-friends with linking length of $r=0.20$ times the interparticle
separation. This means that groups create excluding regions where we can
not randomly locate another group and therefore do not follow a Poisson 
distribution \cite{SmithScocci}. 
Shot-noise is larger in this case, which we call
"super-Poisson". After exclusion, the fraction of available volume is 
$(1-r)^3 \simeq 0.5$ so we expect the Poisson term to be corrected by 
the square root of this fraction, ie roughly 
$\alpha_{noise} \simeq 1/(1-r)^{3/2} \simeq 1.4$.
This is exactly what we find for our model in Eq.(\ref{eq:errth})
for group mocks. We have studied super-Poisson errors in two group simulations (z=0 and z=0.5) with the same linking length $r=0.20$, and obtain the same value of $\alpha_{noise}  \simeq 1.4$, as expected, despite the large difference
in the number density of groups. 

What about real LRG galaxies? Do they follow the Poisson shot-noise or
the groups shot-noise model?
Real LRG data seems to follow more closely the 
group shot-noise as we find that we need $\alpha_{noise}=1.4$ rather
than $\alpha_{noise}=1$ for Poisson term when we fit the JK errors in
the real data at small $\sigma$. As we have seen with mocks, the JK errors
reproduce very well the true errors for small $\sigma$ both the Poisson
(dark matter) and super-Poisson (groups). 
The amplitude of the JK errors on 
small $\sigma$ is quite closer to group mocks
than to dark matter mocks (this can be seen by comparing Fig.\ref{fig:errps5mice}
to Fig.\ref{fig:errps5lrg} below).
This by itself is a very interesting result because it clearly
shows that LRG galaxies have a tendency for exclusion. This suggests that 
a dominant fraction of LRG populate separate dark matter halos (this has already been studied with the same conclusions using semi-analytical models, Almeida et al 2008, or with real LRGs, Wake et al 2008 and Zheng et al 2008). 
It also shows that using dark matter mocks or a standard (Poisson)
point process to simulate LRG  will result in an important underestimation
of errors.

\begin{figure}
	\centering{ \epsfysize=5.5cm\epsfbox{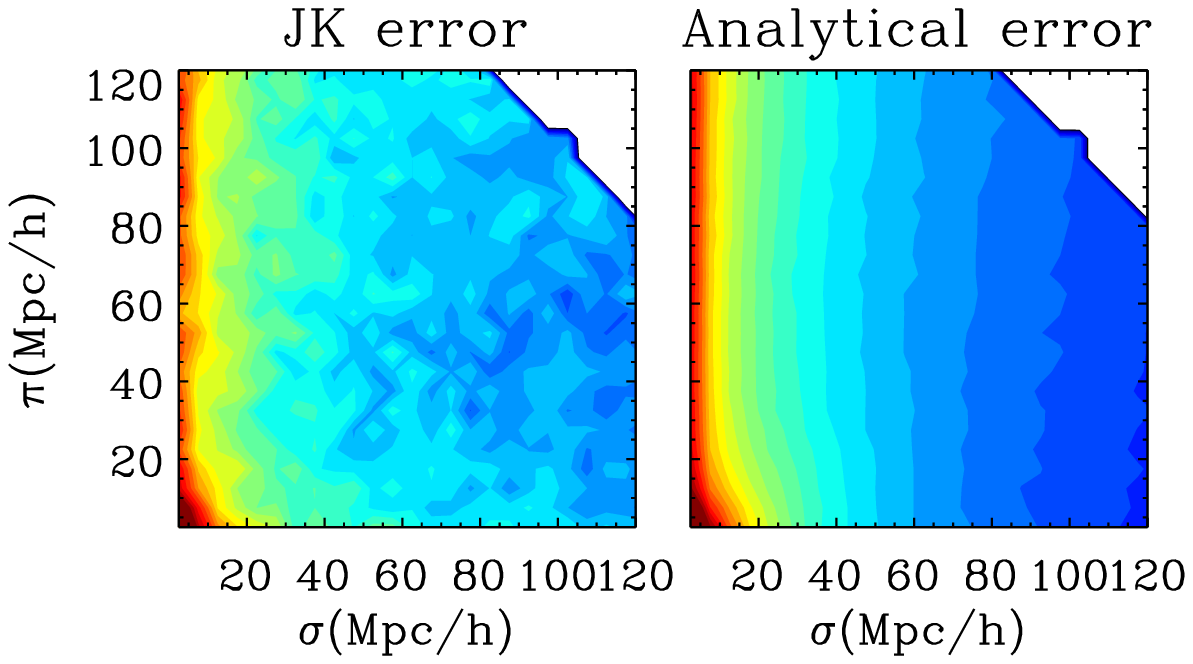}}
	\caption{Same as  Fig.\ref{fig:errps5mice} for real
LRG data.
\label{fig:errps5lrg}
}
\end{figure}

\subsubsection{Errors in LRG}

Fig.\ref{fig:errps5mice} shows the full 2D version of  Fig.\ref{fig:errorvalid2}.
We show here the differences in the diagonal error
between MC, JK and the theory error in Eq.(\ref{eq:errth}), 
when binning the correlation function
with 5Mpc/h. MC error is calculated using MICE simulation and JK is the mean
over all the JK that we have calculated in each mock. JK errors work well for small
$\sigma<20$Mpc/h but that become higher than MC when going to large $\sigma$ for
the reasons explained above. Our "theory" error model
form agrees with MC error at all scales, except in
the radial direction, $\sigma=2.5$Mpc/h, for the case of group mocks 
(where we used a fixed $\alpha_{noise}=1.4$ at all $\sigma$), 
which is slightly higher than the MC or the JK errors.
The model for dark matter mocks, with  $\alpha_{noise}=1$ does well
at all $\sigma$.
This can be seen in Fig.\ref{fig:errorvalid2}. Thus in the
radial direction we need to change  $\alpha_{noise}$ for groups
to $\alpha_{noise}=1.2$, a lower value than $\alpha_{noise}=1.4$. 
The reason for this is most probably the random component of peculiar velocities 
in redshift space, which ramdomizes the radial positions of group 
and reduces the super-Poisson shot-noise contribution, which comes from
self-exclusion in group positions. This is no problem when 
analyzing real data as we can fit 
 $\alpha_{noise}$ to match the JK error at small values of $\sigma$ 
 (with a different value at $\sigma<5$ than at $5<\sigma <20$ Mpc/h)
 and then use our error model for large
 values of $\sigma$ where JK does not work well. We could also directly use the
 MC errors in the group mocks, which seem to agree quite well with real
 LRG data (ie JK errors are very similar in both cases, 
 see Fig.\ref{fig:errps5lrg}).

In Fig.\ref{fig:errps5lrg} we show the JK error obtained from the real LRG data 
using a random catalog 10 times denser than the data (as done with the
mocks), and the analytical form with a fixed $\alpha_{noise}=1.4$
and $\alpha_{signal}=1/95$.
As in the mocks, for large $\sigma$ the JK error 
is bigger than the analytical error, which should be more 
representative  of the true (MC) error. At small scales the model
follows  the JK prediction, except for $\sigma<5$ Mpc/h where we
need to change to $\alpha_{noise}=1.2$ (the figure shows the model
for $\alpha_{noise}=1.4$ at all $\sigma$).
Note the similarity between the errors 
in the data and in the group mocks in Fig.\ref{fig:errps5mice}.  
For our final analysis of the data we use
a random catalog 20 times denser to reduce the error contribution due to 
the shot-noise in the randoms.

 We have done a similar
analysis with the L-BASICC halo simulations from Durham \cite{angulo}, which have
approximately the same bias as real LRG in SDSS, and we find very good agreement with
our fitting formula.  If we use a random catalog 20 times denser, as we do 
for the analysis of the data, the error is lower, just as expected.

\subsubsection{Covariance $\xips$}

\begin{figure}
\centering{ \epsfysize=6cm\epsfbox{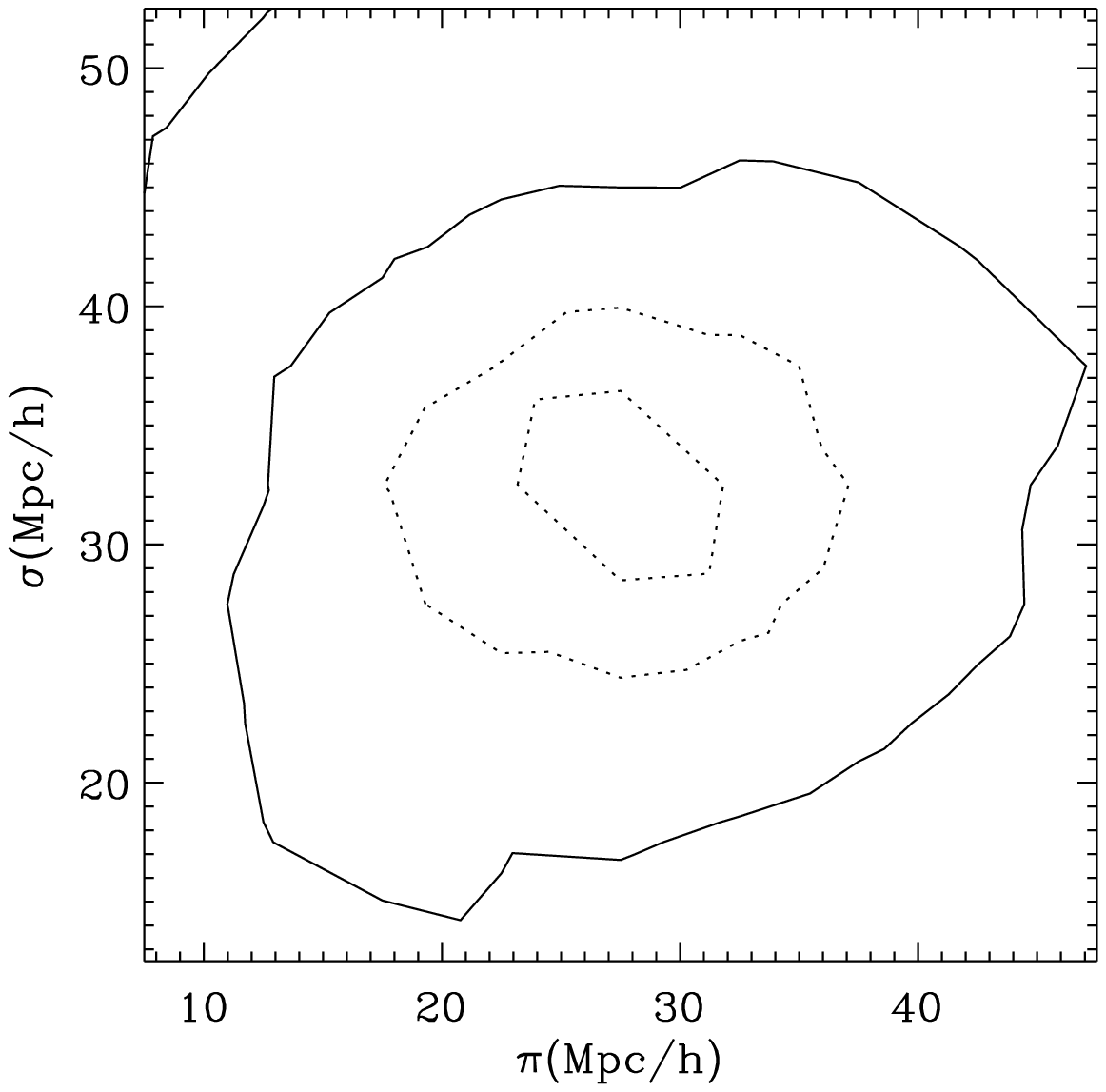}}
%	\centering{ \epsfysize=4.2cm\epsfbox{figures/covpisigmaJK.ps}
%	\epsfysize=4.2cm\epsfbox{figures/covpisigmaJK2.ps}
%	\epsfysize=4.2cm\epsfbox{figures/covpunts2.5-107.5.ps}}
	\caption{Normalized covariance of $\xi(\pi,\sigma)$ for groups in a
	bin centered at position
	$(\sigma=32Mpc/h,\pi=28Mpc/h)$. Contours show fix values of the
	covariance away from the central position,
	 with contours 0.5, 0.3, 0.2 and 0.1. \label{fig:covps5} }
\end{figure}

So far we have shown diagonal errors. 
The normalized covariance is in fact quite small for dark matter
mocks. It has the same approximate shape and amplitude for all the points.
The amplitude of the covariance is just given by the distance between bins: 
at a separation of 5Mpc/h  the normalized covariance is $\le 0.1$. 
This is also true for group mocks in the radial direction. But the
covariance is larger in the case of groups for $\sigma>5Mpc/h$. 
At a radial distance of 5Mpc/h between bins in $\xips$
the normalized covariance is about 0.5 and dies below $0.2$
for bins separated by more than 15 Mpc/h. 
This is illustrated in Fig.\ref{fig:covps5}, which just shows
the covariance in one of the bins. Other cases are very similar.
The inner contour, which corresponds to $0.5$ for groups, is very
similar in shape for dark matter and radial bins, but the amplitude
in this case is 0.1.

\begin{figure*} 
\centering{ \epsfysize=5.cm\epsfbox{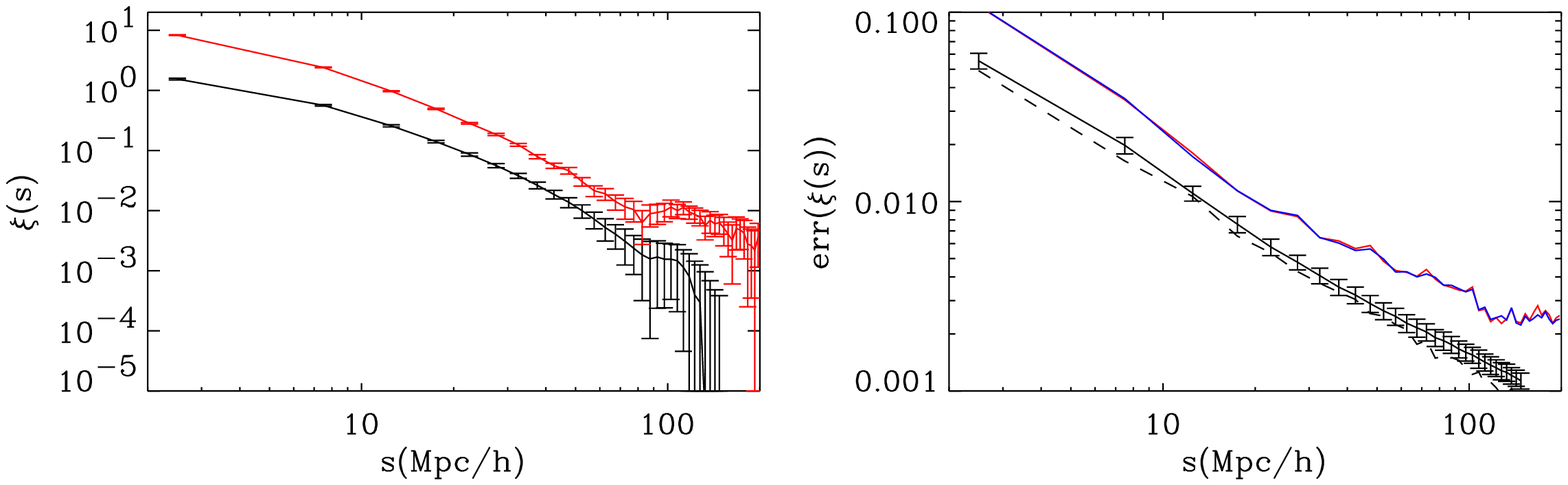}}
\caption{Left: Monopole $\xi(s)$ with JK errors for MICE simulations (black, lower
line) and for real LRG (red, upper line) using a bin of 5Mpc/h. Right: Diagonal
error for MICE dark matter simulations (JK with dispersion solid black line, 
MC dashed black line) and for JK errors in real LRG data 
(red for $N_R=10$ and blue for $N_R=20$)
\label{fig:errcorrs5} } 
\end{figure*}

If we bin the plane $\pi-\sigma$ with 1Mpc/h, the same theory error form 
in Eq.(\ref{eq:errth}) still works on all scales. So does
the JK, at low $\sigma$. But in general errors are higher than when using a wider bin
(smaller bin means smaller number of pairs).
This increase is compensated by a decrease in the covariance, which becomes
practically zero for all elements outside the diagonal.

\subsection{Errors in the monopole $\xis$}\label{sec:errorsmonopole}

The error in the monopole is not easy to predict theoretically, but in this case
JK error works very well for both variance and covariance on all scales. 
In the left panel of Fig.\ref{fig:errcorrs5} we plot  the
monopole when we bin the data with a separation 5Mpc/h: black for the dark matter
mocks and red for LRG data (which will be presented in more detail
in section \ref{sec:results}). In the right panel, we see the difference
between the mean JK error and its dispersion (solid line with errors) and the MC
error (dashed line), and over-plotted the JK error for the real LRG (color). 
As can be seen, the error for LRG is larger than the one in simulations, since the
signal is also higher, due to an overall bias $b$ in the amplitude. 
The small difference between blue and red line in Fig.\ref{fig:errcorrs5} is
due to the number of the particles in the random catalog ($N_R$), 
which does not change
 the estimation of the error in this case.
Results are very similar for group mocks, with higher amplitudes, as
expected. The important point here is that the JK error works well at
all scales.

\begin{figure}
\centering{
\epsfysize=4.5cm\epsfbox{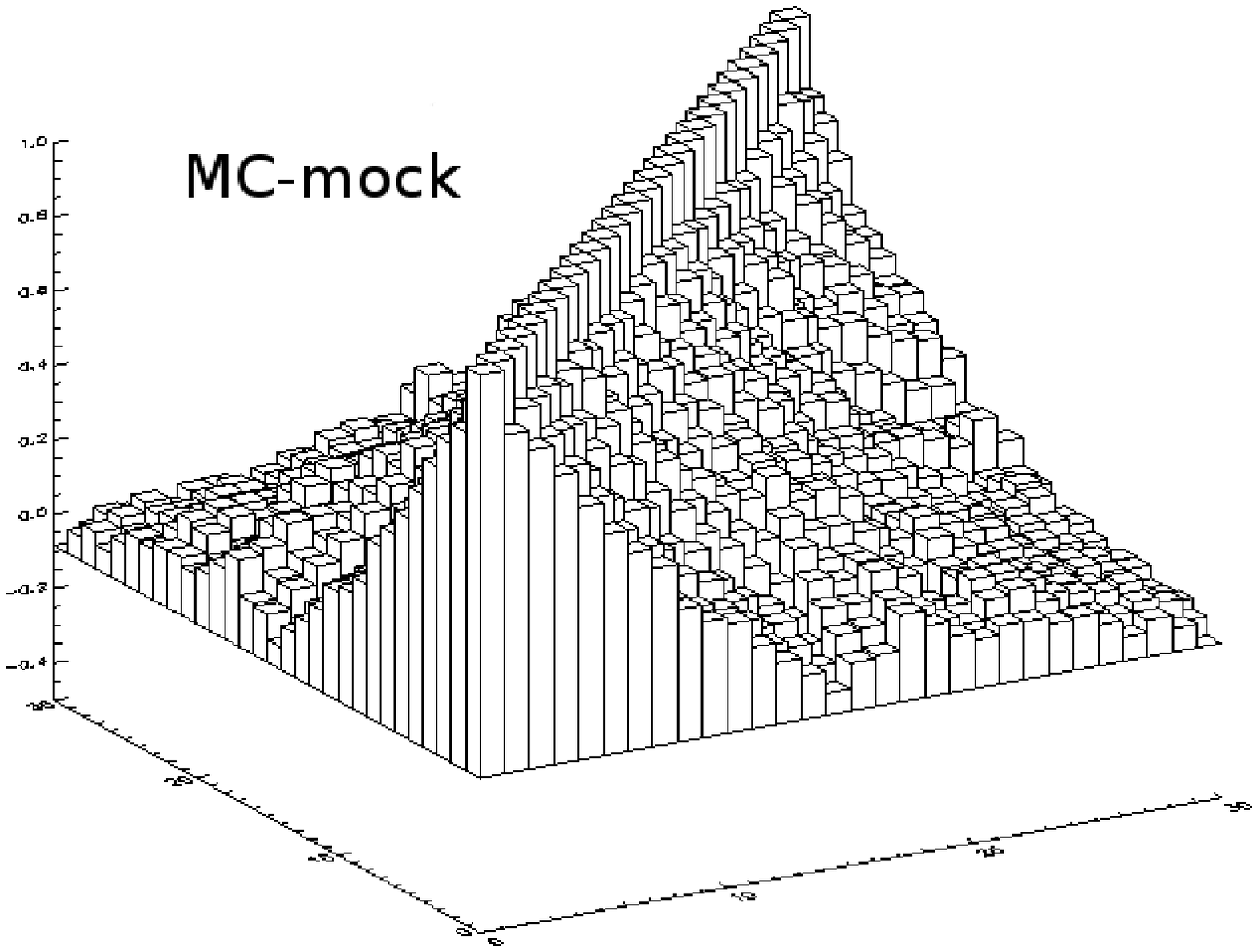}} 
\centering{\epsfysize=4.5cm\epsfbox{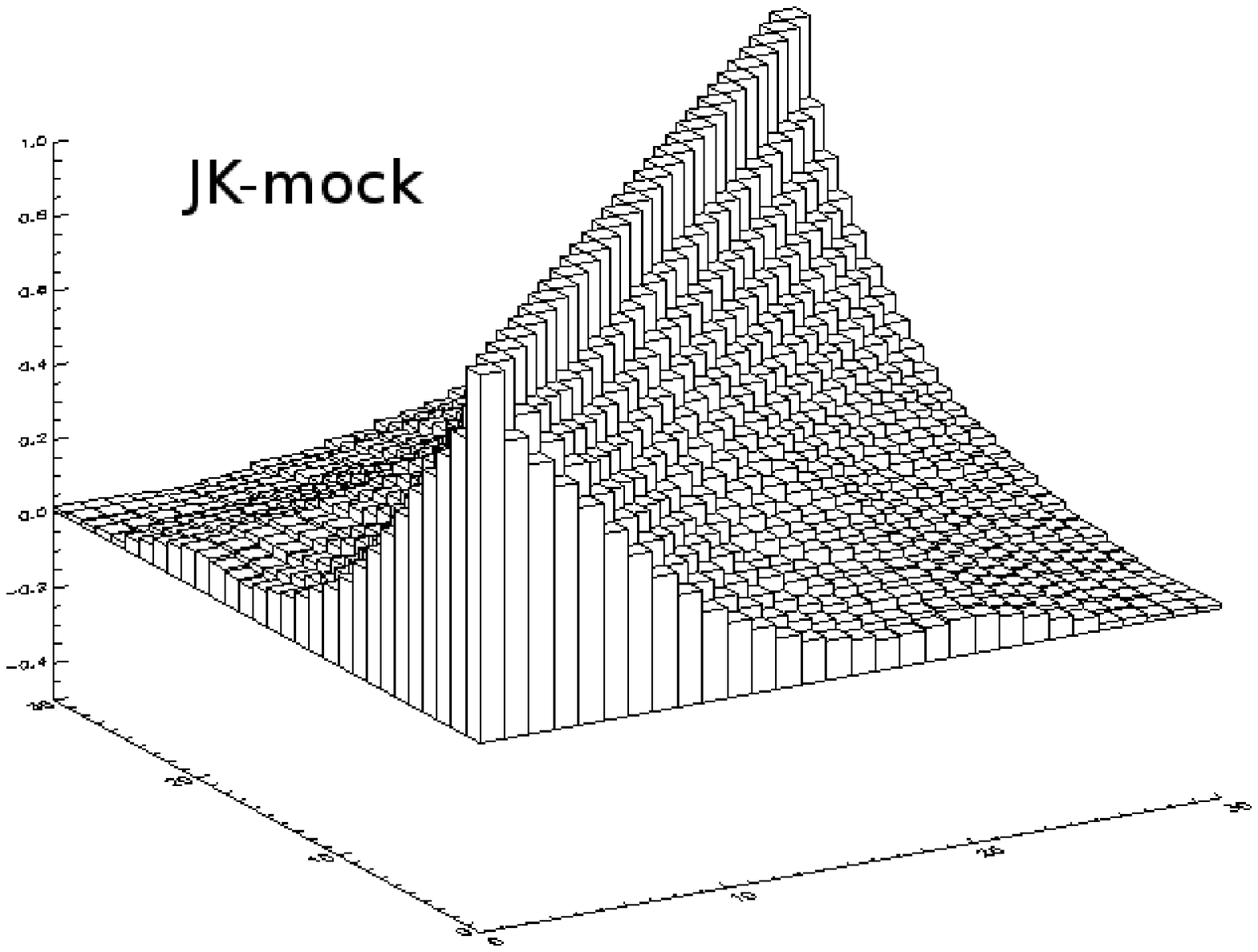}}
\centering{\epsfysize=4.5cm\epsfbox{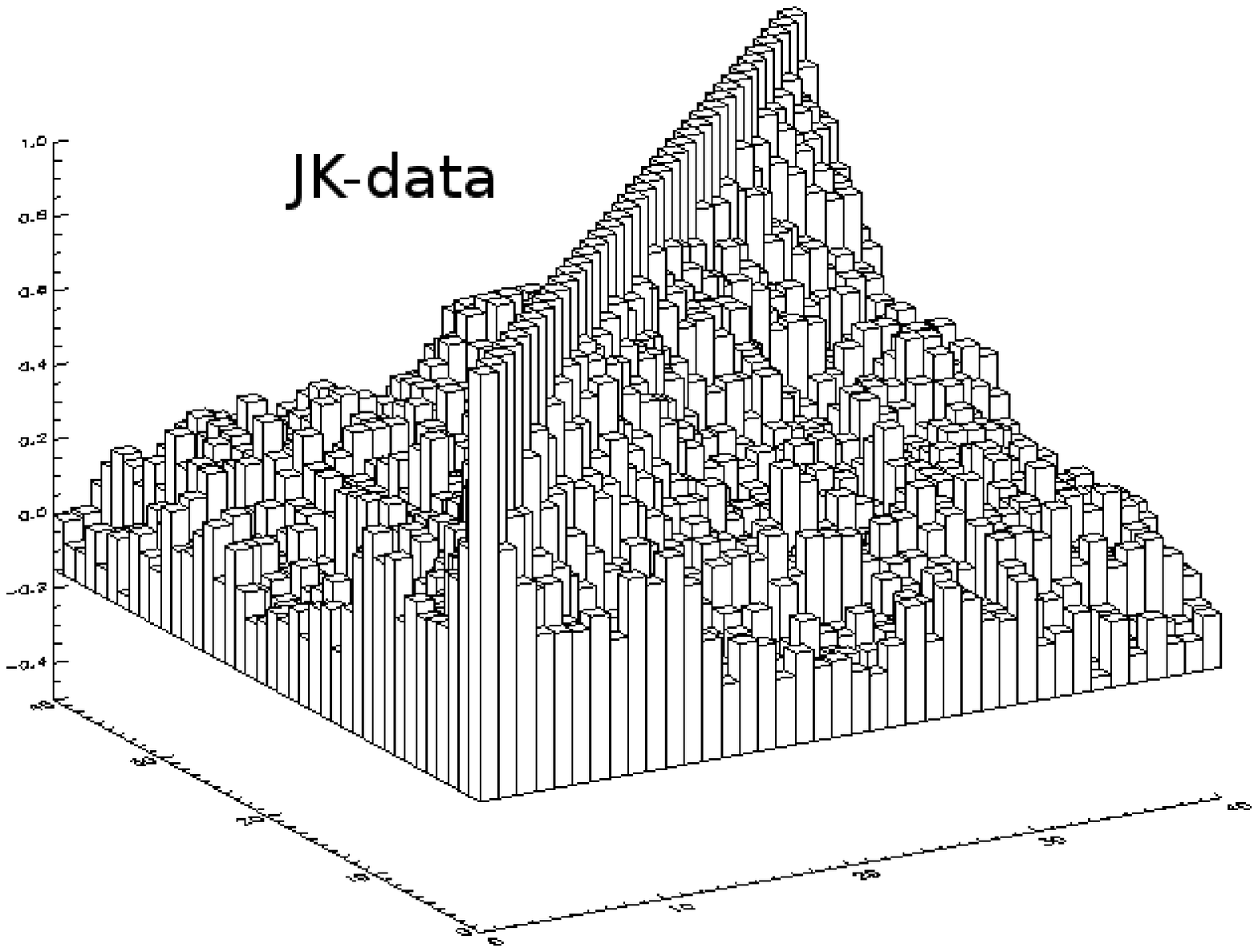}}
\caption{Normalized covariance of MC (top) and JK (middle) for the monopole
$\xi(s)$ in MICE group mocks 
(for binning of 5Mpc/h). Bottom panel correspond to JK error in the real 
LRG data \label{fig:covcorrs5}} 
\end{figure}

\begin{figure*}
\centering{\epsfysize=5cm\epsfbox{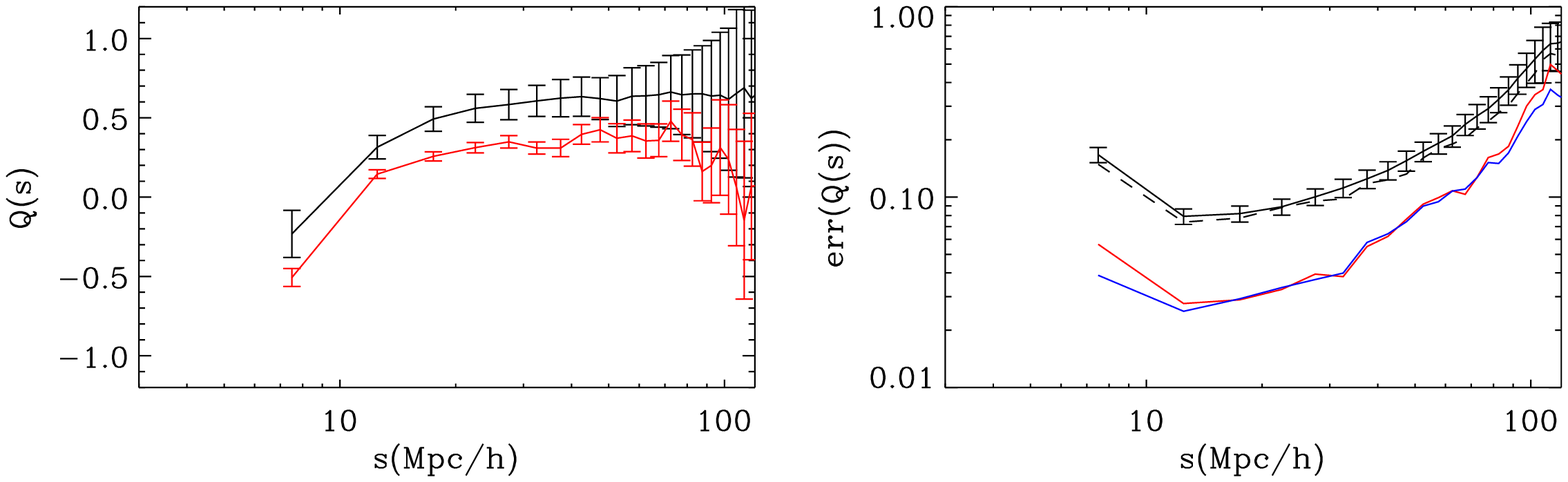}}
\centering{\epsfysize=5cm\epsfbox{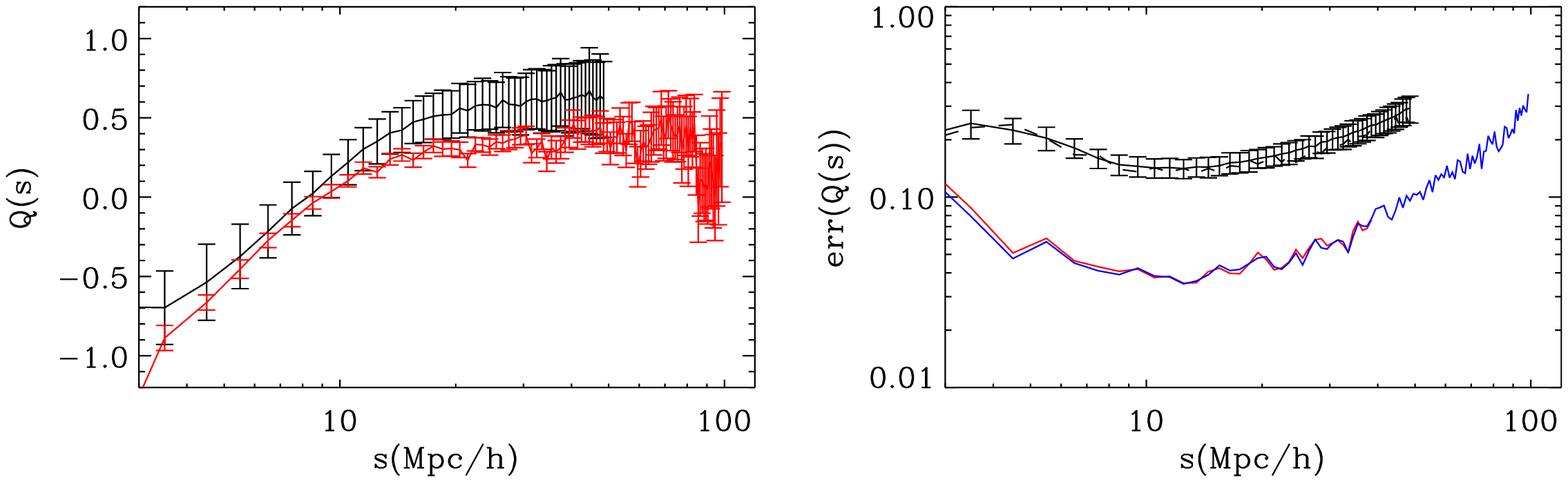}}
\caption{{\it Top Left:} Q(s) with errors for MICE dark matter mocks
 (black higher signal) and for LRG (red lower signal) using 
 a bin of 5Mpc/h. {\it Top Right:} Diagonal error for MICE
simulations (JK with dispersion solid black line, MC dashed black line) and for
LRG data (red and blue for $N_R=10$ and $N_R=20$).
{\it Bottom:} Same as top with a bin of 1Mpc/h.
\label{fig:errquadru5}}
\end{figure*}

The normalized covariance, in Eq.(\ref{eq:ncovMC}),
for the monopole is plotted in Fig.\ref{fig:covcorrs5}. The result is not
completely smooth for the MC case (top panel), because we need more than 216
simulations to have a smooth result.
 But we see that it has the same shape than JK covariance, which we
will use to fit our LRG data. JK covariance is smoother than the MC because we
have taken the mean over 216 x 63 realizations (216 mocks x 63 JK zones), which
seems enough to converge. The bottom panel shows JK errors in the real LRG
data, which is quite similar to the mocks, but looks noisier because
this is just a single realization.

We have also binned the monopole $\xis$ with 1Mpc/h.  JK
error also works in this case, and we also see that the error is higher here than in
the 5Mpc/h bin, while the covariance is smaller, practically equal to zero. 

\subsection{Error in the quadrupole Q(s)}
\label{sec:errorsquadrupole}

For the quadrupole, the JK error also works quite well, as we can see in
Fig.\ref{fig:errquadru5} for 5Mpc/h (top panels) and 1Mpc/h
binning (bottom panels). Again here , the solid line with errors shows the JK error and its dispersion, the
dotted line is the MC error, and the color and lower line is the JK error in LRG
data. The error for LRG data, in the quadrupole, is lower than the simulations
one. This is because  the error here is proportional to the signal, 
and in this case, LRG signal is lower since Q(s) depends on $\beta$ 
which depends inversely on bias $b$. Results for group mocks are
in good agreement with data and yield similar results for the comparison
on JK and MC errors.

\subsection{Validity of the models}\label{sec:validity}

We also use the simulations to test the methods that
we will apply to real data (LRG) in the following sections. We want to study if
we can recover the parameters that were input in the simulations.

\begin{figure} 
\centering{\epsfysize=5.5cm\epsfbox{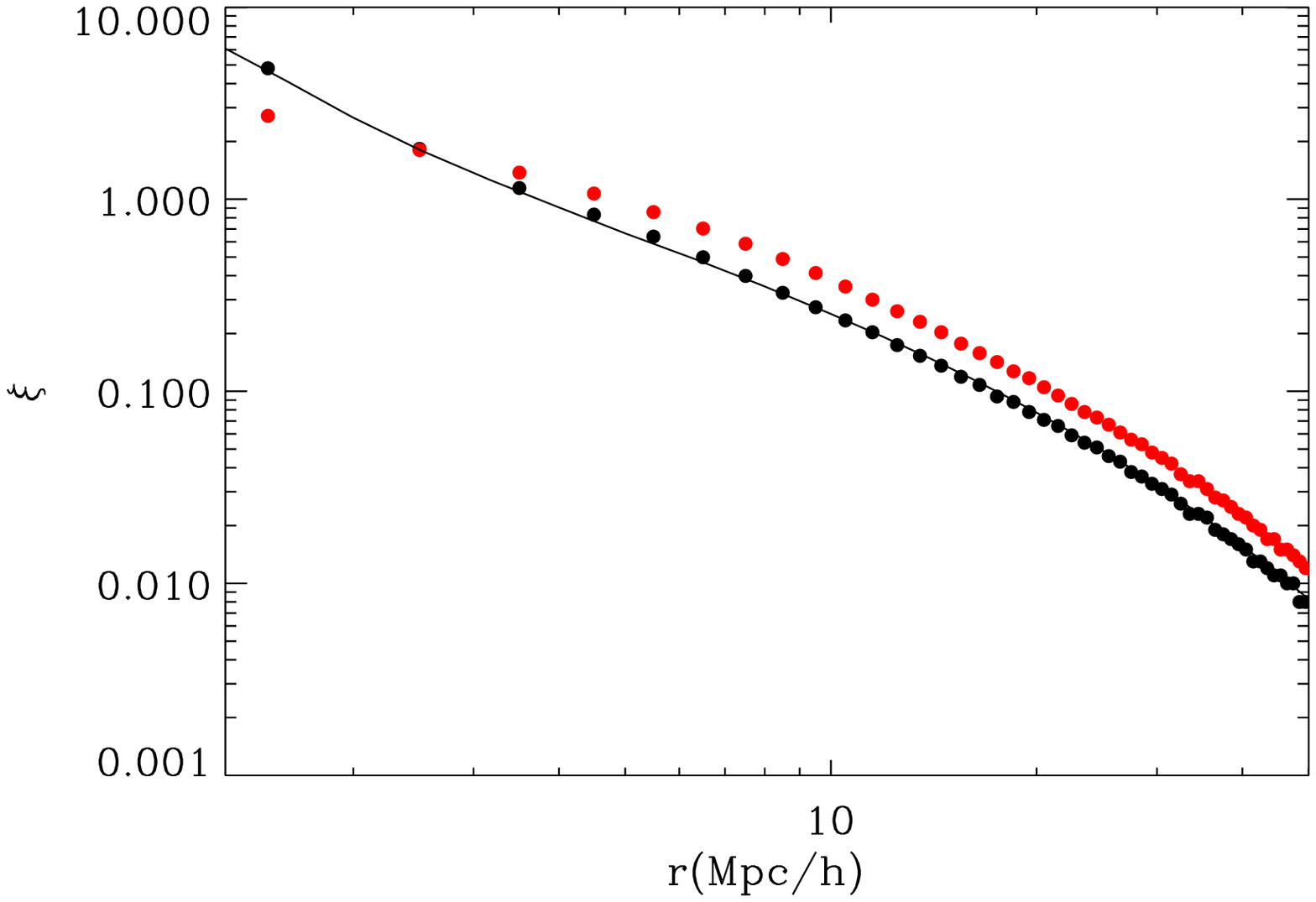}}
\centering{\epsfysize=5.5cm\epsfbox{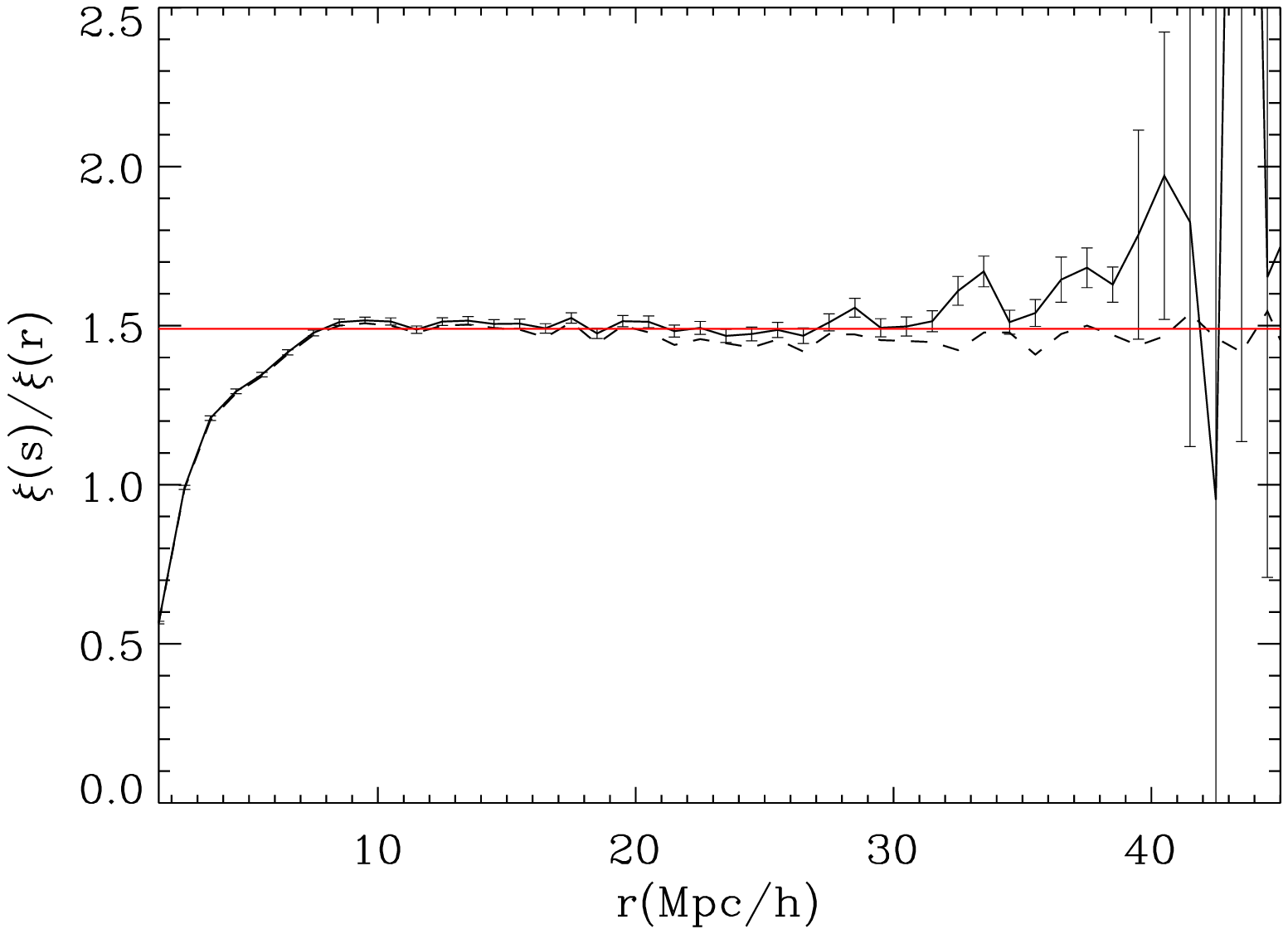}}
\caption{{\it Top panel}: Real-space correlation function $\xir$ (black dots) with its model (solid line) and redshift space correlation function $\xis$ (red dots), which is $\xir$ biased by a constant Kaiser value for large scales.
{\it Bottom panel}: Mean $\xis/\xir$ over MICE dark matter mocks with errors 
(on the mean, scaled as $1/\sqrt{nsim}$), and Kaiser prediction 
for large scales for $\beta=0.62$,
corresponding to the input model (red). In dashed, we have plotted mean($\xis$)/mean($\xir$), a biased estimator,  which is different than the solid line mean($\xis/\xir$). 
The ratio is constant from 8Mpc/h up to 30Mpc/h, where errors start to blow up.    \label{fig:xisxirmice}} 
\end{figure}

\begin{figure} \centering{
\epsfysize=5cm\epsfbox{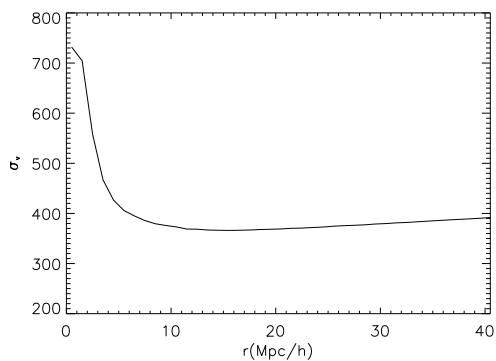}}
\caption{ Dispersion in the pairwise velocity distribution $\sigma_v$ in the MICE
simulation as described in Eq.(\ref{e:fv}) when we change the distance between particles 
\label{fig:pairwise7680}} 
\end{figure}

\subsubsection{Recovering the real-space correlation function}

We first estimate the 2-point real-space $\xir$ and redshift-space
$\xis$ correlation functions
and compare $\xir$ to the input model that describes the simulation (see top panel in Fig.\ref{fig:xisxirmice}). We find very good agreement.
We look at the ratio $\xis/\xir$ between the redshift space and real
space, which should be a function of the distortion parameter $\beta$ at large
scales (and for the distant observer approximation), as in Eq.(\ref{eq:xisr}). In
bottom panel of Fig.\ref{fig:xisxirmice} we have plotted the mean ratio over the simulated
mocks, with its error, and over-plotted in red the expected value for the ratio
at $\beta=0.62$ as in the input model. We also plot the mean($\xis$)/mean($\xir$) in dashed line (a biased estimator), which is similar to the mean($\xis/\xir$) until 30Mpc/h. It seems to converge below 10Mpc/h and starts to fail around 30Mpc/h, where errors rapidily increase, which
is in agreement with other analysis (see Fig.13 in Hawkins et al 2003). We could
obtain $\beta$ from the ratio $\xis/\xir$, but it is difficult in real data
since we do not have direct information of the real-space $\xir$, only through
integration of the anisotropic $\xisp$ through the line-of-sight (see
\S\ref{sec:realspace}). The advantage of using this ratio is that it converges
to a constant value at small scales. But we can only use this below 30 Mpc/h
with real data because the recovered real-space correlation function 
$\xir$ starts to fail on larger scales (see Fig.\ref{fig:perpint}). We
have fitted these scales with the simulations and the results for $\beta$ are
the ones expected. However, we have checked that it is better to obtain 
$\beta$ from the quadrupole Q(s), since it is more independent 
of other cosmological parameters and can be trusted to larger scales.

\subsubsection{Fitting of the quadrupole Q(s) to obtain $\beta$ and $\sigma_v$}

We can calculate the multipoles of $\xips$ to decompose the
anisotropy between the LOS and the
perpendicular direction.  We have tested with models that the monopole $\xis$
and quadrupole $\xi(s)$ and even the combination $\xi(s)/\xis$ depend
strongly not only on $\beta$ and $\sigma_{v}$, but also on other parameters like
the shape of the correlation function (ie $\Omega_m$, $\Omega_b$ and
$n_s$), the non-linear
bias and the overall amplitude. On the contrary, the reduced
quadrupole Q(s), defined in
Eq.(\ref{eq:quadru})  only depends strongly on $\sigma_v$ and $\beta$, but not
on the bias. So when using the quadrupole we do not need an expression for the
non-linear bias or a prior on $\Omega_m$ (the shape of $\xir$) to extract 
the $\beta$ information. We can fix
these parameters, and only change $\beta$ and $\sigma_v$. The asymptotic value
of Q(s) for large scales in Eq.(\ref{eq:quadrub}) is a function of $\beta$, but
it is always biased to higher values because of the random
velocities. So it is dangerous to obtain $\beta$ from this asymptotic 
approximation. We instead generate our model for Q(s) based
on the multipoles of $\xisp$ in Eq.(\ref{eq:moment}) and
Eq.(\ref{eq:quadru}).

\begin{figure}
\centering{\epsfysize=5.cm\epsfbox{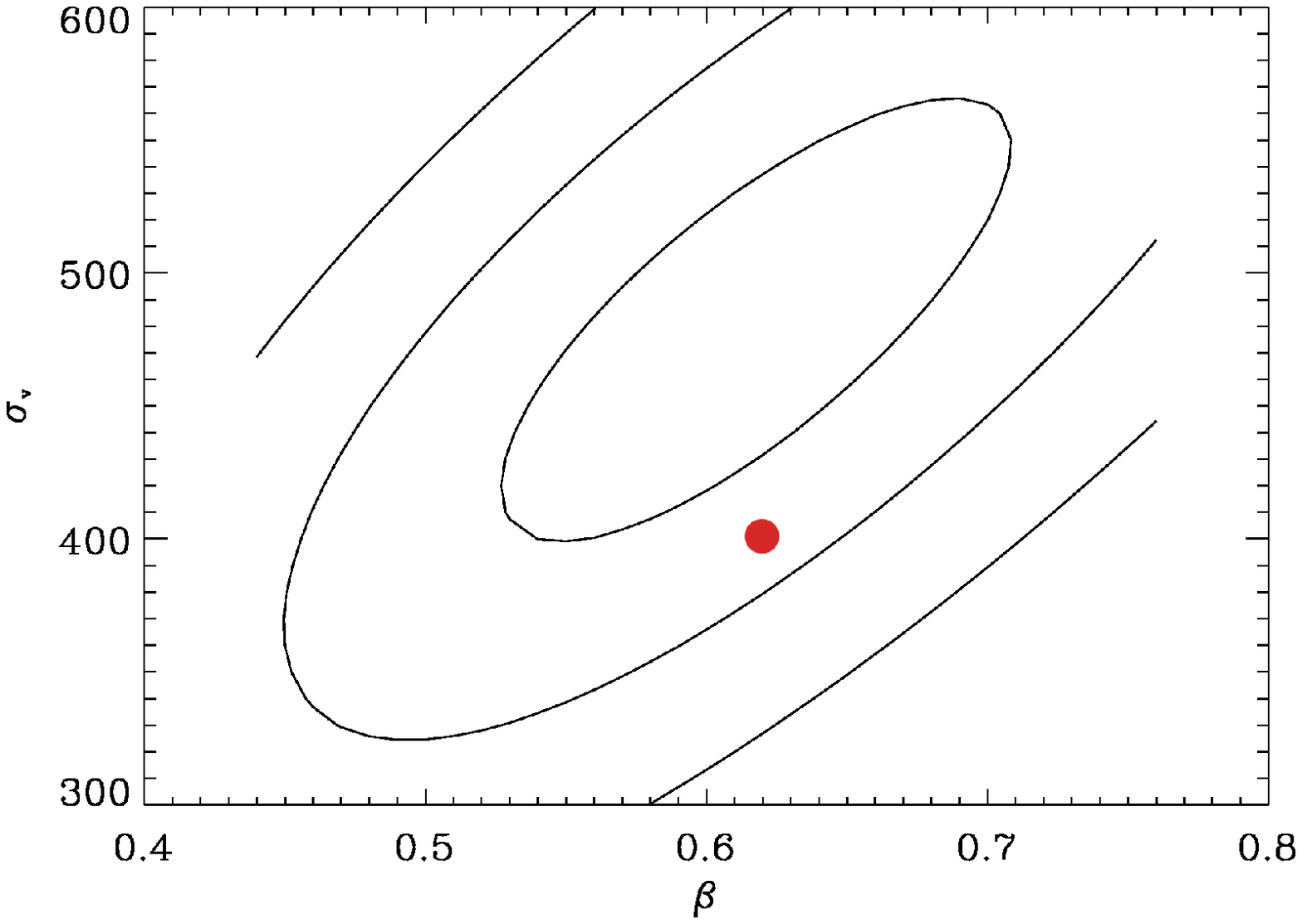}}
\centering{\epsfysize=5.cm\epsfbox{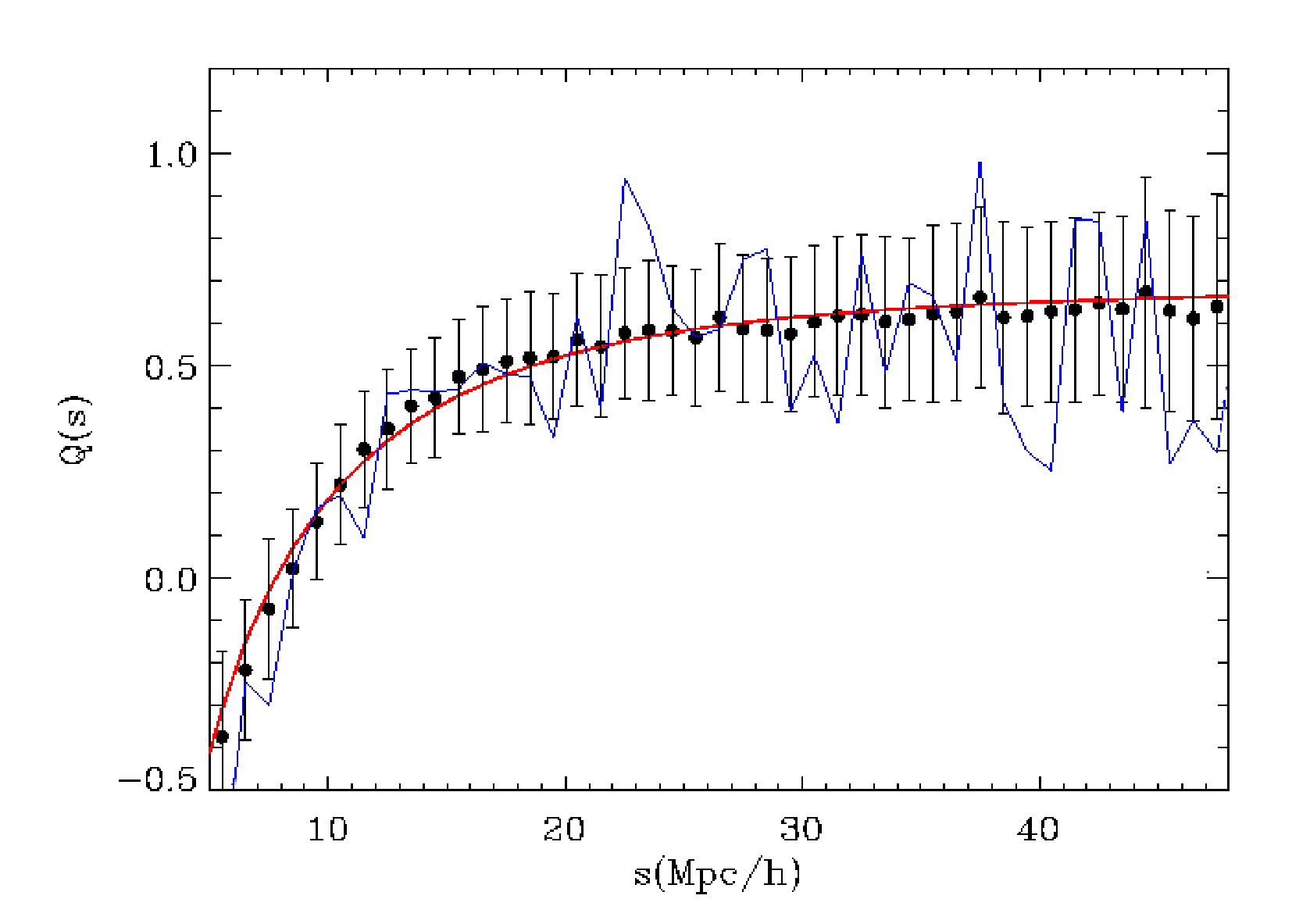}}
\caption{{\it Top panel}: Mean best fit $\beta-\sigma_v$ to MICE quadrupole Q(s). The
red dot shows the value for the input $\beta$ and $\sigma_v$ at large scales.
Contours are $\Delta\chi^2=$ 1,4 and 9 using MC errors for a single mock.
{\it Bottom panel}: Q(s) with MC errors (points), and
best model (red). We also plot in blue the quadrupole for one of the mocks, in order to see the scatter from bin to bin. We recover the input $\beta=0.62$, but $\sigma_v$ is slightly
biased to higher values, probably because we are obtaining an effective $\sigma_v$ which
also accounts for the values at lower scales. \label{fig:quadrumice}}
\end{figure}

\begin{figure*}
%	\centering{\epsfysize=10cm\epsfbox{figures/plotpisigmaMICE1.ps}}
%\centering{\epsfysize=5.5cm\epsfbox{figures/plotpisigmaMICE2.eps}
\centering{\epsfysize=5.0cm\epsfbox{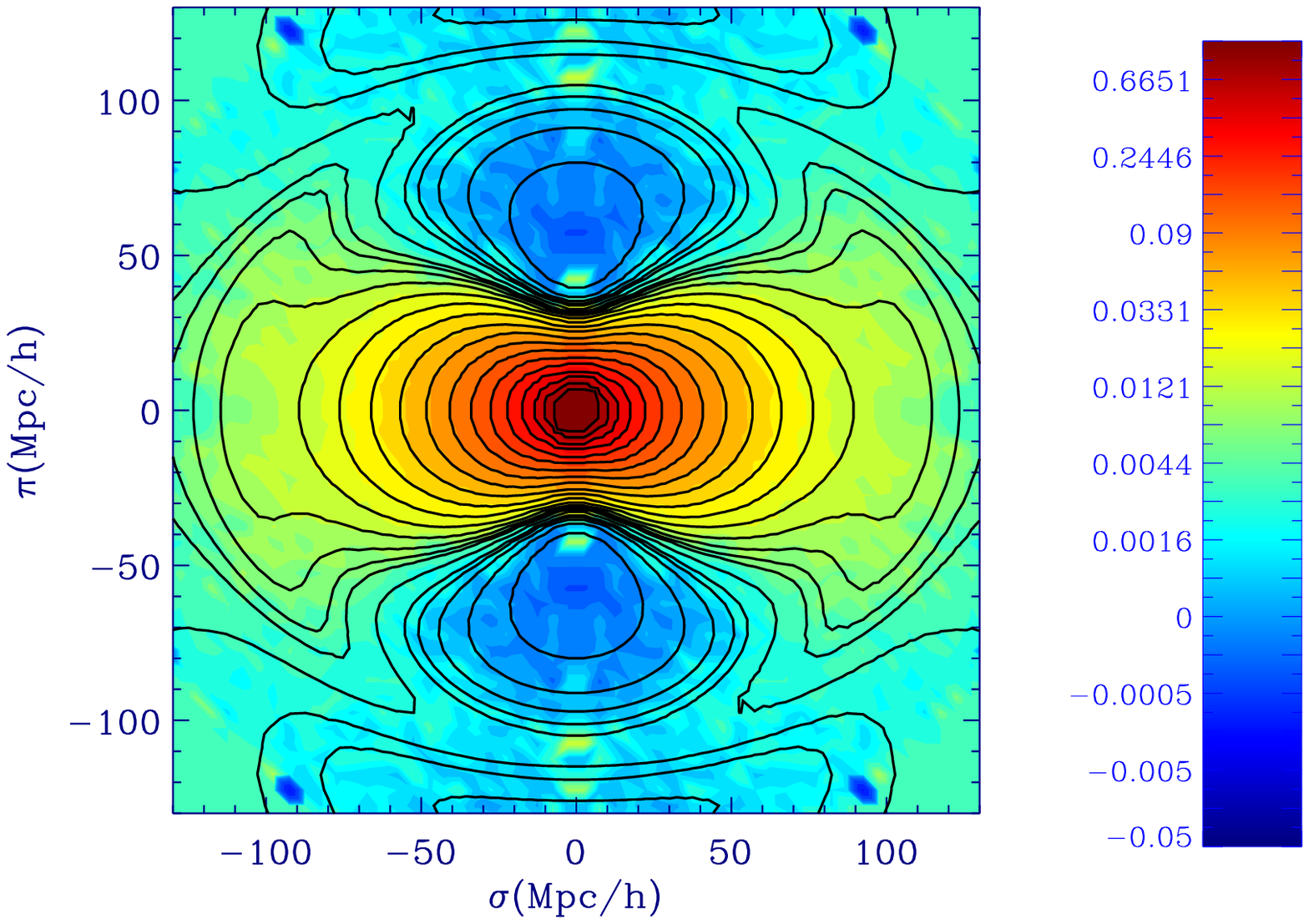}}
\centering{\epsfysize=4.8cm\epsfbox{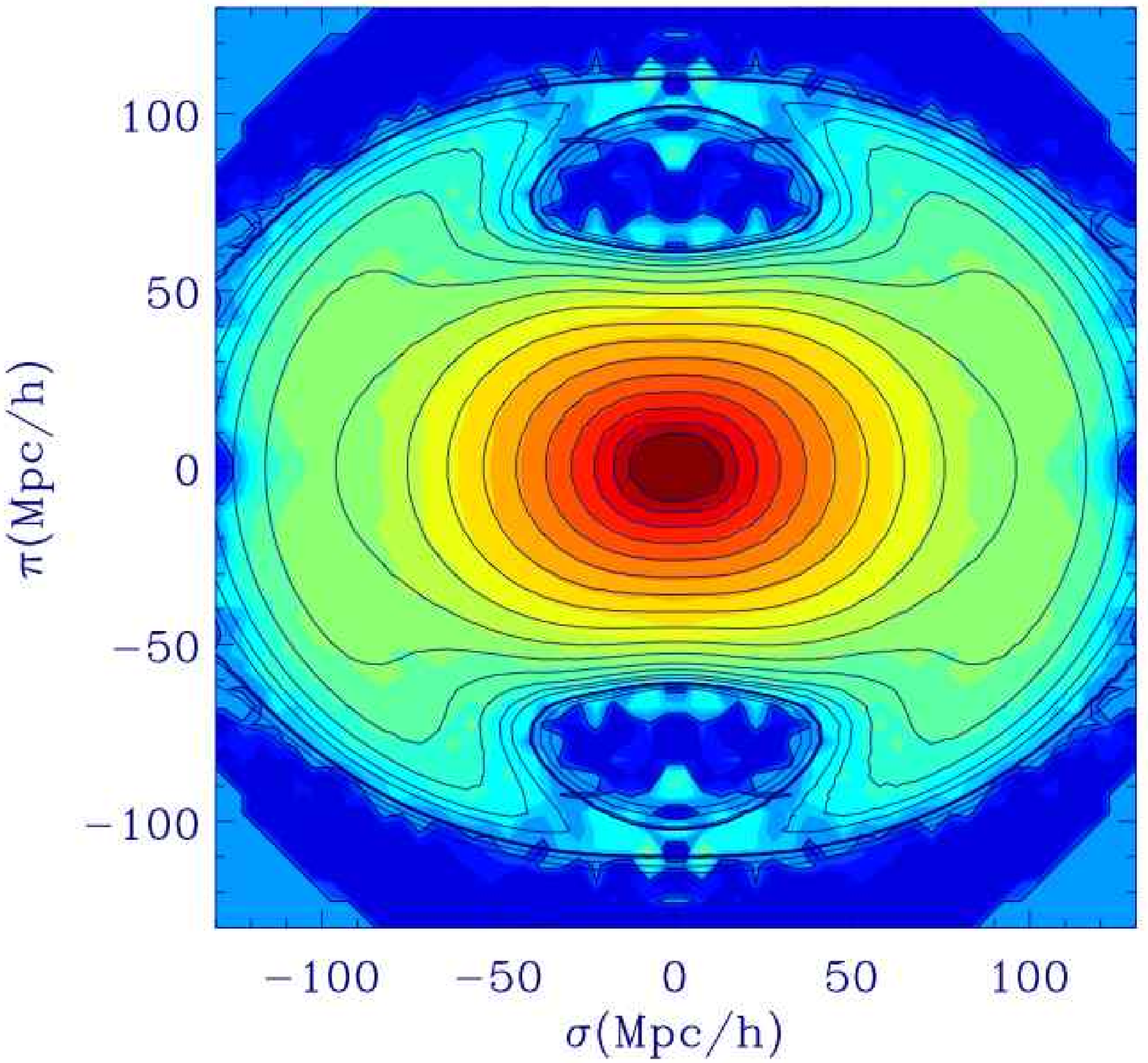}}
\centering{\epsfysize=4.8cm\epsfbox{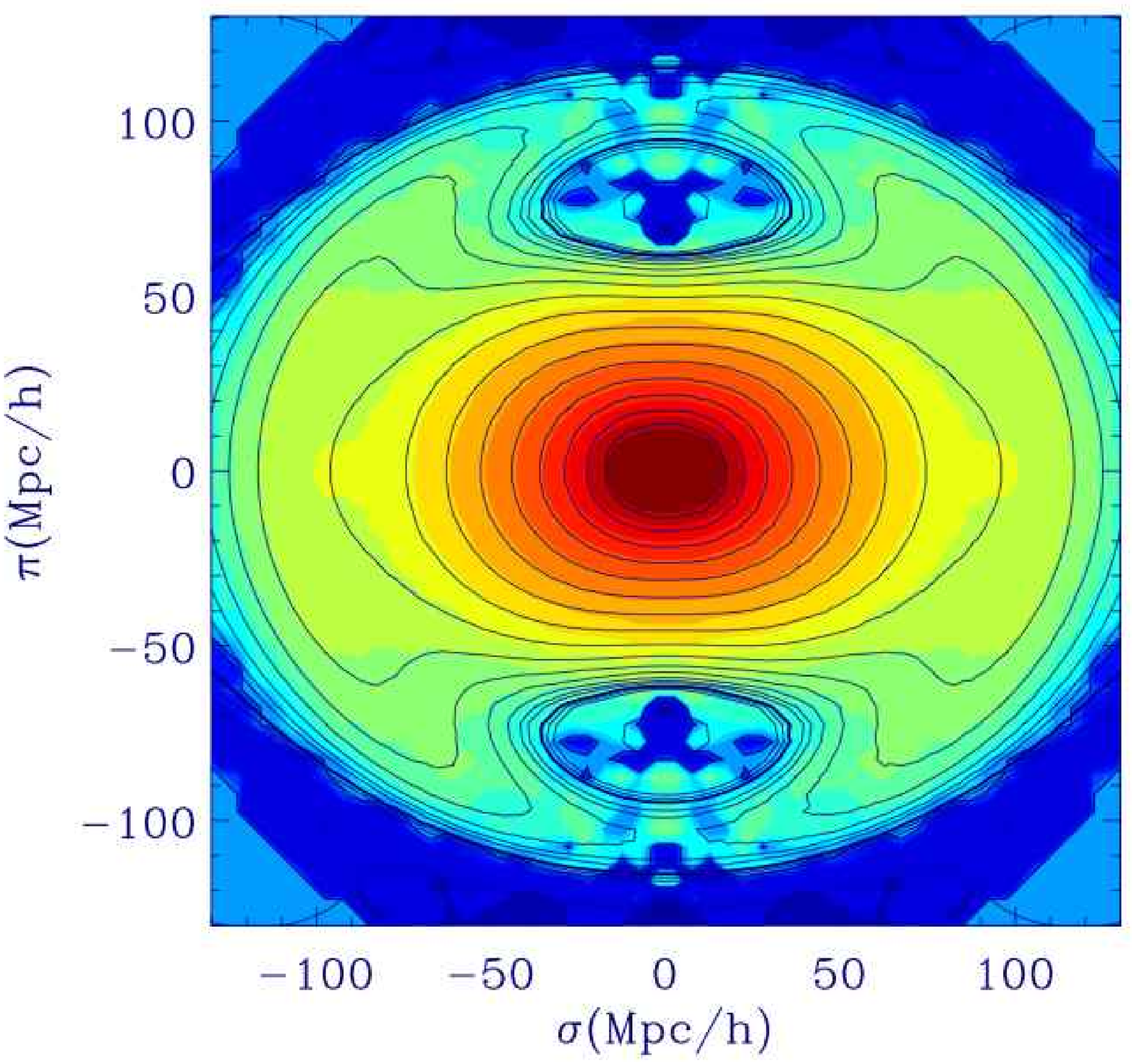}}
\caption{Correlation function  $\xi(\pi,\sigma)$ in different simulations: 
MICE dark matter
mocks  with $b=1$ and $\beta=0.63$ (left); group mocks 
with $b=1.9$ and $\beta=0.25$ (middle); 
and group mocks with $b=3.0$ and $\beta=0.23$ (right).
The contour colors are from the simulations.
 We have overplotted in solid lines the corresponding levels for
models with the same input parameters as the simulation in
each case. \label{fig:corrmice}} 
\end{figure*}

For the dark matter mocks we have that $\beta=0.62$ (see \S\ref{sec:errors}) 
and $\sigma_v$ is approximately 400km/s on large scales, but 
higher when approaching scales smaller than 10 Mpc/h
(see Fig.\ref{fig:pairwise7680}). We fit the quadrupole obtained for each
mock separately. The mean over all the mocks gives the correct exact value of
$\beta=0.62$ and a value of $\sigma_v \simeq$ 450Km/s which is slightly larger
than the asymptotic large scale value of 400 Km/s. This is probably because we are
obtaining an effective $\sigma_v$ which also accounts for the values at lower
scales which are larger, as shown in Fig.\ref{fig:pairwise7680}.
But at $1-\sigma$ (2 dof) $\sigma_v \simeq$ 450Km/s is consistent with
the asymptotic value at large scales, which is plotted as a red dot
in Fig.\ref{fig:quadrumice}.  In Paper-II of this series
we show how to find the
exact value for $\sigma_v$ as a function of scale.

In the top panel in
Fig.\ref{fig:quadrumice}, we show the mean contour  $\beta-\sigma_v$ which is
the average over the individual contours in each simulation, obtained from 
$Q(s)$. In the bottom panel, we show the best fit model 
over-plotted over the mean Q(s)
with the MC errors. The MC errors here correspond to a single mock, 
while errors in the
mean value are $\sqrt{N}$ times smaller (N=216 in our simulations). In the
bottom panel of Fig.\ref{fig:quadrumice}, we plot the errors corresponding to a
single mock. We find similar results for group mocks, which have different
values of $\beta$.
We conclude that the model used for the quadrupole, assuming a
constant $\sigma_v$, is perfect to obtain $\beta$ and an effective $\sigma_v$
which does not have to correspond necessarily to the large scale value, but 
rather to a combination of $\sigma_v$ at different scales.

\subsubsection{Fitting of large scales: $\Omega_m$ and amplitude}

From large scales, we will obtain the shape of the linear correlation function
$\xips$. Here we will only fit
$\Omega_m$ and the amplitude of the correlation. We have fixed $n_s=0.98$, $\Omega_b=0.045$
and $h=0.72$ following recent results of WMAP, SNIa and previous LSS
analysis \footnote{see http://lambda.gsfc.nasa.gov/product/map/dr3/parameters.cfm}.

As we have seen in \S\ref{sec:realspace} we can not rely on the projected
correlation function at large scales, or the recovered real space correlation
function, so the way to extract information about the shape of the real
space correlation function is directly from $\xisp$. We use models
that vary with $\Omega_m$, linear amplitude, $\sigma_v$ and $\beta$.
 The linear amplitude
$Amp$ refers to the factor $b(z)\sigma_8$, where b(z) is the bias at redshift z.
$b(z)$ and $\sigma_8$ are completely degenerated in the correlation function,
also with the growth factor $D(z)$. In fact, what we obtain from observations is
$b(z)\sigma_8 D(z)$, but D(z) is known for each cosmology and the median
redshift of the slice. We have checked with simulations that we can use linear 
bias in the scales that we are interested. We also use the priors
 $\sigma_v$ and $\beta$ found previously from the normalized quadrupole, although
 this does not make a large difference. Then, we need to fit 
$\Omega_m$ and $Amp$. The best place to fit
them, so that is independent on scale, is for intermediate
scales, from 20Mpc/h to 60Mpc/h, to be safe from non-linear bias.
We also stay away from the line-of-sight (LOS) where fingers of God can alter the
information, cutting an angle of 30-40 deg away from the LOS when doing the fit. 
We marginalize over $\beta-\sigma_v$ and obtain a contour for
$\Omega_m-Amp$ which is in very concordance with the input values. Although
$\sigma_v$ obtained from Q(s) is just an effective value, simulations show that
we recover the correct value of $\Omega_m$. We can understand it, 
because $\sigma_v$ does not play an important role in our selected range of
scales in $\xisp$. If we extend to larger
distances, the value for $\Omega_m$ is biased slightly to lower values, probably because
of the wide angle effect. This is important for an angle of 10 deg between galaxies
\cite{matsuwide}, which is more than 70Mpc/h for our worst case
(z=0.15), in the perpendicular direction $\sigma$.

\subsubsection{Model for the 2-point correlation function $\xisp$}

Finally, in Fig.\ref{fig:corrmice}, we plot $\xisp$ at large scales 
for the mean over MICE mocks (colors). We over-plot in solid lines 
the best model (using Eq.(\ref{eq:hamiltonmethod})). In all cases, we use an 
effective $\sigma_v=400km/s$. Note that we use square bins (pixels) of 
5Mpc/h side at all scales and this dilutes the FOG at small 
$\sigma$. The model works very well compared to simulations. 
We can observe in the figures that the overall amplitude is different for each simulation, because it is proportional to the bias. The same bias modifies the distortion parameter $\beta$ (inverselly proportional to the bias) which changes the shape of $\xips$. 
The larger the value of 
$\beta$ the greater squashing effect in the radial ($\pi$) direction.
Also note the closed contours in the radial direction between
$\pi=50-100$ Mpc/h (in dark blue). They correspond to a region of negative 
amplitude caused by the squashing (Kaiser) effect. The larger the value of
$\beta$ the larger the region with negative values. This region is sorounded
by the BAO ring at about 100 Mpc/h. It is remarkable how well
this is followed by simulations, which indicates that this is a very significant
feature. We have shown before that data also follows this feature.

Simulations also show a good agreement with the BAO peak, including some
detection in the radial direction which is enhanced by noise. This will be
studied in more detail in Paper IV of this series.

 We see that the obtained correlation (colored) differs slightly from the
distant observer approximation theory (lines) at large $\sigma$ and $\pi\simeq
0$. The redshift space correlation distortion
in real surveys, which are not located at
infinity, depends on $\pi$ and $\sigma$, but also on the angle between galaxies
$\theta$ and the angle $\gamma_z$ between the direction LOS (at $\theta/2$) and
the vector which goes from galaxy 1 to galaxy 2 (following the notation used in
%\cite{matsuwide}).
Matsubara 2000).
In the  distant observer approximation we assign to the angle
between galaxies the value $\theta=0$. 
%\cite{matsuwide} 
Matsubara (2000) has studied the differences
between the real correlation and the approximation for distant
observers.  In general, the approximation is good for angles $\theta$ below 10
deg, which include all the zone we are using for our analysis in LRG data (the
worst case comes from the closest galaxies at z=0.15, where
$\sigma\simeq 80Mpc/h$ corresponds to $\theta=10deg$). 
The correlation also depends on the
$\gamma_z$ angle. We can see in Fig.9 of 
%\cite{matsuwide}
Matsubara (2000) a comparison between the
real correlation, which depends on the distance between galaxies and both angles
described above, and the distant observer correlation function, used in this
work, which depends only on $\pi$ and $\sigma$. Each position in the 2
dimensional $\pi-\sigma$ is a mixing of different $\theta$ and $\gamma_z$,
because there is a range in redshift, but we can explain qualitatively the lack
of power of the observed correlation respect to the distant observer
approximation theory at large $\sigma$ and $\pi\simeq 0$.

We have also calculated the correlation function limiting the angle 
between galaxies to see in practice how $\xips$ is affected
by wide angle effects (see Appendix B5). As we increase the restriction, we see
how the $\sigma$ direction recovers power in better agreement with
the model predictions. In real data, we rather use all pairs given that
this is quite a small distortion.

\section{Systematic effects}
\label{sec:systematicr}

In this section we look for possible systematic errors that could be imprinted by the 
radial mask through the line-of-sight, the angular mask, or the selection of LRGs
. 
These systematic effects are typically  more important on the largest scales, where
the correlation function becomes smaller.  
In Fig.\ref{fig:baoslices} we show that for some slices there is extra power in $\xis$ 
at scales $s>130Mpc/h$, something which is not expected in the models. This is particularly 
evident for the slice in redshift z=0.3-0.4, where the signal-to-noise is
small ($Amp$ and bias $b$ are smaller in Table \ref{tab:largescales})
and we do not have a significant peak detection. This extra power could be due to
sampling variance but could also be caused by some systematic errors in the
data or the way we analyze it. The extra power seems to be important
at scales larger than the baryonic peak, but we test here if it could
also have some effect over the peak location. This extra power have also been
detected in other analysis of SDSS LRG around similar
scales \cite{blake,padmanabhan2007}. We will focus here on systematics
related to the way we have analyzed the data, rather than systematics in the data
itself. The later has been explored in detail elsewhere \cite{detection}.

\subsection{Radial Selection}

\begin{figure}
\centering{ \epsfysize=6cm\epsfbox{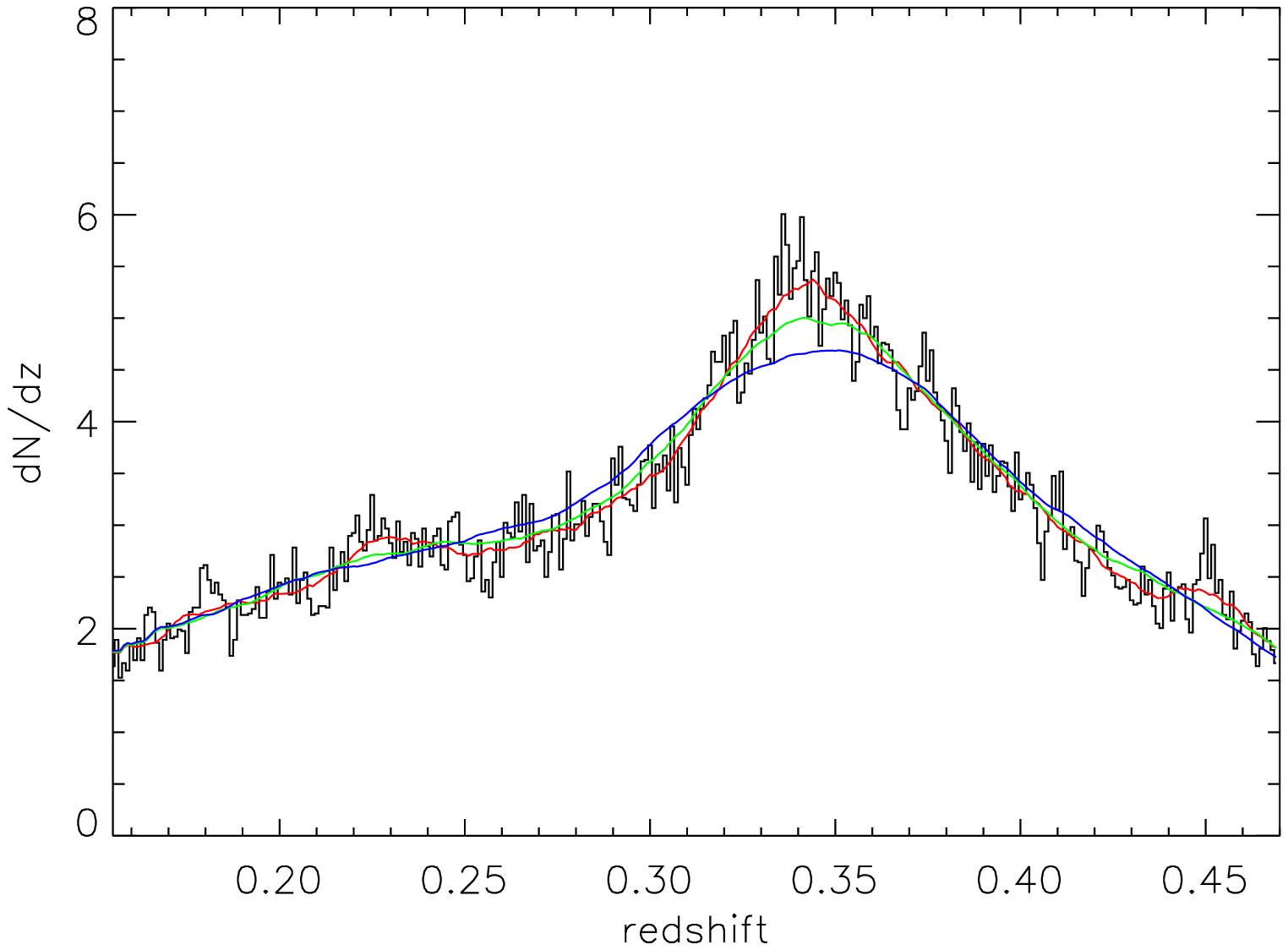}}
  \caption{ \label{fig:dndzrandoms} Selection function for data (black histogram) and smoothed for
the random catalogs with a bin in redshift of z=0.02 (red), z=0.05 (green) and z=0.08 (blue)}
\end{figure}

\begin{figure}
\centering{ \epsfysize=6cm\epsfbox{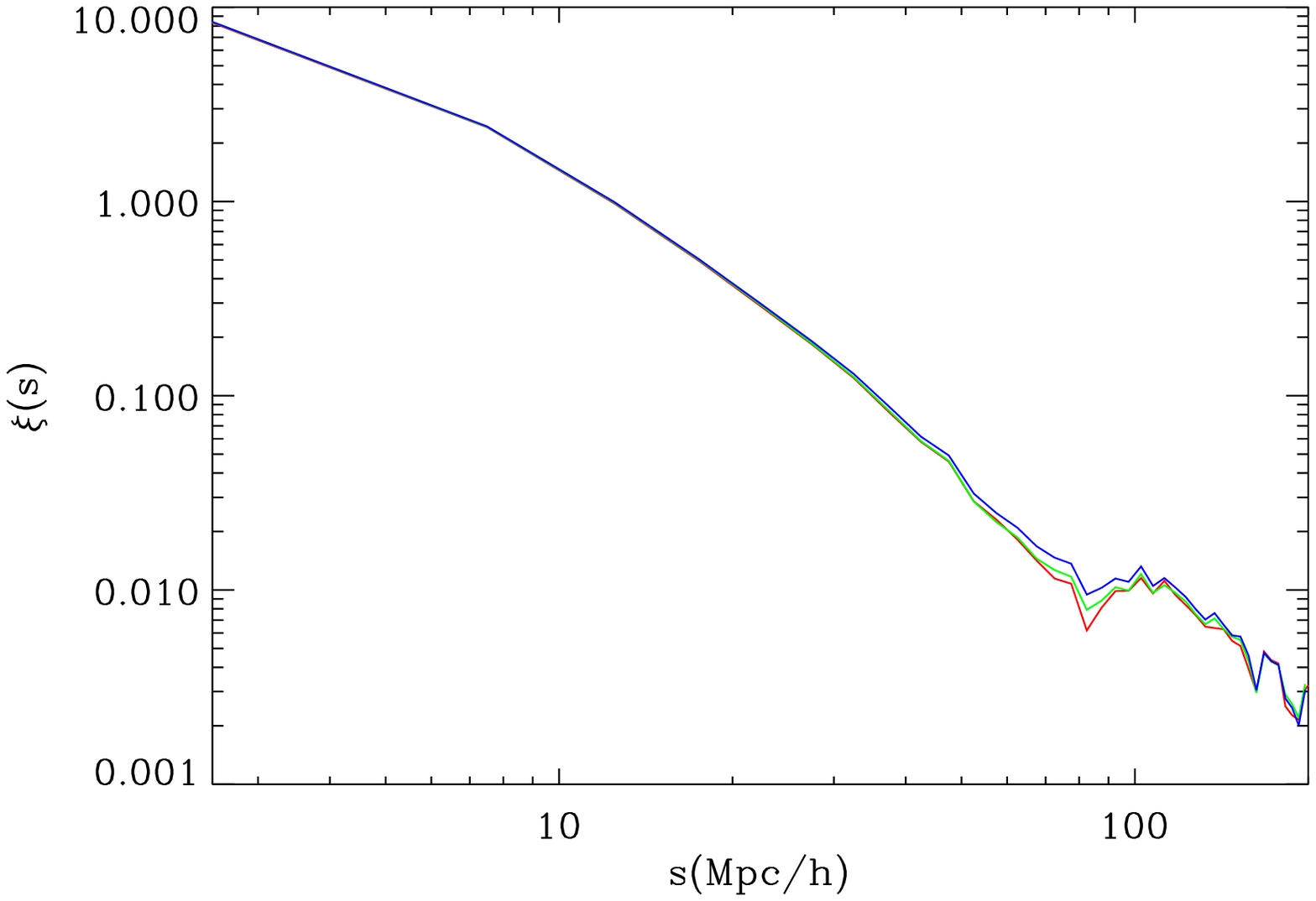}}
  \caption{ \label{fig:monopoleradial} Redshift-space correlation using different redshift 
smoothing in the random selection function of 0.02 (red), 0.05 (green) and 0.08 (blue)}
\end{figure}

\begin{figure}
\centering{ \epsfysize=7cm\epsfbox{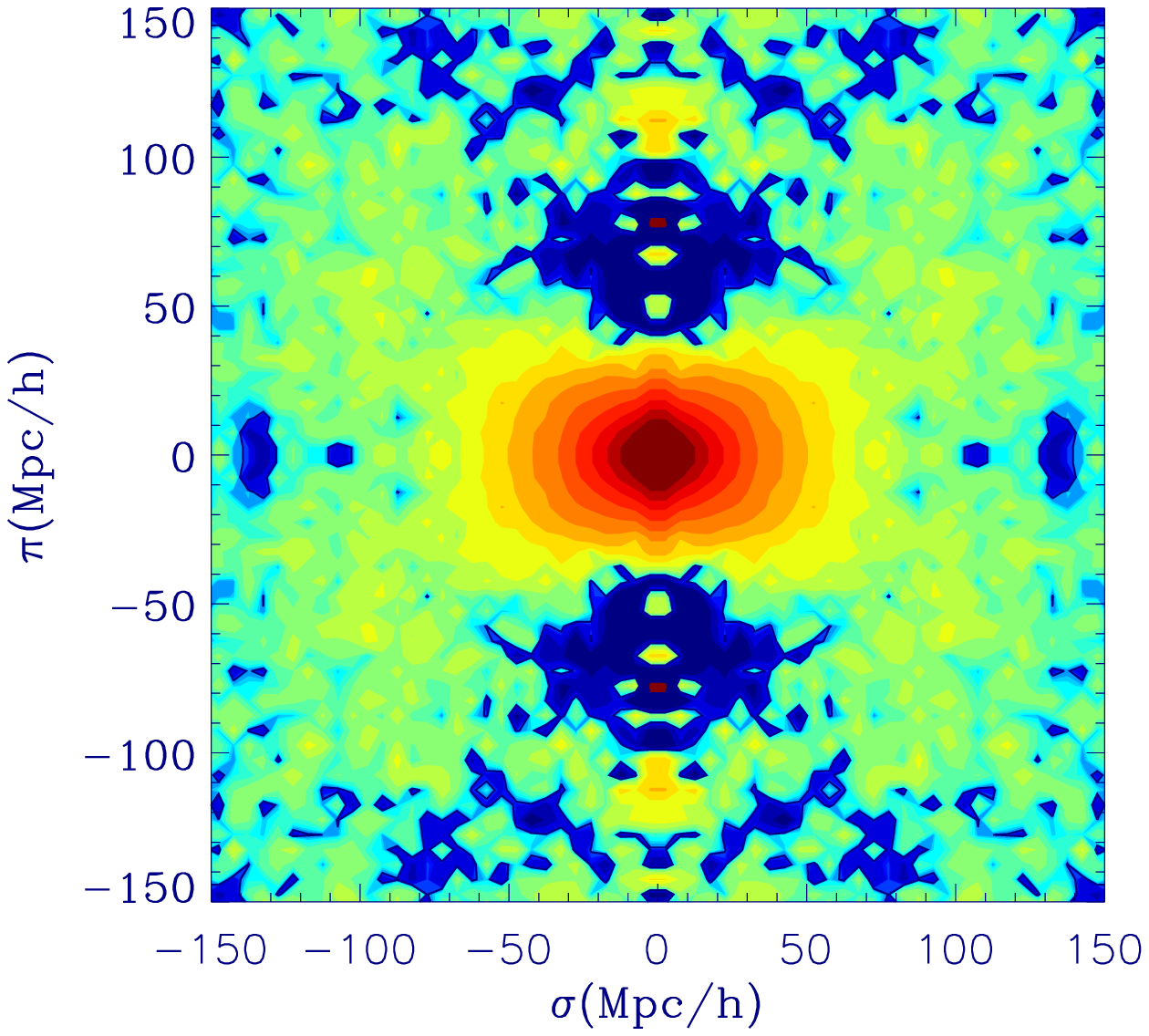}}
\centering{ \epsfysize=7cm\epsfbox{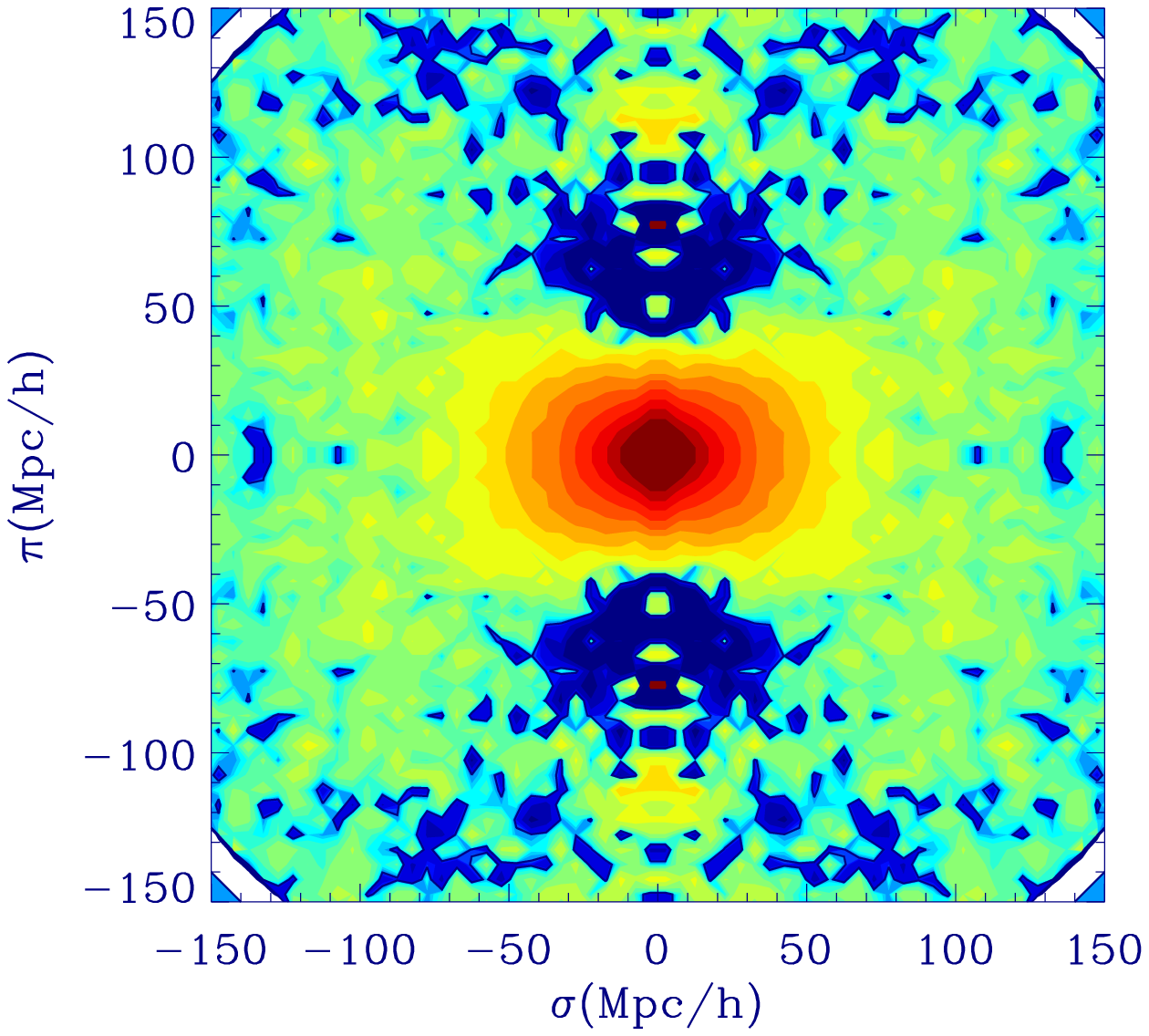}}
\centering{ \epsfysize=7cm\epsfbox{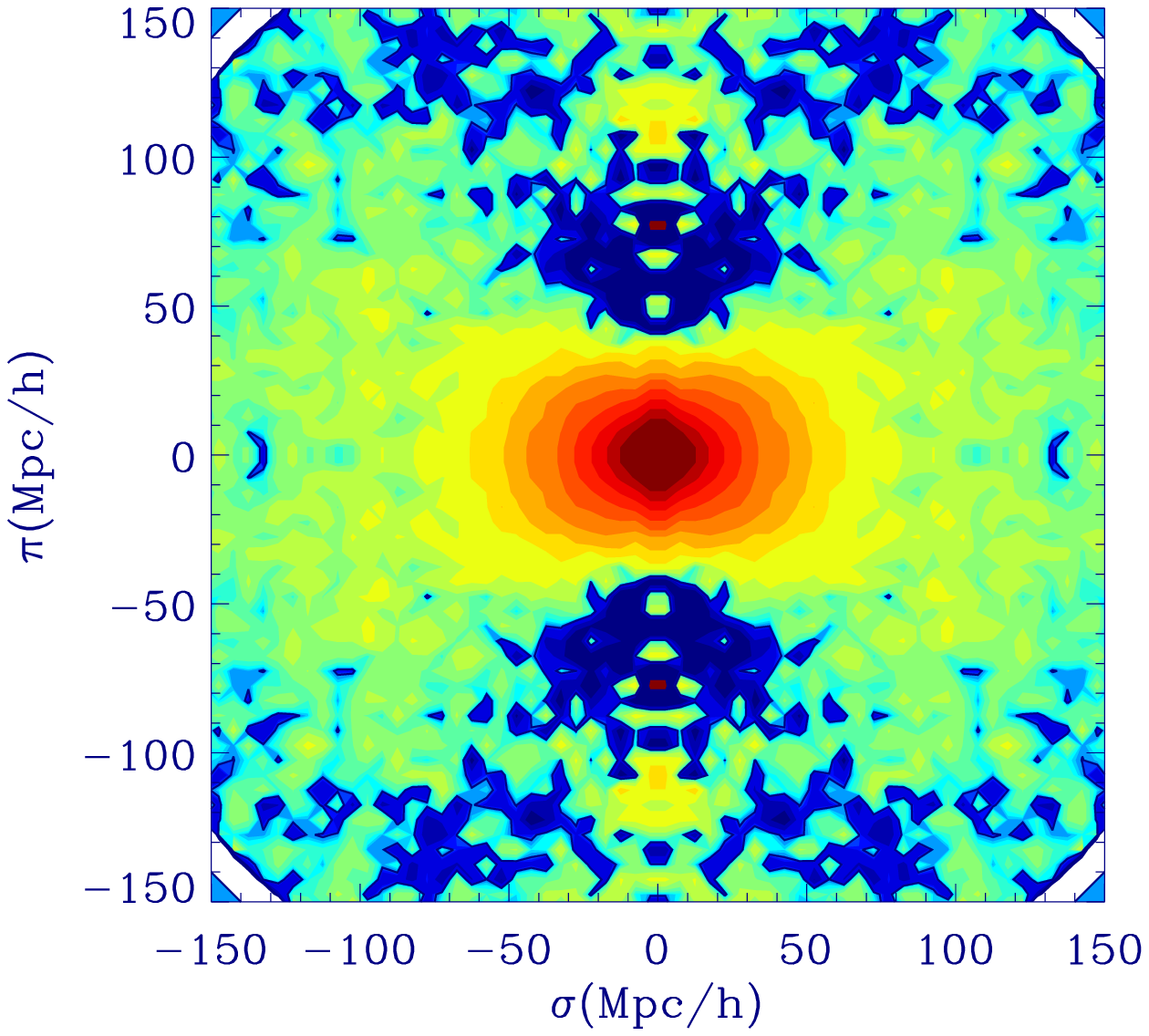}}
  \caption{ \label{fig:pisigmaradial} $\xips$
correlation using different redshift 
smoothing in the random selection function of 0.02 (top), 0.05 (middle) and 0.08 (bottom).}
\end{figure}

First, we test the radial selection function that we use for the random catalogs. 
If we use exactly the same radial density distribution
$N(z)$ as selection function of the data, we suppress the radial modes in
the $\pi$ direction. In Fig.\ref{fig:dndzrandoms} we can see the differences between 
different smoothing windows in the data selection function, and in Fig.\ref{fig:monopoleradial}
and Fig.\ref{fig:pisigmaradial} 
 the redshift-space correlation function for these three smoothing bins. 

We do not see any  significant difference between the three cases indicating that
our analysis is robust with respect to the radial selection.

\subsection{Weighting}

\begin{figure}
\centering{ \epsfysize=6cm\epsfbox{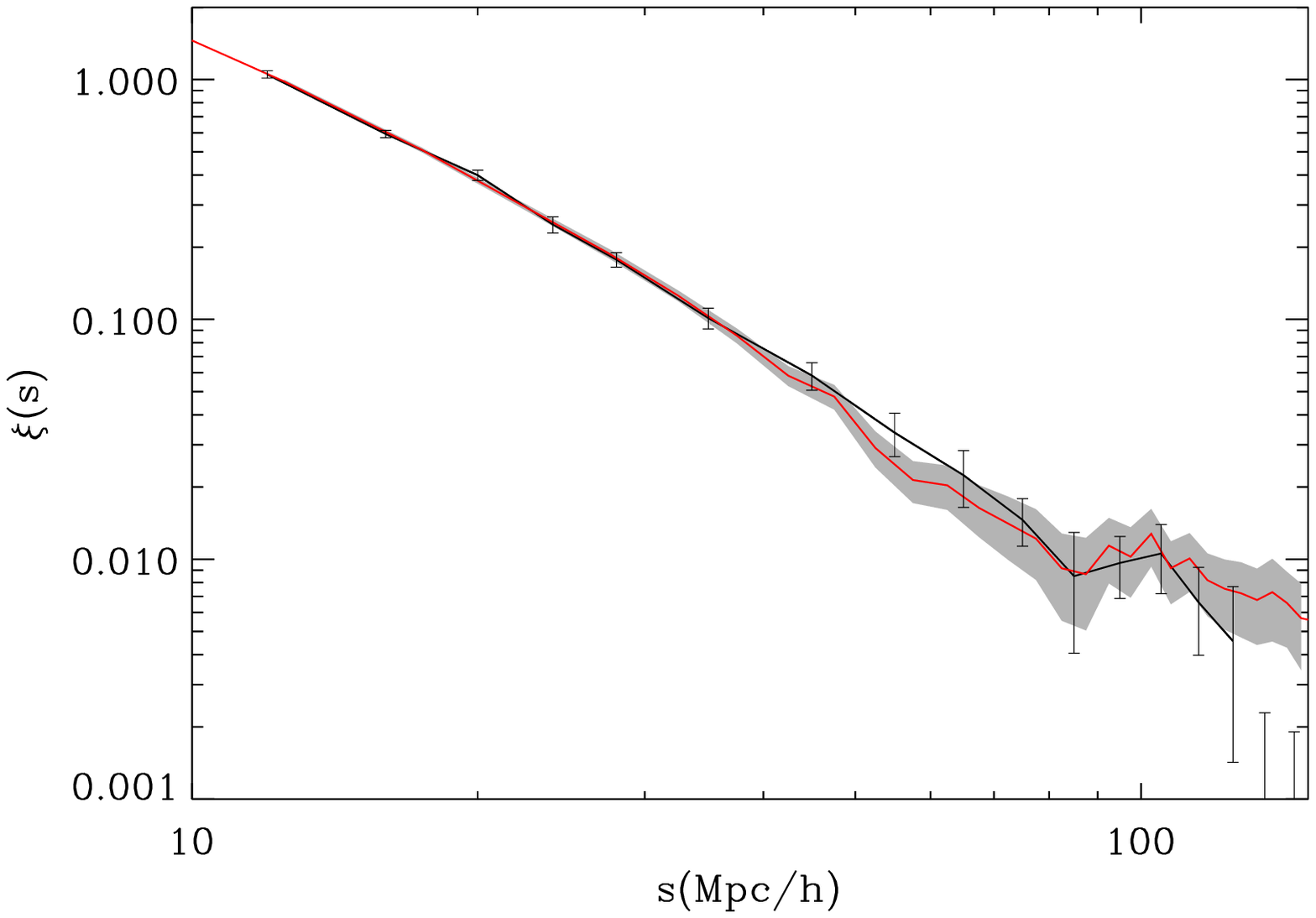}}
  \caption{ \label{fig:compeisenstein} 
Estimation of the redshift-space correlation function for LRG. In black line
with JK errors correspond to the result by Eisenstein et al. 2005.
%\cite{detection}. 
Over-plotted in red is our result, where the shaded region corresponds to our error
estimate.}
\end{figure}

In Fig.\ref{fig:compeisenstein} we compare our results 
 in the monopole (red with shaded region as errors) to that of
%\cite{detection} 
Eisenstein et al (2005) (in black line with errorbars) for all the sample.
Our result is consistent with this previous estimate  despite the increase
in the DR6 area  and the
difference in the selection. However, at larger scales than the baryonic
peak, we observe some extra-power in our estimation. 
We wonder if the difference is just due to sampling, selection 
or to the way we estimate the correlation function (we do not include 
the same weighting as Eisenstein et al 2005)\\
%\cite{detection}).\\

We have also calculated the correlation function using a weighting scheme, 
as the one explained in 
%\cite{detection}.
Eisenstein et al (2005). We weight the sample using a scale-independent weighting
that depends on redshift. When computing the correlation
function, each galaxy and random point is weighted
by 1/(1 + n(z)Pw) \cite{fkp1994}
where n(z) is the comoving number density and $Pw =40,000h^{-3}Mpc^3$. 
We do not allow Pw to change with scale
so as to avoid scale-dependent changes in the effective bias
caused by differential changes in the sample redshift. 
We choose Pw at $k\simeq0.05hMpc^{-1}$ as in 
%\cite{detection}. 
Eisenstein et al (2005).
At z $<$ 0.36, nPw is about 4, while $nPw\simeq1$ at z = 0.47.

\begin{figure}
\centering{ \epsfysize=6cm\epsfbox{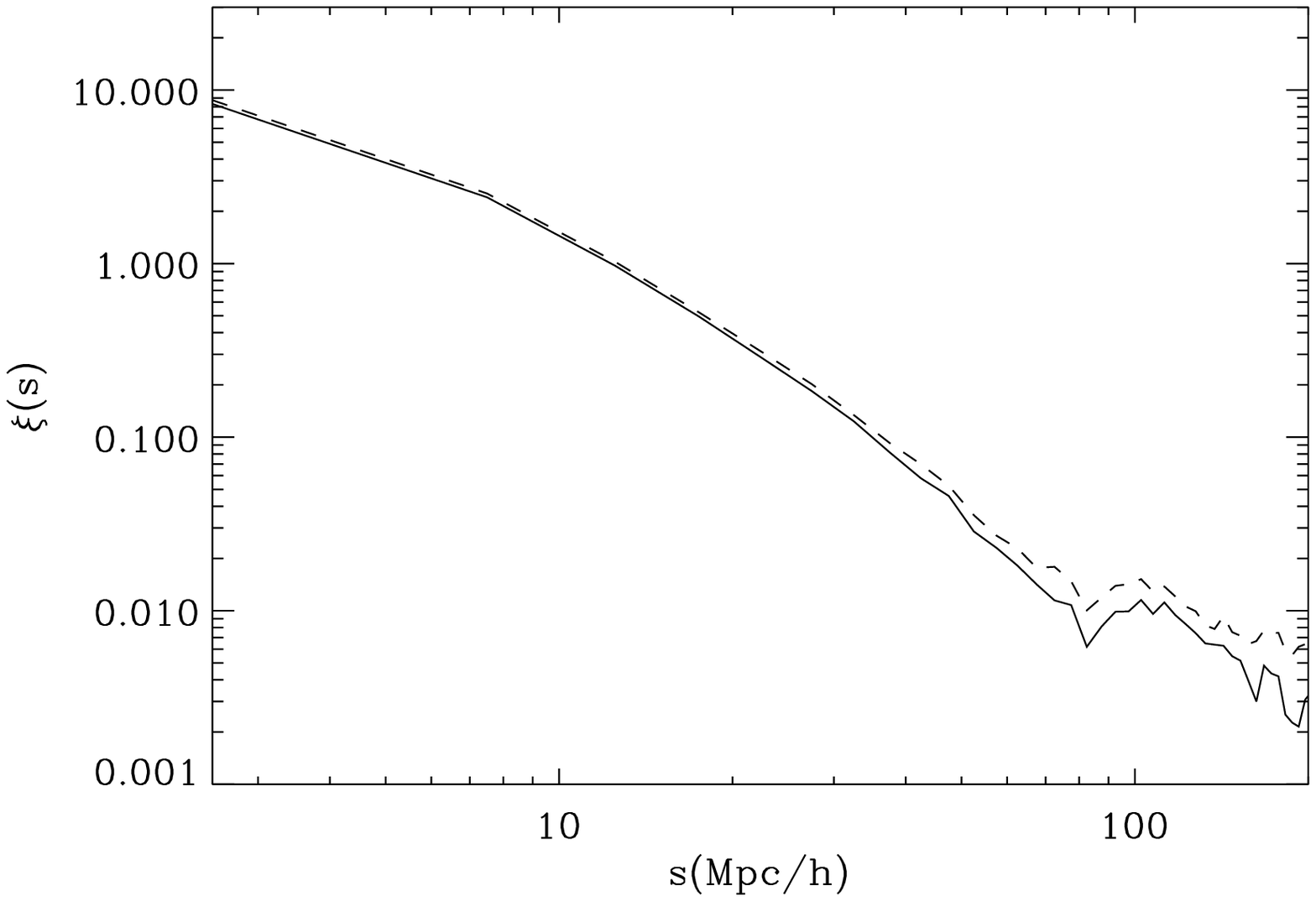}}
  \caption{ \label{fig:monopoleweight} Our previous estimation of the redshift-space correlation 
function for LRG (solid) compared to estimation using a weighting as explained in the text (dashed)}
\end{figure}

In Fig.\ref{fig:monopoleweight} we can see the comparison between the correlation function estimated 
without weighting (solid line) and with weighting (dashed line). Contrary of what we were looking for, 
the extra power is higher in the weighting scheme, which makes sense, since we are now giving more importance to the
higher redshift  pairs, which have a larger bias (see below). 

\begin{figure}
\centering{ \epsfysize=7cm\epsfbox{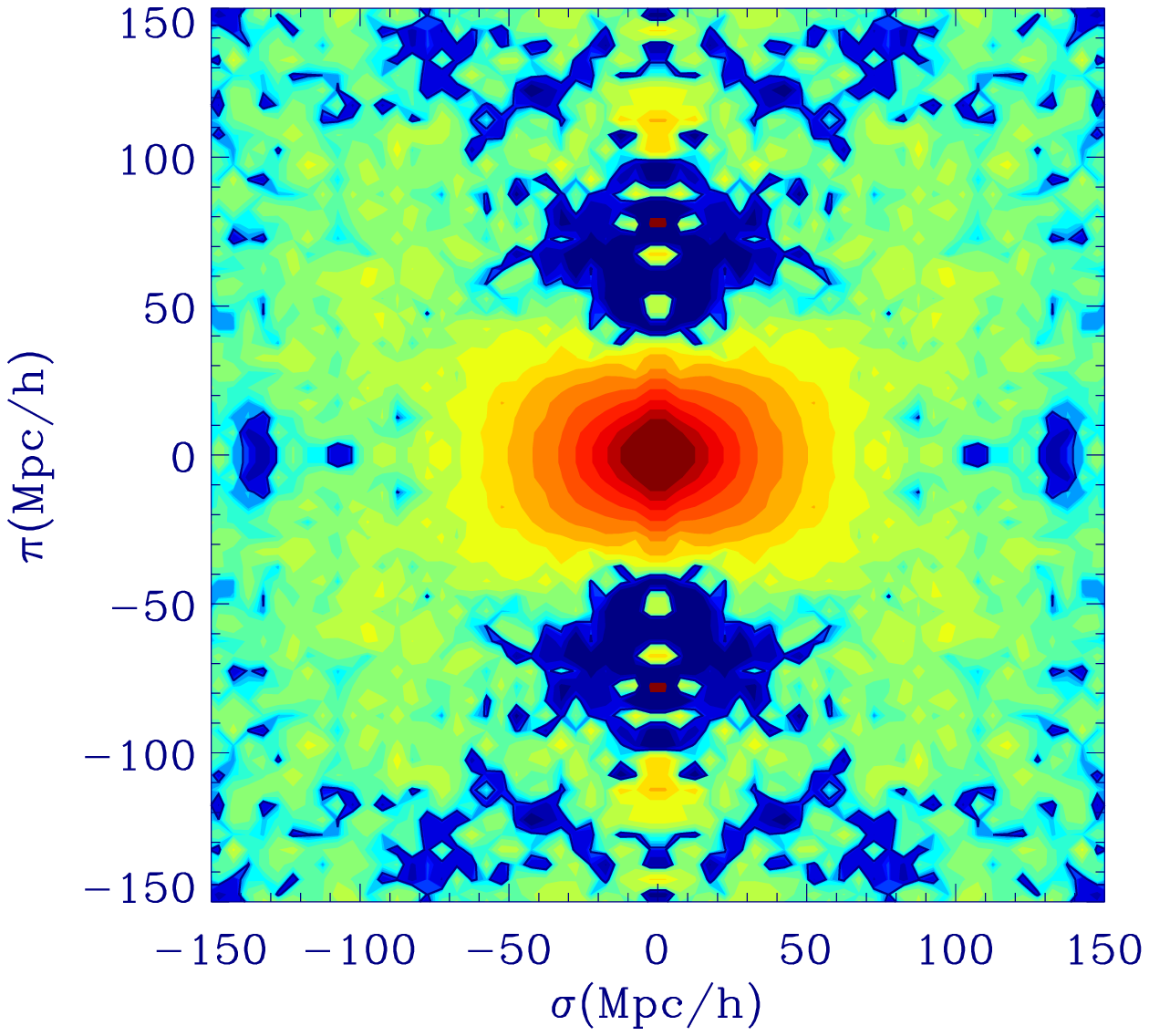}}
\centering{ \epsfysize=7cm\epsfbox{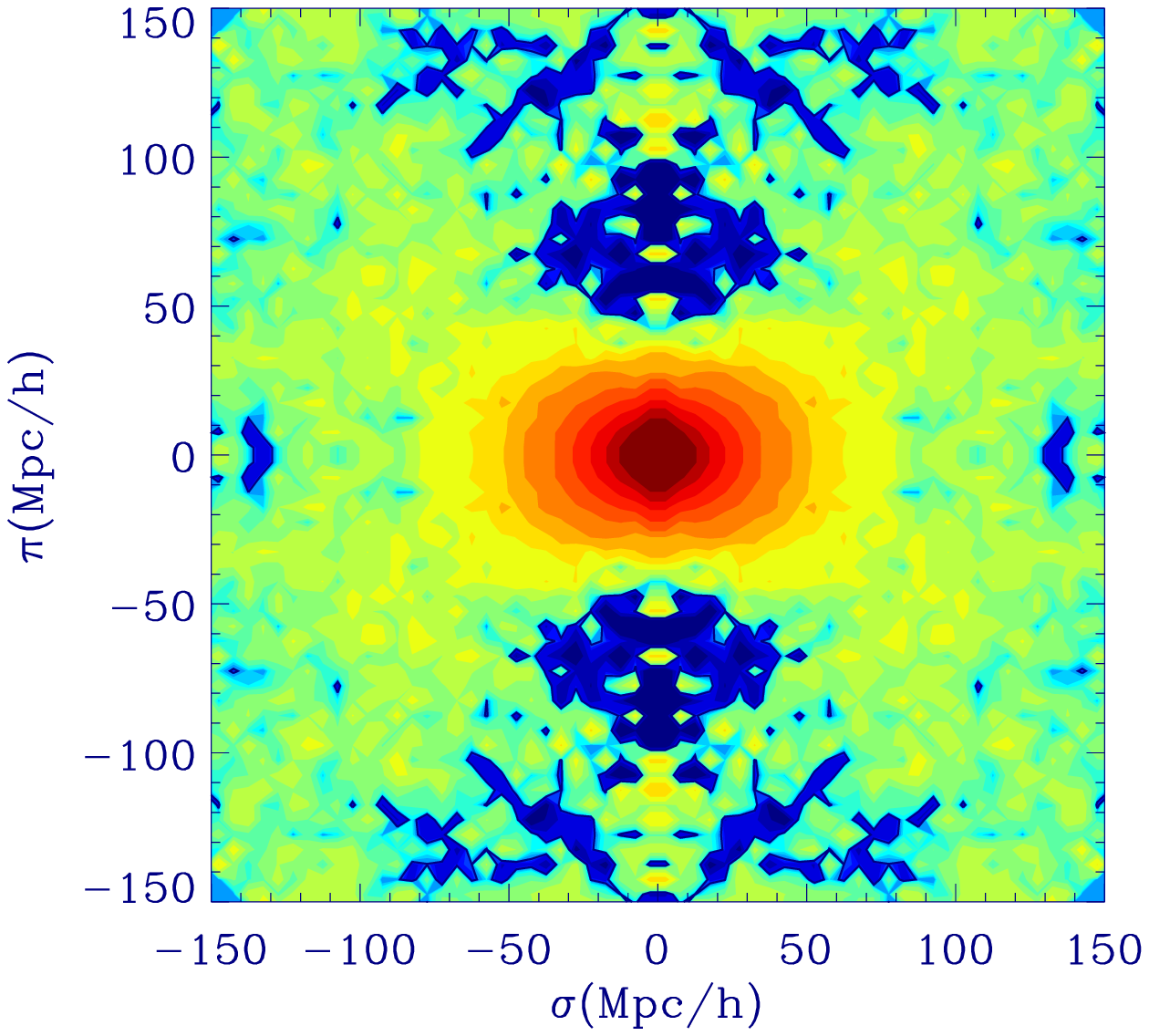}}
  \caption{ \label{fig:corrweight} Our previous estimation of the redshift-space correlation function $\xips$ for LRG 
(top panel) compared to estimation using a weighting as explained in the text (bottom  panel)}
\end{figure}

In Fig.\ref{fig:corrweight} we have plotted the anisotropic correlation 
function $\xips$ for both cases, showing only slight differences. It is 
apparent that the monopole is more sensitive that $\xips$ to displace
systematic differences. We will therefore concentrate on results for the
monopole from now on.

\subsection{Angular Selection}

\begin{figure}
\centering{ \epsfysize=6cm\epsfbox{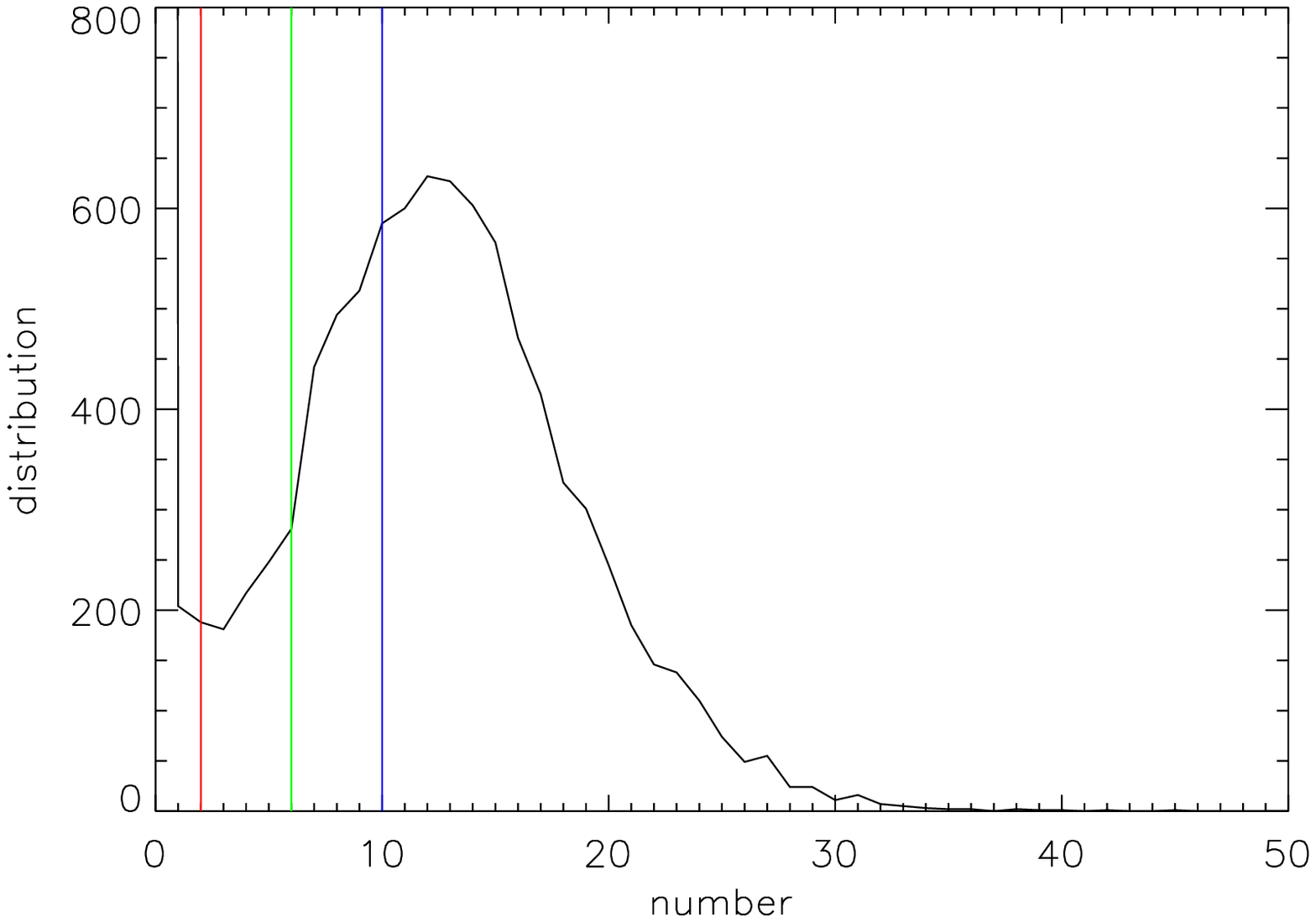}}
  \caption{ \label{fig:distribution64} Distribution function of number of galaxies per pixel (nside=64 using the package Healpix)}
\end{figure}

\begin{figure}
\centering{ \epsfysize=6cm\epsfbox{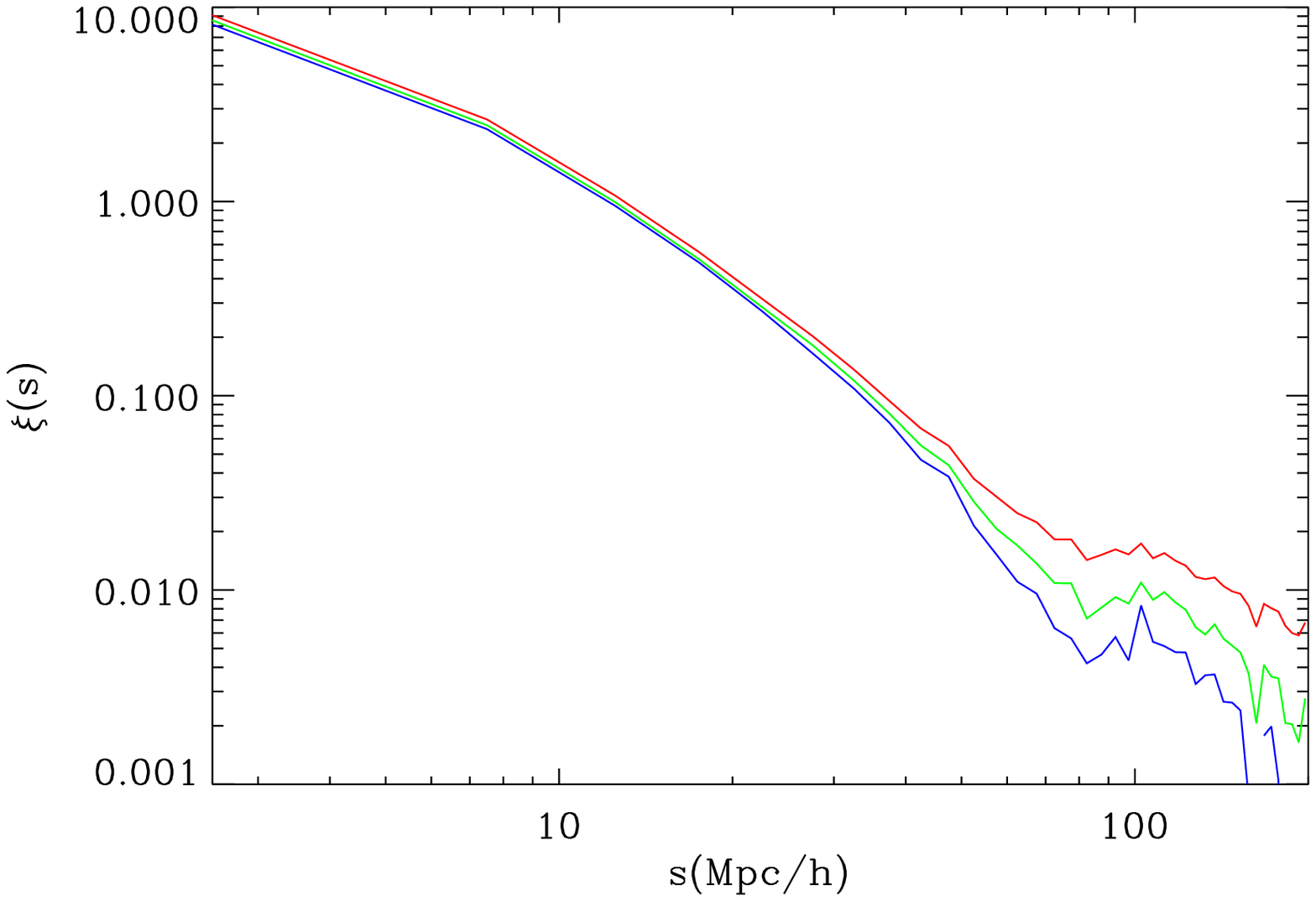}}
  \caption{ \label{fig:monopolemask} We now define our catalog by including all the pixels that have more than $n$ galaxies in the Fig.\ref{fig:distribution64} (n$>$2 red, n$>$6 green, n$>$10 blue) and plot the redshift-space correlation function}
\end{figure}

We next look at the angular selection function, the mask. 
First, we construct a different mask than the original by using 
a Healpix map \cite{healpix}, 
with nside=64 (with pixels of area $\simeq$ 0.8 sq deg). 
In Fig.\ref{fig:distribution64} we plot the distribution function for the number 
of galaxies per pixel. The pixel is large enough to distinguish between real empty 
pixels and artificial ones. We only include pixels that have more than n galaxies, 
where n=2,6 and 10. As we can see in the distribution plot, if we include more 
than 2 galaxies, we are probably including artificially void zones, which will 
create extra power and pencil beams at the direction line-of-sight, while when
 we include only the pixels with more than 10 galaxies, we are probably excluding 
some real voids. In this second case, the density is higher than the real one, 
so the density contrast is lower. We can see these effects in 
Fig.\ref{fig:monopolemask} for the correlation function in redshift space. 
We see that the correlation is lower when we increase the minimum number 
of galaxies (from red to blue), but we can always see the baryonic peak 
and the three lines have the same shape approximately. Our results are similar 
to the case n $>$ 6, which does not include artificial voids and does 
not eliminate real voids. 

\begin{figure*}
\centering{ \epsfysize=6cm\epsfbox{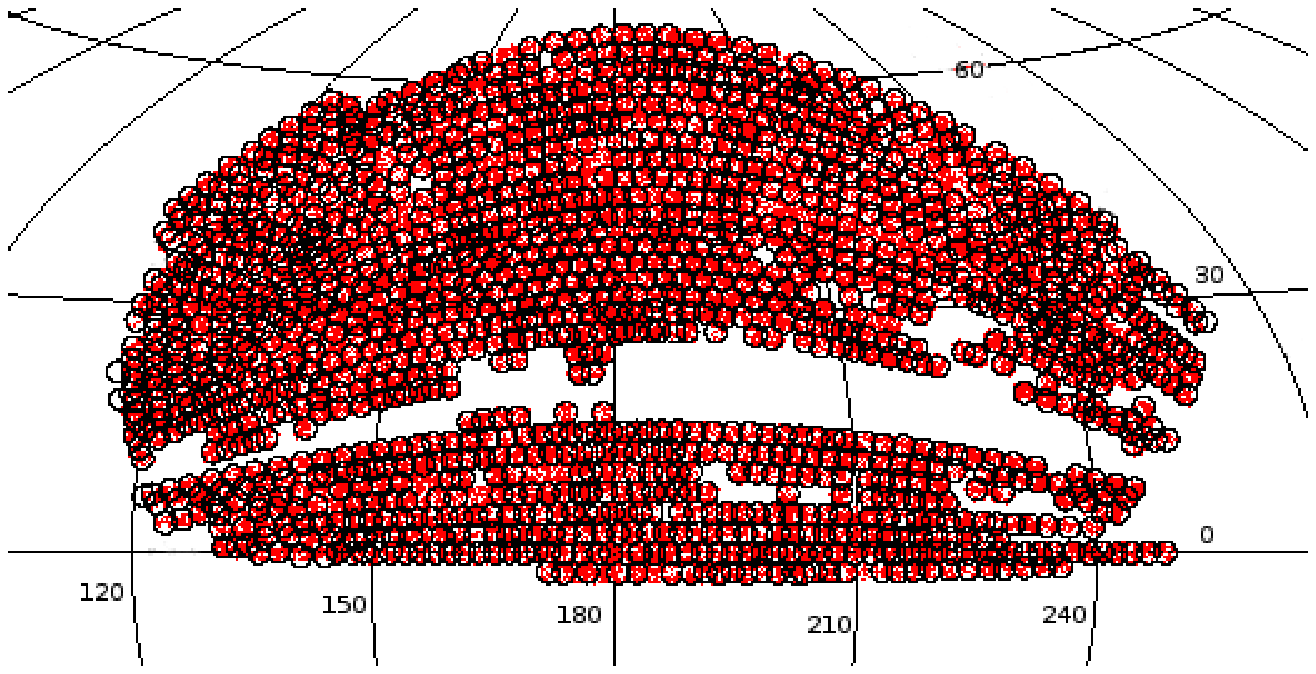}}
  \caption{ \label{fig:platesplot} SDSS DR6 survey with our selection 
of LRG galaxies as red points and the spectroscopic ``good'' plates as black circles }
\end{figure*}

\begin{figure}
\centering{ \epsfysize=6cm\epsfbox{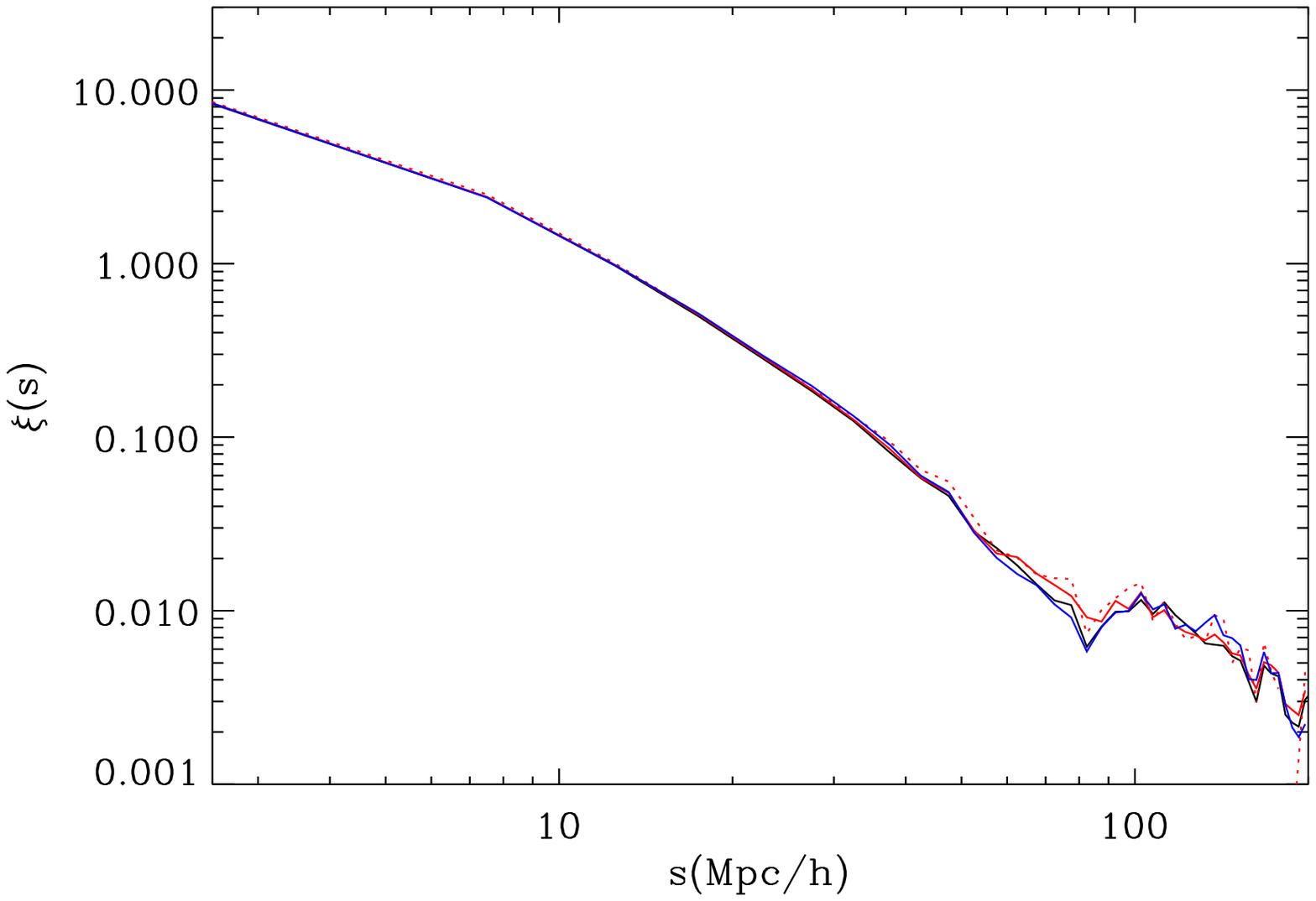}}
  \caption{ \label{fig:monopoleplate} Redshift-space correlation function for our mask (black), maskplate1 as indicated in the text (red), maskplate2 (dotted red) and north stripe (blue)}
\end{figure}

The spectroscopic survey of SDSS is observed using circular plates, 
which contain about 600 fibers each to take spectra. Targets are selected 
from the photometric survey, although the spectroscopic survey is not exactly 
the same as the photometric one. There are some plates that have not been 
observed properly due to known problems, which are explained in the 
web (http://www.sdss.org). We have extracted from our previously 
calculated mask all the galaxies that are not laying inside ``good'' 
plates (maskplate1). In Fig.\ref{fig:platesplot} we can see the plates 
(black circles) and the galaxies (red dots). Moreover, we can also eliminate 
all the galaxies that lay inside a bad plate to ensure that we are taking 
only the really good ones (maskplate2). This second mask reduces the number 
of galaxies significantly, and the correlation is then a lot noisier, but 
we show here the results. In Fig.\ref{fig:monopoleplate} we have plotted 
the redshift-space averaged correlation function for the new mask based 
on spectroscopic plates. Results are very similar to our previous result, 
in black. Moreover, we over-plot the result for the north stripe of SDSS, 
which contains the most significant part of the survey, in blue. The anisotropic 
redshift-space correlation function is really similar in all these cases, 
so that we do not find it useful to plot it.\\

We have also tried to extract all the plates that have a large number of galaxies, 
which have big superclusters, since it is known that big clusters can bias 
the correlation function for regular galaxies \cite{baughetal}. The result is not significantly modified
which could have been expected since LRG in SDSS DR6 cover a much larger volume
that 2DFGRS or SDSS for regular galaxies.\\

\subsection{Volume limited samples}

\begin{figure}
\centering{ \epsfysize=6cm\epsfbox{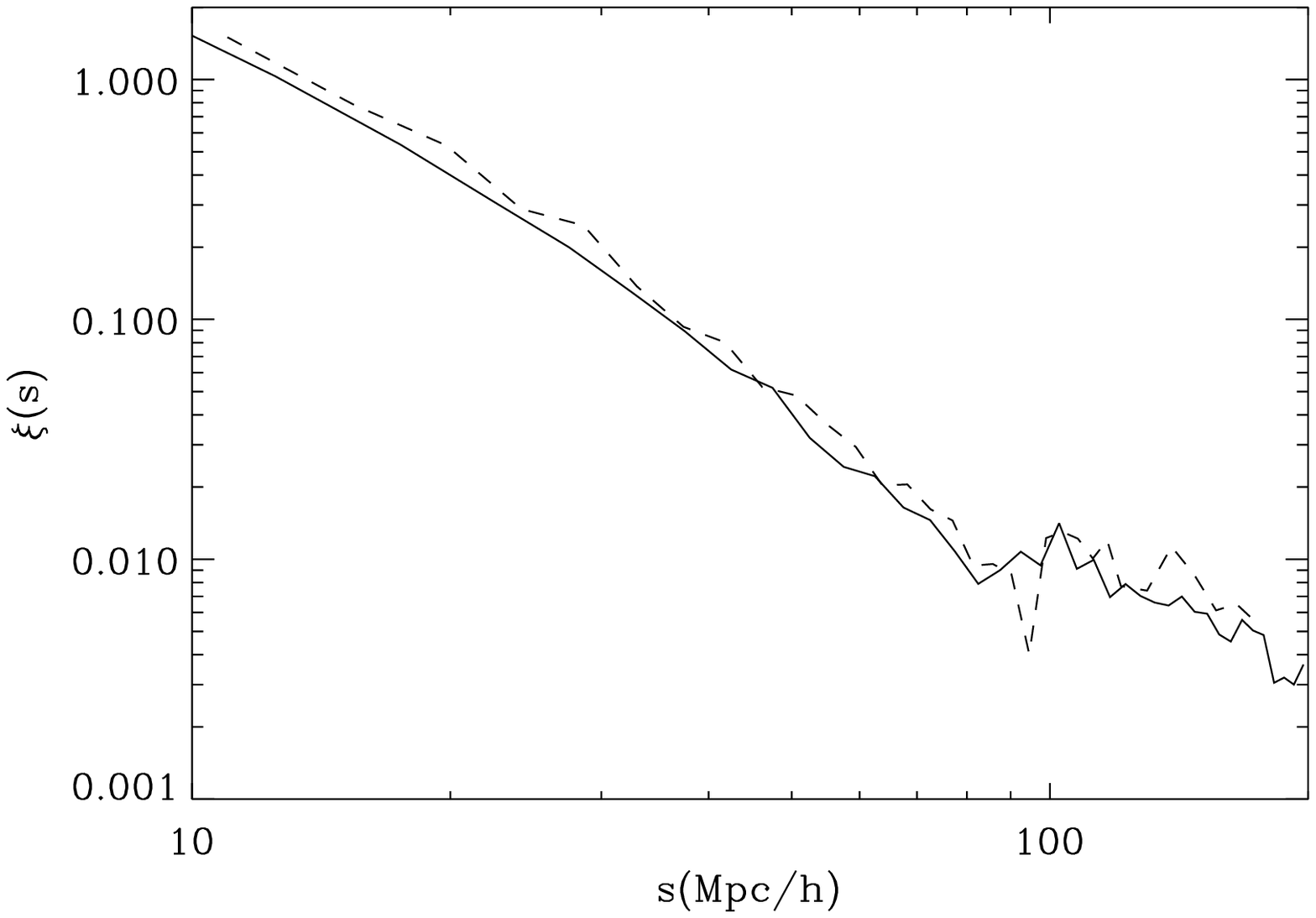}}
  \caption{ \label{fig:methodsdif} Redshift-space correlation function for a volume 
limited slice selection of LRG galaxies with z=[0.15,0.38] and $M_r$=[-22.5,-21.5]. 
Solid line shows the result when using a random catalog. Dotted line shows a new method 
based on pixelization which validated the previous results}
\end{figure}

In volume limited surveys, we can estimate the correlation function with a 
pixelization scheme.  The correlation function can then be estimated as:

\begin{equation}
 \xi(s)=\sum_{ij}\delta_G(s_i)\delta_G(s_j)/N_{pairs}(s)
\end{equation}

where $\delta_G=N_{G}/<N_{G}>-1$ and the sum extends to all pairs i,j 
separated by a distance $s \pm \Delta s$. We have taken a volume limited 
part of the selected galaxies, with redshift z=[0.15,0.38] and absolute 
magnitude $M_r$=[-22.5,-21.5] and have calculated the correlation using 
this method. This is the same sample used in Paper IV of this series.
We have also calculated a similar selection by using the 
traditional method with a random catalog. Results are plotted in 
Fig.\ref{fig:methodsdif}, dotted for the pixel method and solid
for the randoms. This is a good test to validate our results, since both methods
are quite different. As we can see in the figure, the 
two estimations are very similar. The slight shift in amplitude at 
small scales is due to the pixel
smoothing, a cube of 10 Mpc/h on the side for this case. This shift can
be modeled with a top-hat window function reproducing the exact
 amplitude difference. There is a very good match
using smaller pixel size. Also note how these estimations, based on a volume limited
sample with about half of the LRG galaxies, agree quite well 
with previous results (ie in Fig.\ref{fig:monopoleplate}) which 
includes all galaxies at a price of a more complicated selection function.
Despite all this difference, we see no significant
 change in $\xis$ at the largest scales.
\\

\begin{figure}
\centering{ \epsfysize=6cm\epsfbox{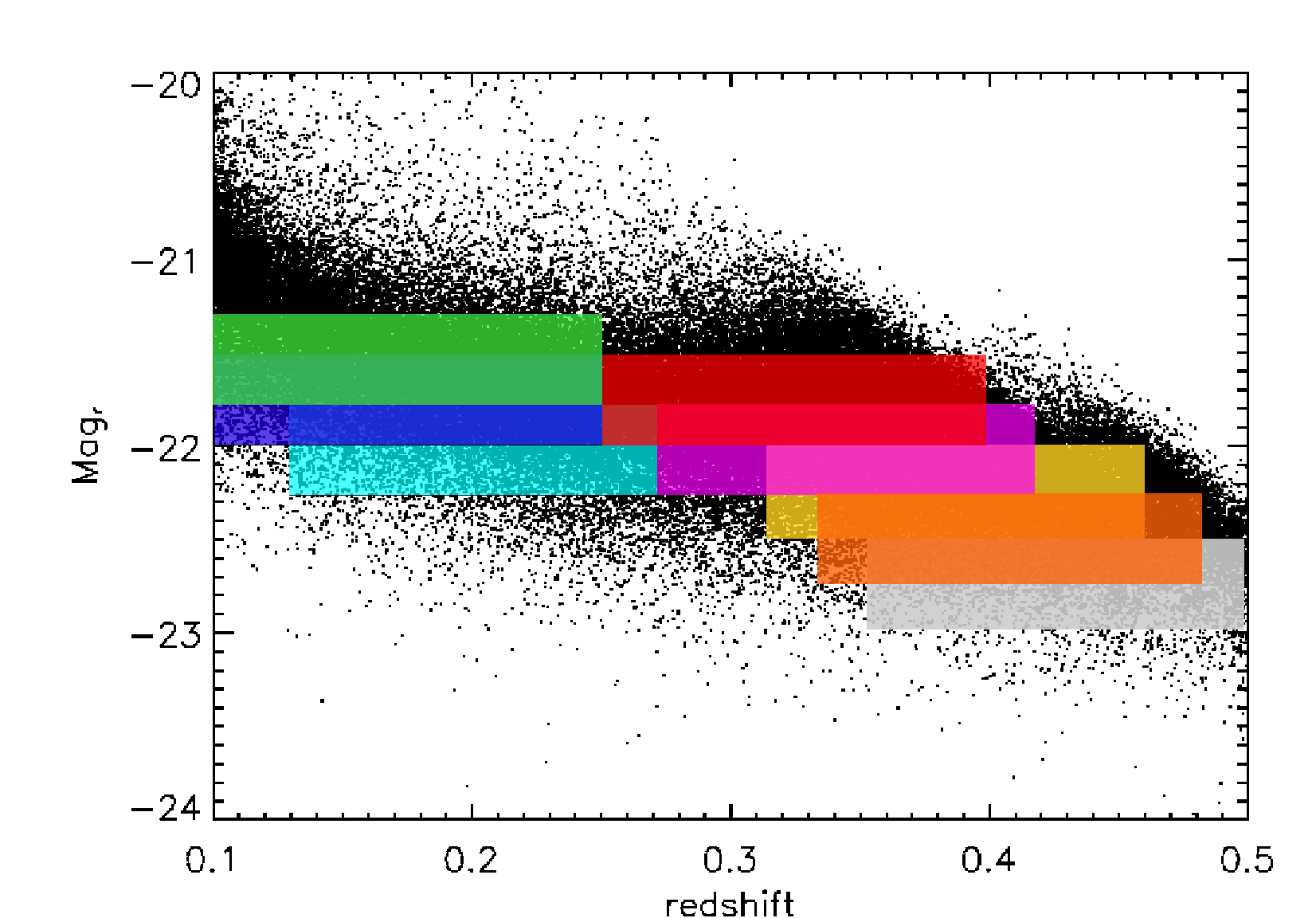}}
  \caption{ \label{fig:slicesplot1} We have divided the catalog in different approximately volume limited slices as indicated in table \ref{tab:slicesLRG}. Here we over-plot the slices in the plane $M_{r}-z$ }
\end{figure}

\begin{figure}
\centering{ \epsfysize=6cm\epsfbox{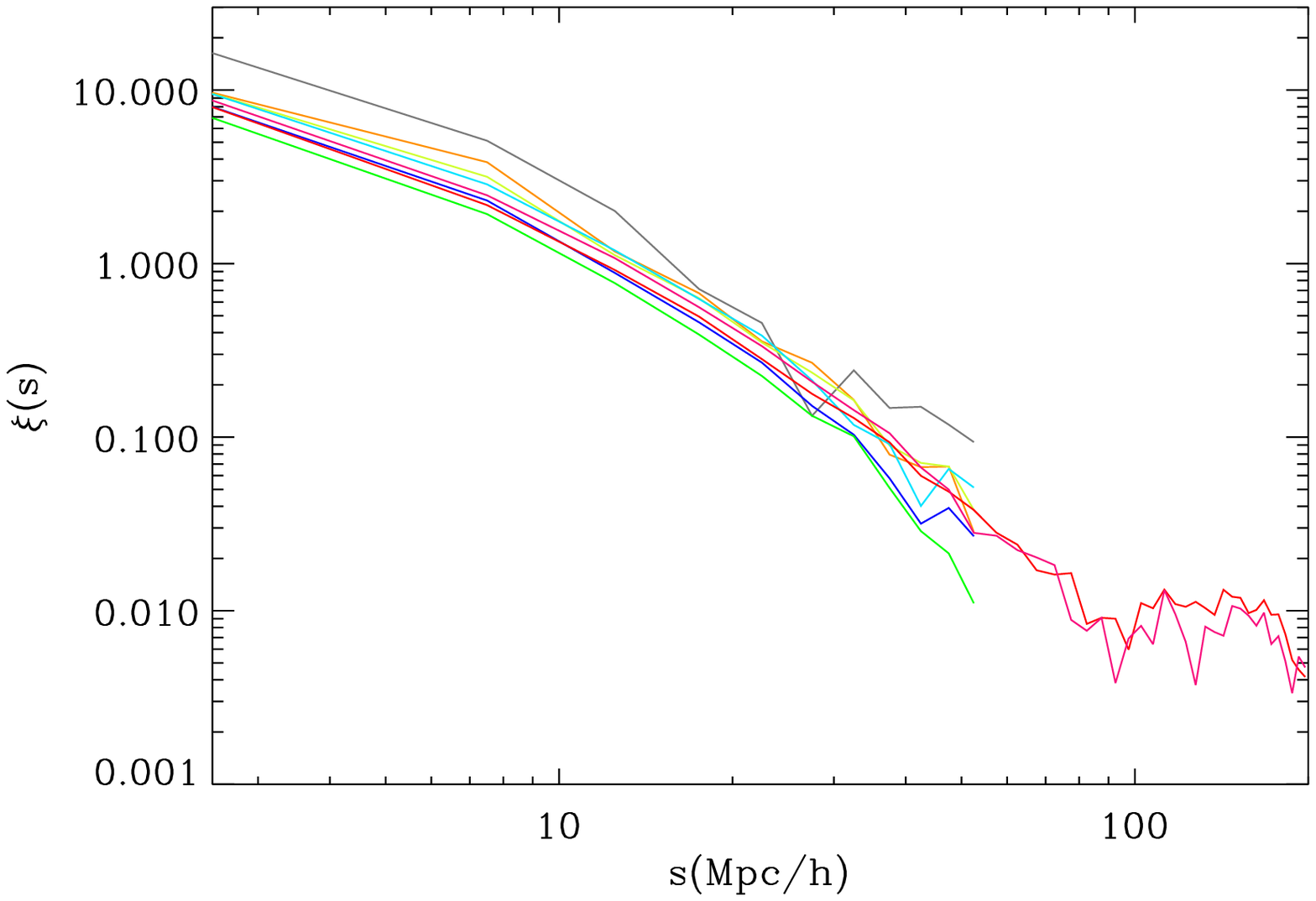}}
  \caption{ \label{fig:slicesplot2} Redshift-space correlation function for the different slices plotted in Fig.\ref{fig:slicesplot1}}
\end{figure}

We have also divided the catalog in different volume limited slices as indicated 
in table \ref{tab:slicesLRG} plotted in Fig.\ref{fig:slicesplot1}. The most 
and least luminous  slices (in the bottom right and top left)
contain fewer galaxies and have too much noise
at the baryonic peak, but we see that the red and pink slices, 
at intermediate redshifts and 
magnitudes, contribute to the extra-power that we see at large scales (Fig.\ref{fig:slicesplot2} 
with the same colors as Fig.\ref{fig:slicesplot1}). Note how these subsamples cover
a sharp feature a z=0.35 in the selection function (eg see Fig.\ref{fig:dndzrandoms}) 
which is clearly an artifact of the selection function because of the double
selection cut in LRG. This feature is diluted when we consider volume
limited samples as we discard the less luminous galaxies $M_r>-21.5$
which contribute most to this peak. As shown above, the correlation $\xis$ for a volume
limited sample in the range $M_r$=[-22.5,-21.5], shown in Fig.\ref{fig:methodsdif}
is in excellent agreement with the one for the whole sample in
Fig.\ref{fig:compeisenstein}. Thus we conclude that our measurements
are quite robust and are not affected by the selection cuts.

\begin{table}\label{tab:slicesLRG}
\caption{Slices in the plane $M_r - z$}
\begin{center}

\begin{tabular}{|c|c|c|c|c|}
\hline
Sample     &     M-range    & Mean M & mean z   &   z-range\\
\hline
S22.50     &  -22.50 -23.00 &  -22.75 &  0.43   & 0.35-0.50\\
S22.25     &  -22.25 -22.75 &  -22.50 &  0.40  &  0.33-0.48\\
S22.00     &  -22.00 -22.50 &  -22.25 &  0.38  &  0.31-0.46\\
S21.75h    &  -21.75 -22.25 &  -22.00 &  0.35  &  0.27-0.42\\
S21.75     &  -21.75 -22.25 &  -22.00 &  0.20  &  0.12-0.27\\
S21.50h    &  -21.50 -22.00 &  -21.75 &  0.32  &  0.25-0.40\\
S21.50     &  -21.50 -22.00 &  -21.75 &  0.18  &  0.10-0.25\\
S21.25     &  -21.25 -21.75 &  -21.50 &  0.18  &  0.10-0.25\\
\hline
\end{tabular}
\end{center}
\end{table}

 We have not plotted larger scales for the 
other slices because they are quite noisy. However, at intermediate scales, we can see clearly 
the bias due to different intrinsic luminosity, more biased when more luminous, although bias
 seems to be independent on scale, in the scales used for our analysis.\\ 

\subsection{Wide angle approximation}\label{sec:wideangleeffect}

\begin{figure}
\centering{ \epsfysize=7.cm\epsfbox{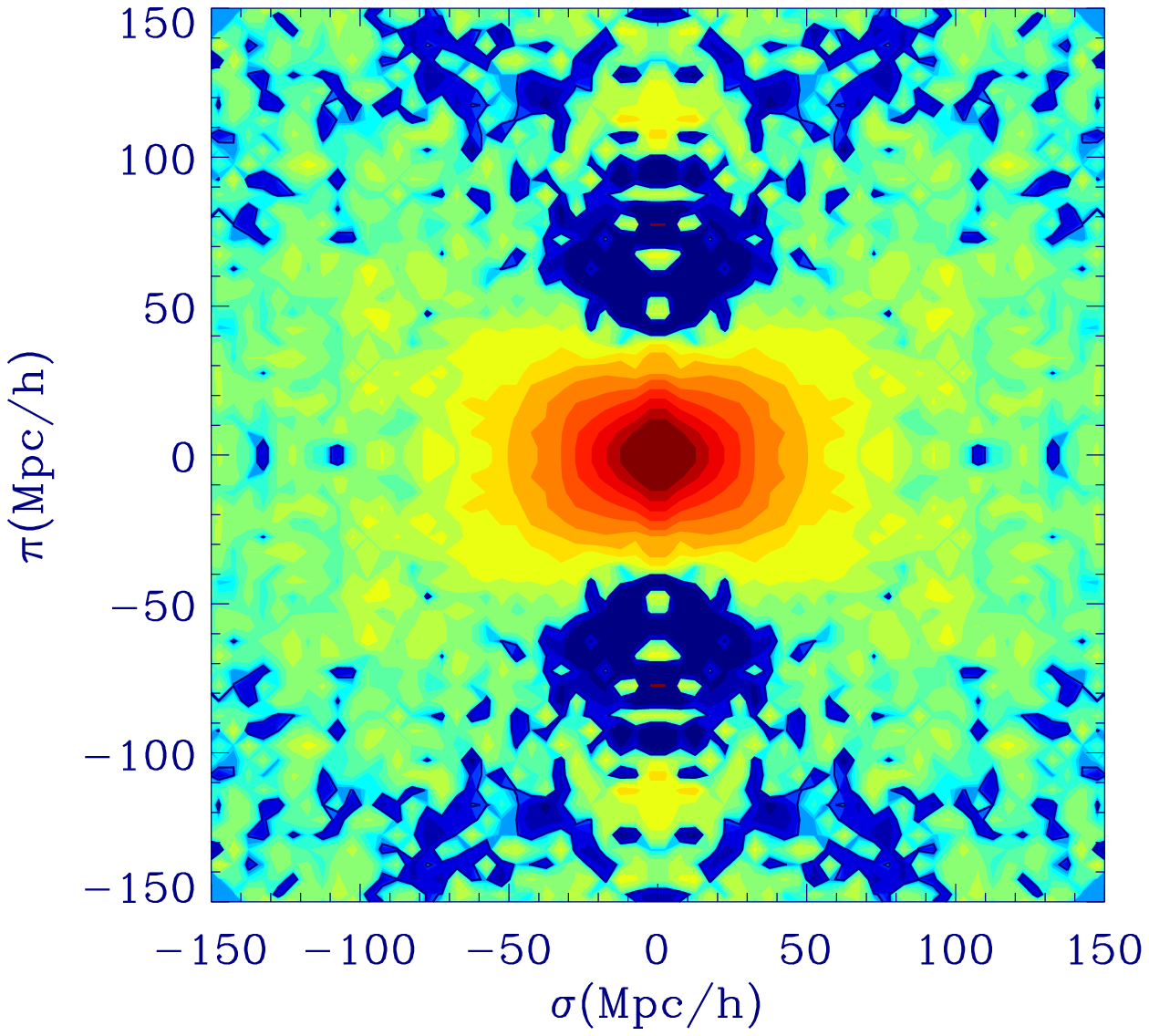}}
\centering{ \epsfysize=7.cm\epsfbox{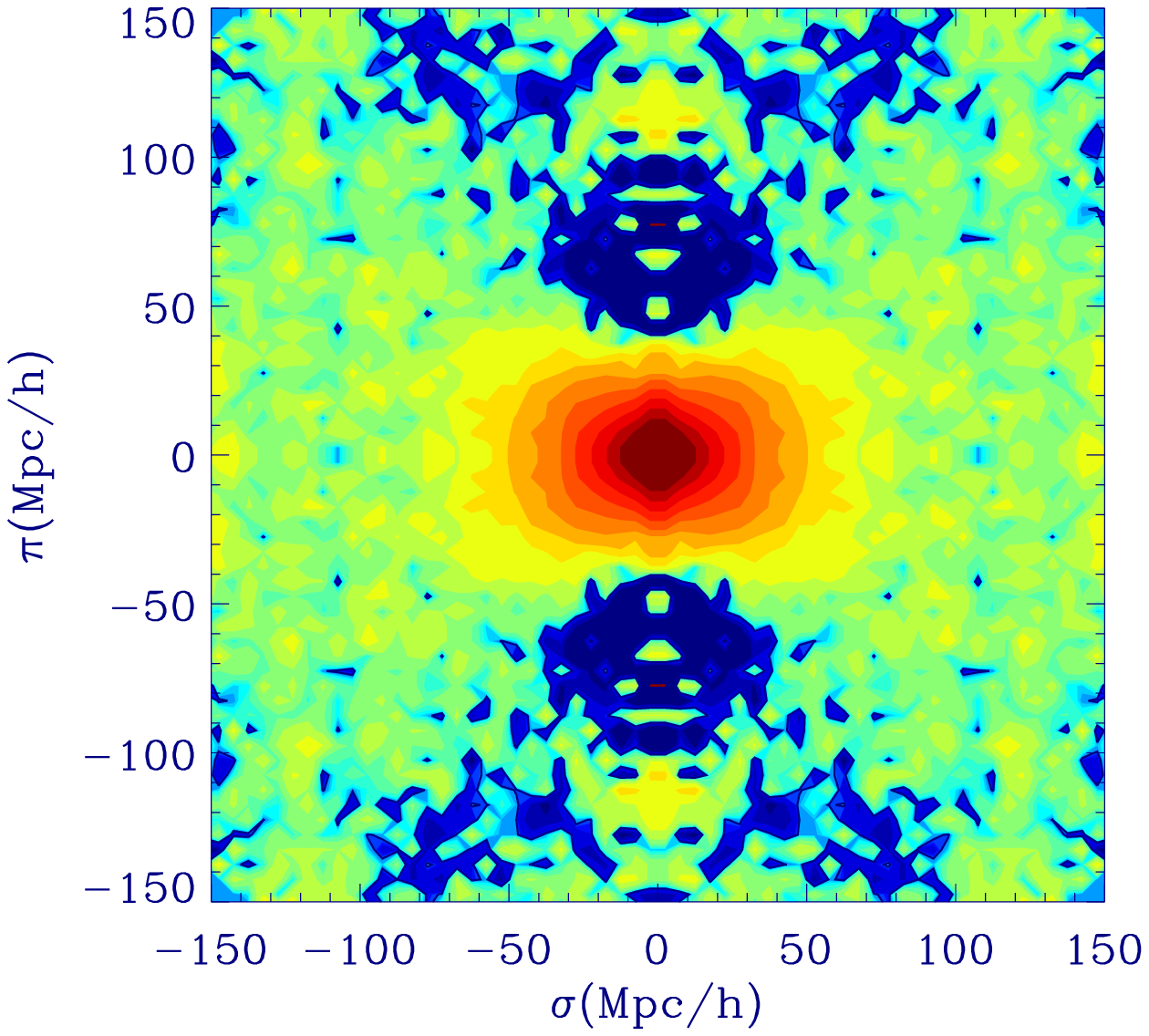}}
  \caption{ \label{fig:wideangleplot} Redshift-space correlation function $\xips$ including all the galaxy pairs (top panel)
and limiting the angle between galaxies to $\theta<10deg$ (bottom panel)} 
\end{figure}

We have calculated the correlation function limiting the angle between galaxies to see 
if wide angle effects that are apparent in  $\xips$ disappear. In Fig.\ref{fig:wideangleplot}, we see the anisotropic 
$\xips$ without limits in the angle (top panel) compare to the one
only accepting galaxies with $\theta<10deg$ 
(bottom panel). As we increase the restriction, we see
 how the $\sigma$ direction recovers power, which explains the lack of power we saw in the 
$\sigma$ direction when angles are too big to apply the distant observer approximation. 
The angle between galaxies explain part of the distortions due to wide angle effects, 
specially the ones that are concentrated at small $\pi$ and large $\sigma$, which affect 
the first slice considered, from z=0.15-0.3. The angle $\gamma_z$,  between the direction 
LOS (at $\theta/2$) and the vector which goes from galaxy 1 to galaxy 2 (following the 
notation used in 
%\cite{mmaatsuwide})
Matsubara 2000), is also important and can also imprint some 
modifications at larger $\pi$ and large $\sigma$.\\

We conclude that the measurements on the BAO scale are quite robust and
the extra power at the largest scales is probably the result of sampling
fluctuations and shot noise. For redshifts around z=0.35 we find extra noise
on large scales, probably due to an artifact in the selection function.
 In any case, none of the systematics we have
explored modify the peak detection in a significant way.

\end{document}